\documentclass[reprint,preprintnumbers,aps,prd,amsmath,amssymb,nobibnotes,nofootinbib,onecolumn]{revtex4-2}
\usepackage{xcolor}
\usepackage{subcaption}
\usepackage{amsmath}
\usepackage{amssymb}
\usepackage{graphicx}
\usepackage{hyperref}
\usepackage{slashed}
\usepackage[nolist]{acronym}
\usepackage{booktabs}
\usepackage{multirow}
\usepackage{rotating} 
\usepackage{csquotes}

\usepackage{tikz}
\usepackage[compat=1.1.0]{tikz-feynman}

\newcommand{\shat}{\hat{s}}

\newcommand{\U}[1]{\ensuremath{\mathrm{U}(#1)}}

\newcommand{\Zp}{Z^\prime}
\newcommand{\MZp}{M_{\Zp}}
\newcommand{\GZp}{\Gamma_{\Zp}}
\newcommand{\Up}{U(1)^\prime}
\newcommand{\SUL}{SU(2)_L}

\newcommand{\haithem}{\textsc{hAIthem}}

\usepackage{etoolbox}
\makeatletter
\newif\ifin@appendix
\let\orig@appendix\appendix
\renewcommand\appendix{%
  \addtocontents{toc}{\protect\in@appendixtrue}%
  \orig@appendix}
\makeatother
\makeatletter
\let\orig@contentsline\contentsline
\long\def\contentsline#1#2#3#4{%
  \ifstrequal{#1}{section}{\orig@contentsline{#1}{#2}{#3}{#4}}{%
  \ifstrequal{#1}{title}{\orig@contentsline{#1}{#2}{#3}{#4}}{}}}
\makeatother

\begin{document}

\title{Searching for BSM Experimental Signatures\\ with Large Lagrangian Models}

\author{Ibrahim Elsharkawy}
\affiliation{Department of Physics, University of Toronto and Vector Institute, Toronto, ON, Canada}
\affiliation{NERSC, Lawrence Berkeley National Laboratory, Berkeley, California, USA}
\author{Victoria Knapp-Perez} 
\affiliation{Department of Physics and Astronomy, University of California, Irvine, CA 92697}
\affiliation{Halluminate, San Francisco, California, USA, 94107}
\author{Wahid Bhimji}
\affiliation{NERSC, Lawrence Berkeley National Laboratory, Berkeley, California, USA}
\author{Aishik Ghosh}
\affiliation{Georgia Institute of Technology, Atlanta, GA 30332}
\affiliation{Lawrence Berkeley National Laboratory, Berkeley, CA 94720}

\begin{abstract}
The search for physics Beyond the Standard Model (BSM) is generally limited not by the supply of theory descriptions but by the lack of discriminating experimental observations. A case in point is dark matter, where the overwhelming gravitational evidence only goes so far in distinguishing between models within a vast theory space. Exploring the space of testable model \emph{signatures} may help identify overlooked experimental observables and indicate the utility of future experiments. A challenge is designing a search through model signatures outside what is found in the literature. Our primary contribution is \haithem{}, a framework that combines the self-guided exploration of reinforcement learning (RL) with the broad literature-derived knowledge of LLMs. We build an RL agent that learns to find which portions of a theory's high-dimensional parameter space are not excluded under some subset of constraints by playing a Battleship-style ``game'' against a suite of phenomenology tools. The agent is built as a \emph{Large Lagrangian Model} (LLaM), an autoregressive transformer that reads a tokenized Lagrangian, is pretrained at scale (here on $\sim 1$ billion tokens from $\sim 10{,}000$ Lagrangians), and is fine-tuned in a live environment.  The framework then constructs a decision tree that separates RL-found regions using observables computed with established tools, and passes the remaining degenerate regions to a set of LLM agents that compete to produce realistic signatures. In this proof of concept, RL-search outperforms an evolutionary-algorithm baseline, finding more viable regions with greater physical diversity. In a restricted space of single dark scalar multiplet models, we find that \haithem{} proposes interesting combinations of previously studied observables, such as the application of a halo-independent kinematic ratio to paleo-detectors.
\end{abstract}

\maketitle
\tableofcontents

\section{Introduction}
\label{sec:intro}

The current state of the search for beyond Standard Model (BSM) physics is arguably one of experimental drought and theoretical abundance. No new elementary particle has been confirmed since the Higgs in 2012. The LHC has set null limits across thousands of channels \cite{ATLAS:2024fdw,CMS:2024zqs}, dark matter direct and indirect detection experiments have yet to discover new physics \cite{Fermi-LAT:2015att,IceCube:2016dgk,LZ:2024zvo,XENON:2025vwd,HESS:2022ygk}, searches for proton decay \cite{Super-Kamiokande:2020wjk}, and neutrinoless double-beta decay \cite{KamLAND-Zen:2022tow} have yet to find a signal, and 
anomalies of the past decade (such as the muon anomalous magnetic moment \cite{Muong-2:2025xyk, Aliberti:2025beg}, the $W$-boson mass \cite{CDF:2022hxs,ATLAS:2024erm}, and the 750 GeV excess \cite{CMS:2016xbb,ATLAS:2016gzy}) seem to be approaching Standard Model values under improved measurements or revised theory calculations. 
Despite this, over the same period, the number of BSM proposals has grown significantly \cite{Bertone:2018krk}, and a new generation of agentic model-building will likely deepen this disparity. Recent work, such as AMBer~\cite{Baretz:2025zsv}, CDRL~\cite{Jha:2026qhs}, \textsc{FERMIACC}~\cite{Agrawal:2026lvg} and \textsc{ALBERT}~\cite{Alexander:2026lpw} have designed reinforcement learning (RL) or used language models empowered with deterministic tools to generate plausible BSM hypotheses at a rate no human can match. Thus, there is an ever growing need for feasible yet underexplored experimental signatures that may probe or exclude the surplus of models.

\begin{figure}
    \centering
    \includegraphics[width=.9\linewidth]{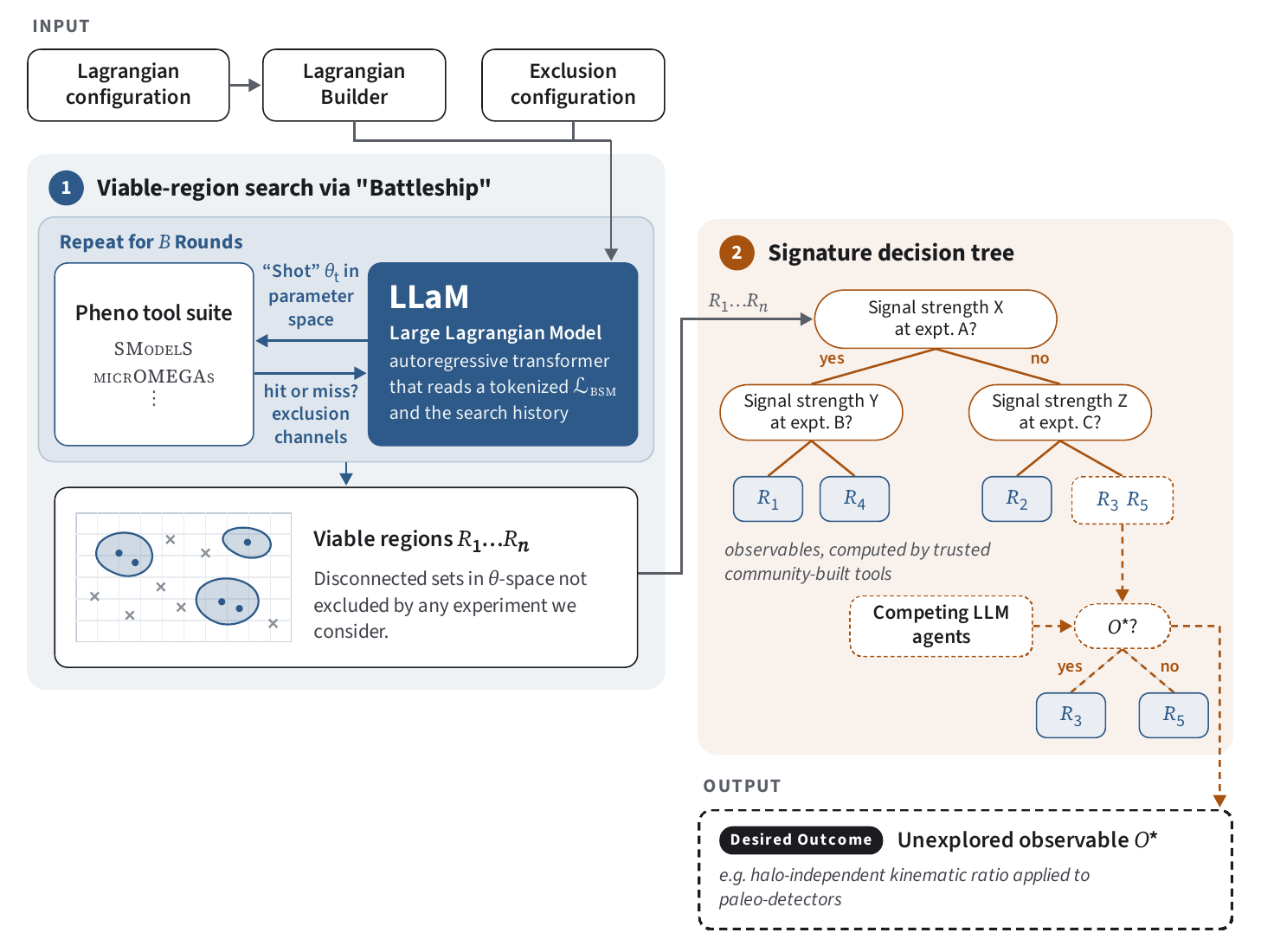}
    \caption{Illustration of the \haithem{} framework. Symbolic Lagrangians are built and fed to the Large Lagrangian Model along with an encoding of what exclusions we consider. The Large Lagrangian Model searches through parameter space over $B$ tunrs, where each round it probes parameter space $N$ times. The non-excluded regions it finds (non-excluded under our set up, not in nature) are then used to build a decision tree first with observables we compute with community built tools. Remaining degenerate nodes are fed to a set of LLM-agents who compete in producing a realistic signature.}
    \label{fig:haith-flow}
\end{figure}

Motivated by this challenge, we propose \haithem{} (\textbf{H}idden-sector \textbf{AI} for \textbf{T}estable \textbf{H}ypothesis \textbf{E}nu\textbf{M}eration), a framework backronymed after Ibn Al-Haytham (965-1040 CE), whose work cemented controlled experiment as the arbiter of competing theory \cite{Sabra:1989optics,Ibn-al-haythem-reflect}. We decompose the problem of searching for BSM signatures (for a given class of Lagrangians) into two subproblems, summarized in Figure \ref{fig:haith-flow}: 
\begin{enumerate}
    \item The search for viable parameter space regions (non-excluded in our setup, not necessarily in nature) for a given BSM model under some subset of constraints. 
    \item The construction of signature “decision trees” that attempt to distinguish between viable regions with observables computed with established pheno tools for yet-to-be-measured signatures, along with LLM-agent-proposed observables.
\end{enumerate}

Dark matter phenomenology is a prototypical example of an overabundance of viable BSM Lagrangians (Lagrangians that reproduce dark matter relic density $\Omega h^2 \sim 0.12$ \cite{Planck:2018vyg} and are not ruled out via direct, indirect, collider, or other experiments) and the lack of positive experimental results. We thus restrict our scope to a subset of dark matter model building as a tractable regime for methodology development.

The first of these sub-problems, mapping the viable region of a given Lagrangian, is near intractable for general models with many parameters. Viable regions are sparse, often disconnected, irregular sub-manifolds of high-dimensional parameter spaces. Naive grid scans scale exponentially in the number of parameters, and random sampling wastes the bulk of evaluations on excluded territory. Genetic algorithms such as \textsc{Gambit}'s differential-evolution-based \cite{Storn:1997uea} \textsc{Diver} have shown promise, outperforming nested sampling and MCMC in ten or more dimensions \cite{GAMBIT:2017yxo,GAMBITDarkMatterWorkgroup:2017fax,Martinez:2017lzg}. This success, however, also illustrates the broader challenges facing scans of high-dimensional BSM parameter spaces. Hyperparameters of the search algorithm need to be optimized for each Lagrangian being scanned \cite{Martinez:2017lzg}. Further, published applications have reached Lagrangian parameter dimension $d=18$, requiring $\mathcal{O}(10,000)$ parameter space probes and repeated scans to ensure proper coverage~\cite{Martinez:2017lzg, Chrzaszcz:2019inj}, a cost expected to grow quickly at higher dimensions.

A key insight is that the search may be more efficient if context-aware, while being treated sequentially with long-term memory. This is precisely the structure for which meta-Reinforcement Learning (RL) is well suited ~\cite{Sutton:2018rl,duan2016rl2fastreinforcementlearning,wang2017learningreinforcementlearn}. One agent is trained to return a policy over actions across a distribution of different environments (for us, new Lagrangians and sets of experimental exclusions), conditioned on those environments and with the ability to generalize to new environments. We thus cast the search as a single-player Battleship. At each step the agent fires a shot, a candidate parameter point, into the parameter space of a given Lagrangian, and trusted phenomenology  tools return a deterministic hit verdict along with the channels of exclusion. The ``ships'' (connected sets of points on the non-excluded manifold) are unknown in number, shape, and location, and the agent is rewarded for finding them efficiently and thoroughly. Each BSM model is a new board, and the agent builds experience finding viable regions by playing many games on such boards. Critically, we treat non-exclusion as a binary criterion rather than a continuous likelihood, which is sufficient for our decision tree needs.

\begin{figure}
    \centering
    \includegraphics[width=\linewidth]{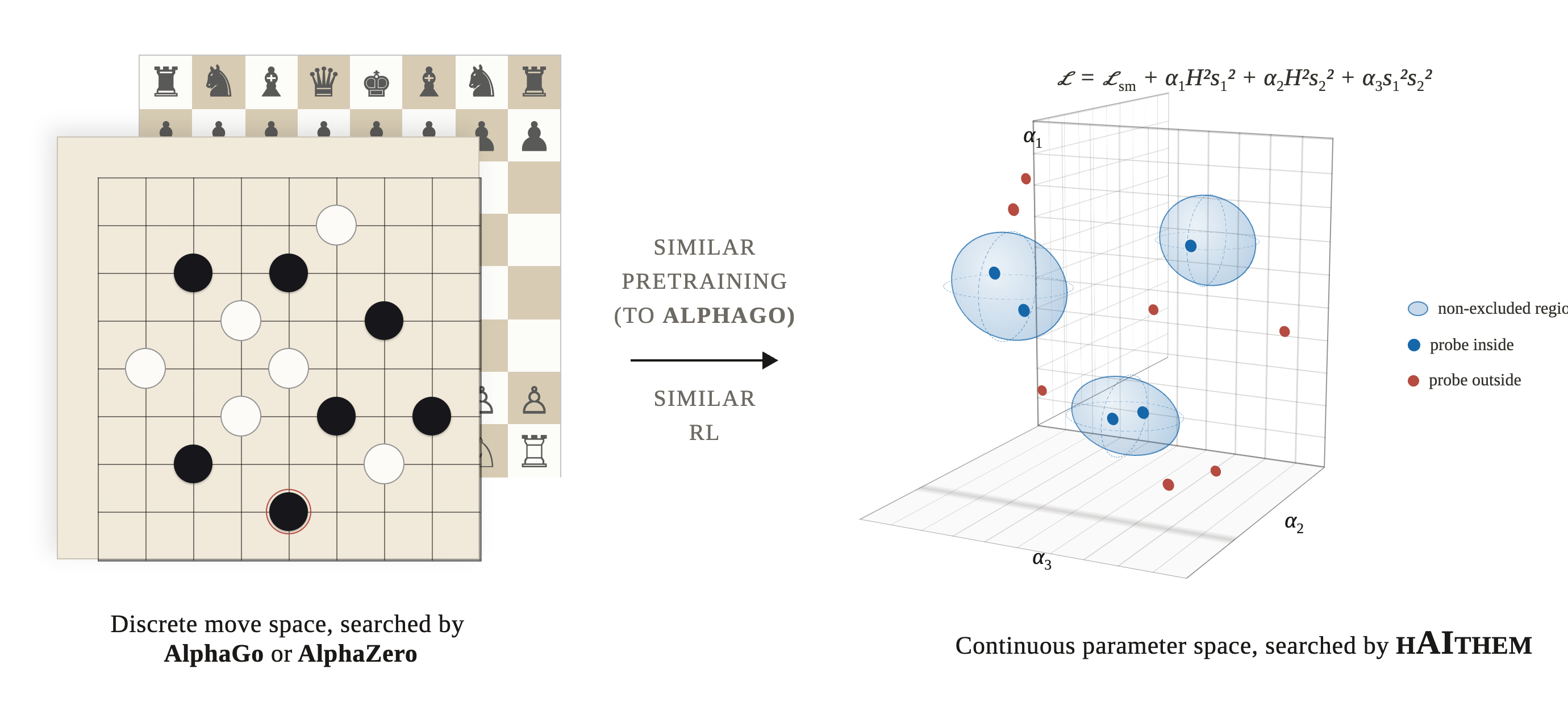}
    \caption{The similarities between \textsc{AlphaGo} and \textsc{AlphaZero} (left) and \haithem{} (right). Both search optimal moves conditioned on the current play turn and past actions. \textsc{AlphaGo} and \haithem{} are similar in being pretrained at scale and fine-tuned with RL and, like \textsc{AlphaZero}, \haithem{} is not trained on human-generated data. Major differences include the non-discrete nature of the parameter space and the lack of a competitor.}
    \label{fig:haith-alpha}
\end{figure}

For a given environment, this is similar in structure to \textsc{AlphaGo} and \textsc{AlphaZero} \cite{silvergo,SilverGo2,doi:10.1126/science.aar6404}. We have a space too large for systematic search, made tractable by an RL agent trained with a physically grounded “game” engine, see Figure \ref{fig:haith-alpha}. We thus find the most effective method to build such an agent is to adopt the paradigm that defines modern large language models and \textsc{AlphaGo}. That is, large-scale pretraining on an offline dataset, followed by Reinforcement Learning fine-tuning in a live environment. We refer to the resulting class of models as \textbf{L}arge \textbf{La}grangian \textbf{M}odels (\textsc{LLaM}s) \footnote{Unfortunately, the space of acronyms is tight. An LLaM is not an LLM nor LLaMA \cite{touvron2023llamaopenefficientfoundation}. The LLaM also shares a qualitative spirit with \cite{Koay:2025bmu}.}. \haithem{} contains the first realization of an LLaM. 

The second subproblem's decision tree formulation (see Figure \ref{fig:haith-flow}) coupled to the LLaM is built to encourage the search for signatures beyond the literature. Each node in the tree is an observable with a binary outcome that splits LLaM-found regions (for example, whether we expect an observation for a given experiment), and the leaves of the tree correspond to parameter-space regions that share an experimental signature defined by the set of nodes that comprise the path to the leaf. We first construct the tree with observables computed with trusted tools, depth-ordering observables by their ability to split the set of regions we consider. In this construction, degenerate leaves are leaves with more than one parameter-space region sharing every experimental signature we compute with trusted phenomenology tools. These leaves are then handed to a set of LLM agents that attempt to propose realistic observables that continue the splitting. 

Once specific Lagrangians and parameter values (eg., masses, couplings) are fixed, the tree construction turns the problem of breaking these degeneracies into a concrete task rather than an open-ended query for an LLM. The LLM agents must compete to give realistic signatures along with sharp binary verdicts as to how degenerate regions differ in their proposal. Additionally, they are forced to think beyond observables already computed earlier in the tree (reuse is explicitly prohibited) and are allowed null answers (no signature found). We further find that the decision tree formulation leads to a natural grouping of Lagrangian-viable regions into experimentally distinguishable classes, similar to the inverse map of supersymmetry at the LHC proposed in \cite{Arkani-Hamed:2005qjb}.

The paper is organized as follows. Section \ref{sec:problem} formalizes our Large Lagrangian Model setup, including the dark matter search space we consider, the experimental constraints that define viability, and describes our automation of symmetries to symbolic Lagrangians to required observables. We also describe our model architecture, how we encode and condition on a DM Lagrangian and the search history, and the Reinforcement Learning and pretraining process. Section \ref{sec:DecisionTree} then develops the decision-tree methodology that converts viable-region maps into decision-tree splits by deterministic or LLM-proposed signatures. Section~\ref{sec:results} benchmarks the LLaM and then builds a decision tree for the set of all single-scalar multiplet dark matter Lagrangians our space allows, studying the LLM-agent proposals. Section \ref{sec:conclusion} concludes with a summary of outcomes and the outlook on \haithem{}.

\section{Searching Parameter Space with Large Lagrangian Models}
\label{sec:problem}

To search signature space, we first search model parameter space for non-excluded regions and then use our decision tree construction. We treat the problem of searching for these non-excluded regions of Lagrangian parameter space as a single-player game of Battleship played by an RL agent over some subset of possible dark matter Lagrangians and their parameters. In training, the RL agent sees a new ``board'' that corresponds to a combination of:
 \begin{enumerate}
     \item A randomly sampled Lagrangian,
     \item Randomly sampled viability cuts (which experimental constraints to consider),
     \item Randomly sampled parameter ranges (where in parameter space is the agent allowed to look),
     \item Randomly sampled budget size (how many times is the agent allowed to probe).
 \end{enumerate}
 The agent then ``plays'' the \emph{episode} by proposing a sequence of parameter points in the continuous \emph{gauge} basis parameter space of that  Lagrangian (eg. masses, couplings). The ``ships'' the agent seeks are regions of this space that are \emph{viable}, i.e. non-excluded by experimental constraints that are active for a given episode.
 
To develop the \haithem{} methodology, we restrict our space of dark matter Lagrangians to minimal additions. That is, Lagrangians built by adding to the SM (potentially many) dark matter electroweak multiplets stabilized by a discrete symmetry and optionally coupled to a dark  $U(1)'$ gauge sector. A candidate can be a real or complex scalar field, a Dirac or a Majorana fermion,
and can be a singlet, doublet, or triplet of $\mathrm{SU}(2)_L$. We consider all expected renormalizable terms, and additionally, for fermions, whose $\mathbb{Z}_n$-charged singlets would have no renormalizable coupling to the SM, we include the dimension-five Higgs portal $H^\dagger H\bar\chi\chi$ and a dimension-six flavor-diagonal four-fermion term, as in \cite{Fedderke:2014wda,Lopez-Honorez:2012tov}, the minimal set where all singlet candidates have a portal. This set up ensures a portal with the SM Higgs, and additional portals are generated via the $\mathrm{SU}(2)_L$ charge, and the dark  $U(1)'$ kinetic mixing. 

At the start of each episode, one Lagrangian is sampled at random from this space, which the agent is handed and tasked with sampling its continuous parameters. A single model is specified by the following discrete choices:

\begin{enumerate}
\item Number of DM fields $n_{\rm DM}\in\{1,2,3,4,5\}$.
\item Per-field quantum numbers:
\begin{enumerate}
    \item Lorentz spin $\in\{\text{scalar, Majorana, Dirac}\}$
    \item $\mathrm{SU}(2)_L$ rep $\in\{\text{singlet, doublet, triplet}\}$
    \item hypercharge $Y\in\{0,\tfrac{1}{2}\}$
    \item if a (scalar) field is complex or not $\in\{\text{real, complex}\}$
    \item Number of copies (akin to generations in the SM) $\in\{1,2,3\}$. We restrict what coupling parameters we consider for copies by enforcing a permutation symmetry. 
\end{enumerate}

\item Discrete stabilizing symmetry $\mathbb{Z}_n$ with $n\in\{2,3,4,5\}$, under which each DM particle carries a non-zero charge and the SM is neutral\footnote{Different copies can have different $\mathbb{Z}_n$ charge, and the two components of a complex scalar field can have different $\mathbb{Z}_n$ charge. This results in two variants of the complex scalar models, one with a secluded component and one without. }.
\item Whether or not a model has an $\mathrm{U}(1)'$ dark gauge sector with a massive $Z'$ mediator (with a mass generated through a Stückelberg-like term $\tfrac12 M_{Z'}^2 Z'_\mu Z'^\mu$ \cite{Ruegg:2003ps,Kors:2004dx}). 
\begin{enumerate}
\item If there is a  $\mathrm{U}(1)'$ dark gauge symmetry, each dark field receives a random dark charge $Q'\in\{-1,0,+1\}$.
\end{enumerate}
 When $\mathrm{U}(1)'$ is active, the mediator mass $M_{Z'}$, the gauge coupling $g_{Z'}$, and a kinetic mixing parameter $\epsilon$, are taken as additional parameters for the agent to act on.
\end{enumerate}
Crucially, the agent sees all Lagrangians and proposes parameter points (masses, couplings, kinetic mixing, etc) in the \emph{gauge basis}, before electroweak symmetry breaking (EWSB). The component expansion of $\mathrm{SU}(2)_L$ multiplets, the canonical normalization of the kinetic mixing operator if the dark $U(1)'$ is active and the diagonalization of mass matrices to post EWSB-mass-basis states occur downstream, see Appendix \ref{app:massmix}. The agent therefore learns to act on the (smaller and simpler) gauge-invariant content of the theory rather than over the full post-EWSB mass spectrum, given type-dependent parameter ranges, see Appendix \ref{app:physsetting}.

Naive combinatorics quantifies the space to be of order $\mathcal{O}(10^{30})$ Lagrangians\footnote{The number of phenomenologically distinct Lagrangians is much lower.} (and thus even more boards), and contains many common dark matter models. For example, the real singlet scalar (Higgs-portal dark matter), the singlet Majorana fermion, and the $Y=\tfrac12$ doublet and $Y=0$ triplet \cite{Cirelli:2005uq, Cirelli:2007xd}. $\mathbb{Z}_n$ with $n\ge 3$, multiple copies of the same field, or the addition of the dark $\mathrm{U}(1)'$, extend our space into more complex multi-component models. We choose this space as a tractable yet sufficiently complex playground to develop the method. A natural next step would be to scale this approach by incorporating a broader set of dark matter models or other BSM models with more parameters.

\subsection{Considered Experimental Constraints}
\label{sec:ExperimentalConstraints}
We say a parameter point is \emph{viable} if it survives the (active) set of constraints in Table \ref{tab:viability} (for each training episode, the active set of constraints are randomly sampled). The choice of constraints is a result of our use of \textsc{micrOMEGAs} \cite{Alguero:2023zol} for cosmological and direct-detection probes and \textsc{SModelS} \cite{Alguero:2020grj} for simple LHC bounds. It is crucial to note we do not consider all experimental constraints that probe our model space in this proof of concept. Thus, a point that is viable here is not necessarily viable in nature. Additional phenomenology software can be added as the corresponding computational tools mature and our search space expands. 

Each active constraint is treated as a hard exclusion. A parameter point that
violates any \emph{active} cut is non-viable. The set of
active constraints is itself part of the episode configuration and is fed to the model. The
relic-density cut is always enforced with an episode-dependent allowable error width governed by randomly sampled $\tau\in[1,50]$. The correspondence to allowed relic abundance is found in Appendix \ref{app:physsetting}.

\begin{table}
\centering
\caption{Observables that define viability. The relic-density cut is always active with an episode-dependent allowable error given by $\tau$. Each remaining constraint is independently active with probability 1/2 per episode, with the resulting mask provided to the model.}
\label{tab:viability}
\begin{tabular}{lllll}
\hline
 Probe & Observable & Cut  & Package \\
\hline
 Relic density      & $\Omega h^2$          & $\tau=1\longrightarrow \Omega h^2< 0.118 $ or $\Omega h^2 > 0.126$ \cite{Planck:2018vyg,Alguero:2023zol} & \textsc{micrOMEGAs} \\
 Direct detection       & $\sigma_{\rm SI},\sigma^p_{\rm SD}$ & $r>1$ at 90\% CL~\cite{Alguero:2023zol} & \textsc{micrOMEGAs} \\
 Invisible Higgs             & $\mathrm{BR}(h\to\mathrm{inv})$ & $>0.11$~\cite{ATLAS:2023tkt,ATLAS:2020kdi} & \textsc{micrOMEGAs} \\
 Fermi-LAT           & $\langle\sigma v\rangle_{c^\star}$ & annihilation-channel and $m_{\rm DM}$ dependent bound  \cite{Fermi-LAT:2015att,Alguero:2023zol} & \textsc{micrOMEGAs} \\
 IceCube       & $\mu_{\rm solar}$   & conservative $b\bar b$ bound as a function of $m_{\rm DM}$ \cite{IceCube:2025fcu,IceCube:2016dgk,Alguero:2023zol} & \textsc{micrOMEGAs} \\
 LHC pair production    & $r_{\rm max}^{\rm LHC}$ & $r_{\rm max}^{\rm LHC}>1$ & \textsc{SModelS} \\
\hline
\end{tabular}
\end{table}

\subsection{Automating Lagrangian Generation and Observable Computation}
\label{sec:pipeline}
\begin{figure}[t]
    \centering
    \includegraphics[width=1\linewidth]{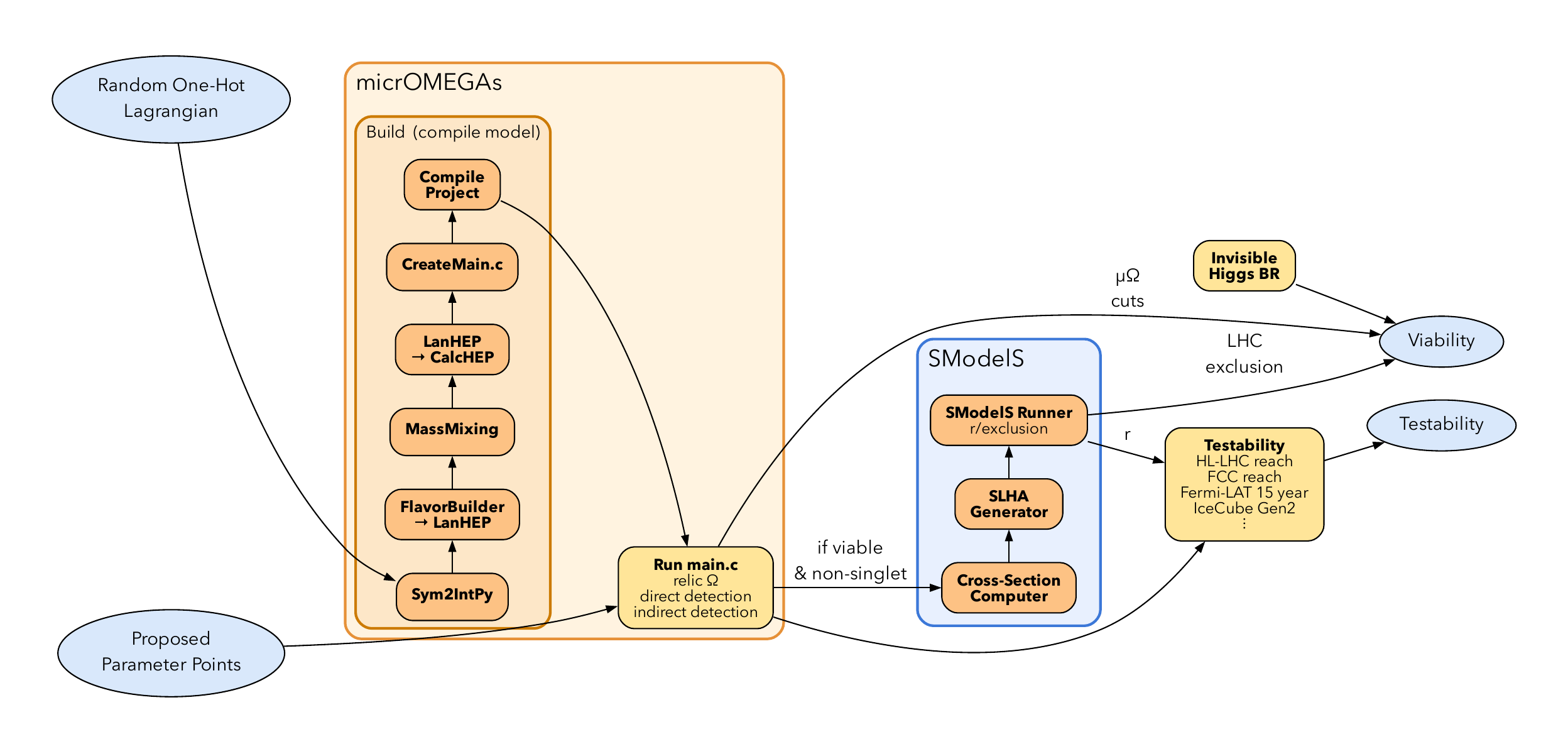}
    \caption{Viability Computation Pipeline}
    \label{fig:pipeline-fig}
\end{figure}
Running our RL agent on a candidate dark-matter model requires an automated forward map from a discrete Lagrangian description (i.e., choices of gauge and stabilizing symmetries and field content defined in Section \ref{sec:problem}) to a symbolic Lagrangian with all allowed terms and proper SM couplings, and another forward map from that Lagrangian, to the physical observables (given point in its continuous parameter space in the gauge basis teh RL agent outputs). This required significant software engineering to facilitate the interactions between pre-existing and custom-built packages and to ensure computation can scale to any number of nodes on a supercomputer.

We implement this map as illustrated in Figure \ref{fig:pipeline-fig}. The first stage turns the symmetry and field choices into a symbolic Lagrangian with a custom Mathematica-to-Python converted package \textsc{Sym2IntPy} \cite{Fonseca:2017lem} and a custom code \textsc{FlavorBuilder} first used in \cite{Baretz:2025zsv}. The second turns that Lagrangian into a built executable and, at each sampled parameter point, computes the relic abundance, direct- and indirect-detection signals, collider constraints, and projected future-experiment reach with \textsc{micrOMEGAs}, \textsc{SModelS}, and custom code, turning them into the per-point viability and testability scores that constitute the agent's reward signal. See Appendix \ref{app:comp-pipeline} for more details. In this proof of concept, we do not include Sommerfeld enhancements or NLO corrections.

\subsection{The Large Lagrangian Model}

\label{sec:Architecture}
\begin{figure}[t]
    \centering
    \includegraphics[width=\linewidth]{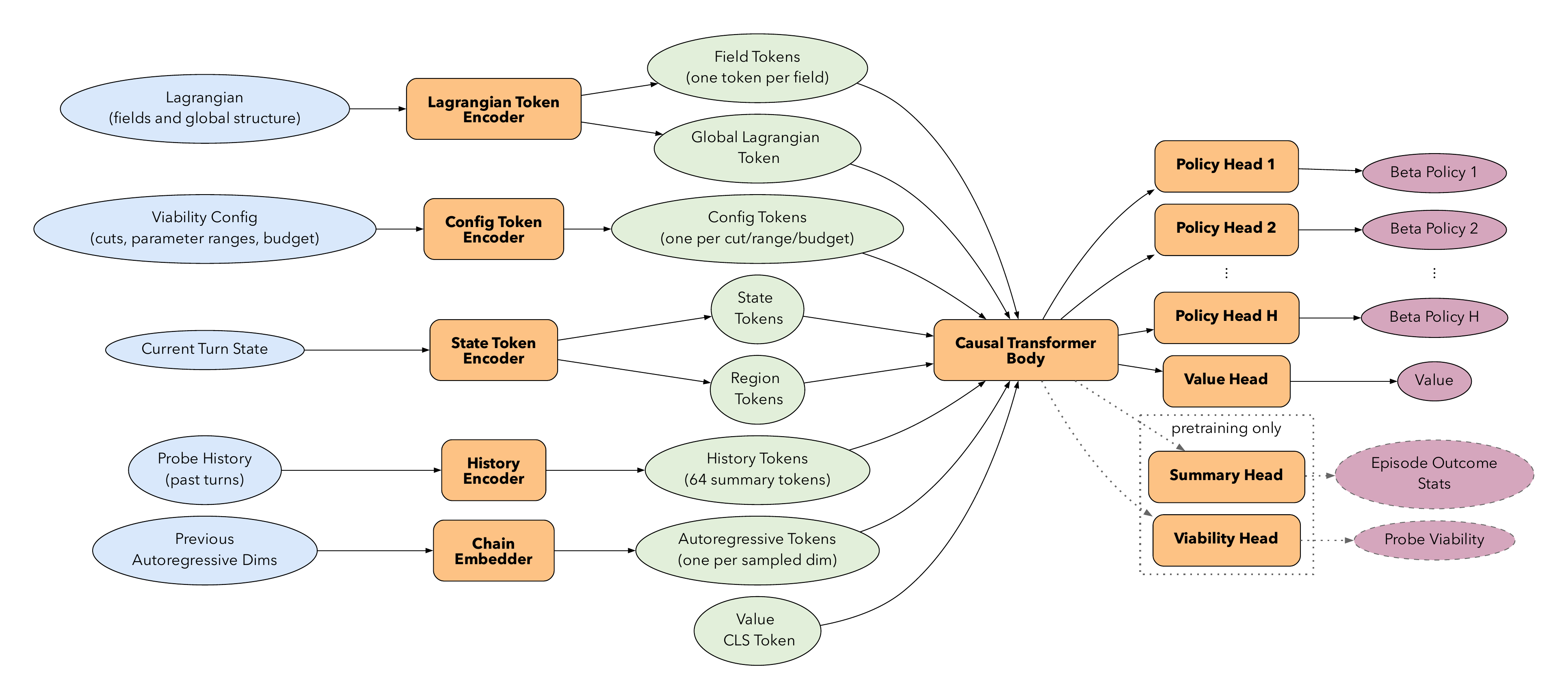}
    \caption{The Large Lagrangian Model. The Lagrangian, viability
    configuration and search history are each encoded into tokens
    (green) that a shared transformer body processes jointly. $H$
    policy heads read the autoregressive tokens, each outputting its
    own per-dimension Beta conditionals. The value head reads a
    learned token to estimate the per-turn value. Two auxiliary heads are added and used during pretraining.}
    \label{fig:model_arch}
\end{figure}
\begin{figure}[t]
    \centering
    \includegraphics[width=\linewidth]{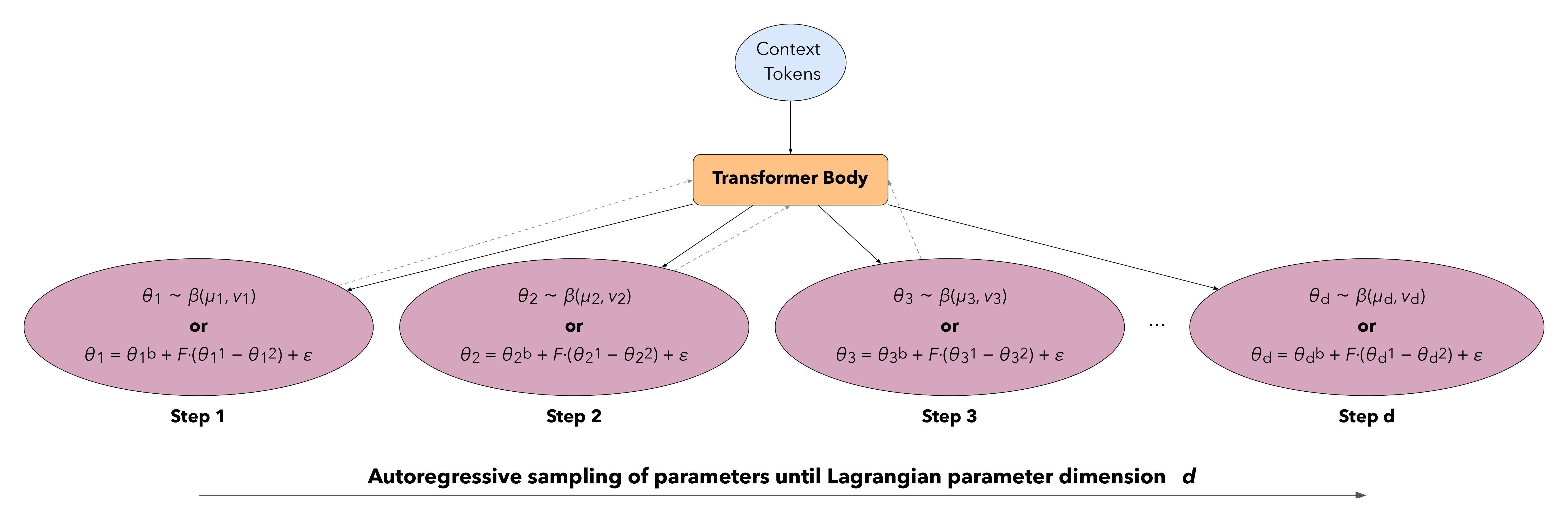}
    \caption{A depiction of autoregressive sampling. Each parameter is drawn from a model forward pass and fed back as context for the next forward pass for the next parameter. The model chooses the beta distribution or differential evolution style policy once per probe and fixes the choice for all parameter dimensions.}
    \label{fig:ar_samp}
\end{figure}

The LLaM is constructed with a transformer body attached to various input embedding modules and output heads (See Figure \ref{fig:model_arch}). It is given three types of input:
\begin{enumerate}
    \item The Lagrangian, fixed for the RL episode,
    \item The episode's \emph{viability configuration} (the active experimental
    cuts, scan ranges, turn budget, and relic-density cut $\tau$),
    \item A dynamic encoding of the search history and search state.
\end{enumerate}
Every input is tokenized after being sliced into pieces and embedded with a small two-layer MLP. After which, all tokens interact through a shared attention body (see Figure ~\ref{fig:model_arch}). The model then outputs an estimate for the turn's reward through the value head, and outputs $H=4$ distributions over Lagrangian parameter space through the policy heads. It does so autoregressively, parameter by parameter, see Figure \ref{fig:ar_samp}. We define two model sizes and give details in Table \ref{tab:model_sizes} and we describe training and model hyperparameters in more detail in Appendix \ref{app:modeltraininghparams}. 

\begin{table}[h]
\centering
\begin{tabular}{l c c}
\toprule
 & small & medium \\
\midrule
Hidden width $d$                  & 256  & 512  \\
Attention Blocks       & 4    & 12   \\
Attention Heads & 8    & 16   \\
Total parameters                 & $\sim 4.8$M & $\sim 44.0$M \\
\bottomrule
\end{tabular}
\caption{Two model-size configurations. We describe training and model hyperparameters in Appendix \ref{app:modeltraininghparams}.}
\label{tab:model_sizes}
\end{table}
\begin{figure}
    \centering
    \includegraphics[width=\linewidth]{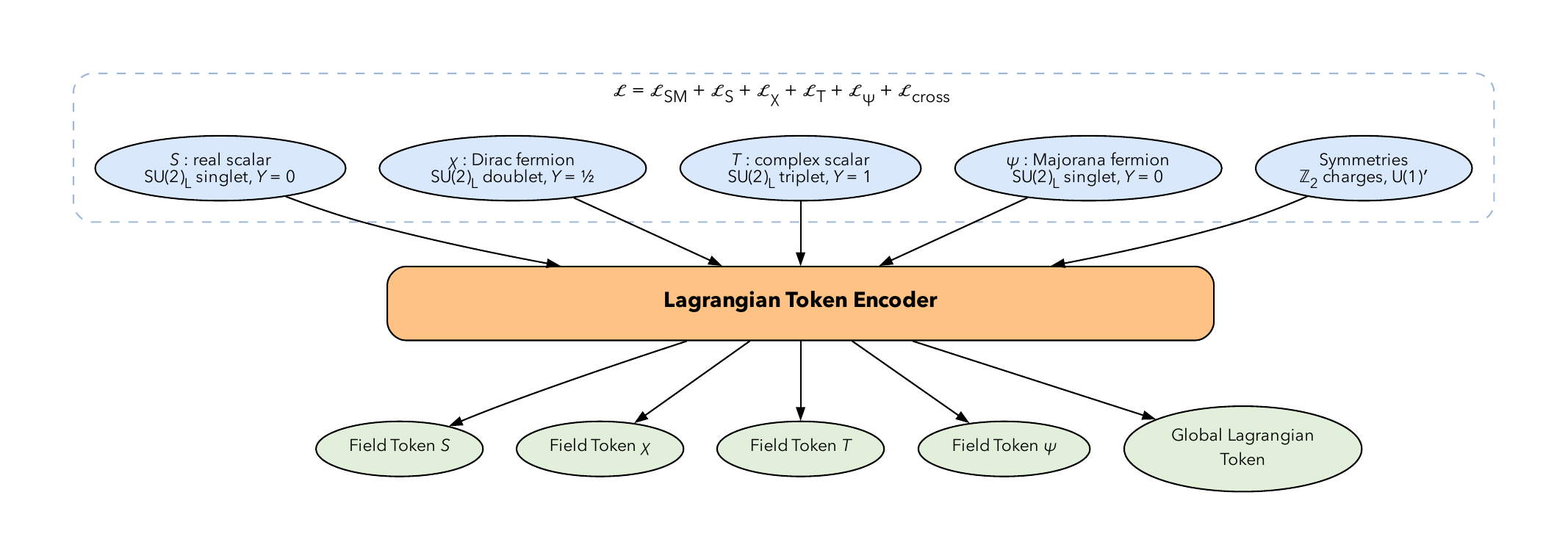}
    \caption{Tokenization of an example Lagrangian composed of a real scalar, Dirac fermion, a complex scalar, and a Majorana fermion. Each dark-sector field is embedded into its own field token from its gauge-eigenbasis quantum numbers, while the symmetry content and per-field summaries
    (dashed) are embedded into a single global Lagrangian token.
    $\mathcal{L}_{\rm SM}$ is fixed across all models and never encoded.}
    \label{fig:Lagrangian_tokens}
\end{figure}
\paragraph{Autoregressive Beta policy and the Differential-Evolution Policy Path.}

The model comprises $H=4$ policy heads and each policy head outputs a distribution over the Lagrangian's $d$ continuous parameters, normalized to the unit cube $[0,1]^d$. Given that we need a distribution that lives in $[0,1]^d$, we choose a Beta distribution in mean-concentration form. That is, each head for each parameter outputs a $(m,\nu)$ such that each probe $x$ is drawn from,
\begin{equation}
x \sim \mathrm{Beta}(\alpha,\beta), \qquad
\alpha = m\nu, \quad \beta = (1-m)\nu .
\label{eq:betaparam}
\end{equation}
The $(m,\nu)$ parameterization cleanly separates \emph{where} a head is looking from \emph{how sharply} it looks. Independent per-dimension Betas cannot represent parameter-parameter correlations, which is why we sample the $d$ parameters \emph{autoregressively}. Dimensions are visited in a fixed
order (field masses first, couplings, then kinetic mixing). See Figure \ref{fig:ar_samp} for a diagram depicting the autoregressive sampling. 

In addition to the Beta Policy, and motivated by the efficacy of differential evolution (DE) parameter space searchers \cite{GAMBIT:2017yxo,GAMBITDarkMatterWorkgroup:2017fax,Martinez:2017lzg}, each head contains a second output path alongside the autoregressive beta policy. That is, for each probe, the \emph{model chooses} either to output a beta distribution \emph{or} to perform differential-style evolution on previously sampled probes. If differential-style evolution is chosen, the model chooses a base probe in one parameter dimension $\theta_i^b$ from previously found \emph{viable} probes, and a pair of other not necessarily viable previous probes $\theta_i^1$ and $\theta_i^2$. With a probability of $c=0.7$, $\theta_i^b$ is then mutated by,
\begin{equation}
    \theta_i=\theta_i^b+F\cdot(\theta_i^1-\theta_i^2) + \epsilon \quad \rm   or \quad \theta_i = \theta_i^b
\end{equation}
where $F\in[.5,.9]$ is some random scalar and where $\epsilon\in[-0.02, 0.02]$ is a small jitter. The process described is nearly identical to standard differential evolution, with the major addition of \emph{learned} rather than random choices for $\theta_i^b,\theta_i^1$ and $\theta_i^2$. 

\subsection{Reinforcement Learning in a Live Environment}
\label{sec:RL-Finetuning}
\begin{figure}
    \centering
    \includegraphics[width=\linewidth]{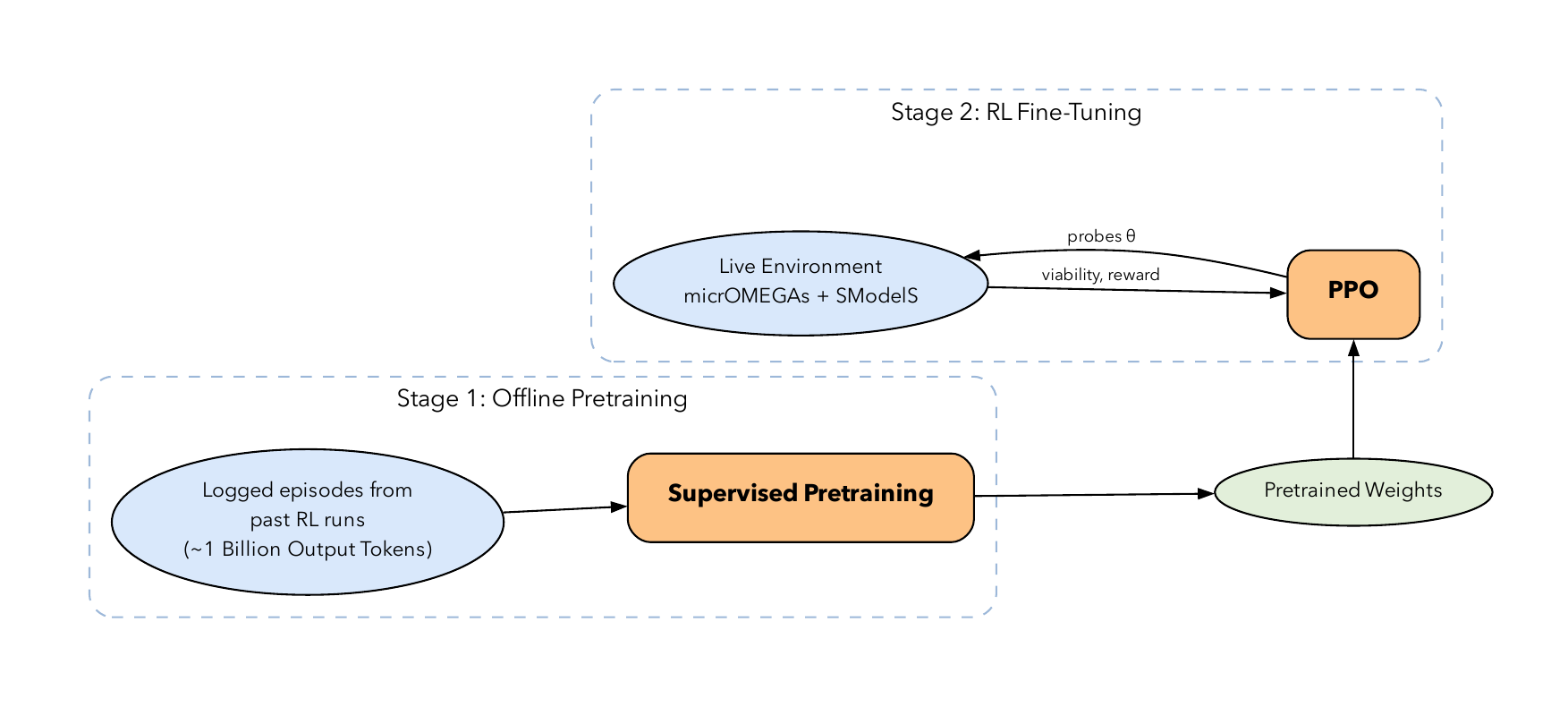}
    \caption{The two-stage training workflow. The policy is first
    pretrained offline (Section ~\ref{sec:pretraining}) on $\sim$10,000
    distinct Lagrangians and $\mathcal{O}(1B)$ output tokens from earlier RL runs, using the supervised objectives of Equation \eqref{eq:pretrain_total}. The resulting weights warm-start PPO fine-tuning (Section ~\ref{sec:RL-Finetuning}), which trains on the live \textsc{micrOMEGAs} and \textsc{SModelS} environment.}
    \label{fig:pretrain_workflow}
\end{figure}
We use an asynchronous implementation of Proximal Policy Optimization, APPO, \cite{DBLP:journals/corr/SchulmanWDRK17}. We find in practice that training only with PPO in a live environment is insufficient and computationally ineffective in fully train a Large Lagrangian Model, especially as we scale model size. To remedy this, we do offline pretraining on a dataset composed of O(1B) output tokens and 10,000 distinct Lagrangians detailed in Section \ref{sec:pretraining} before our live RL training. As in \cite{Elsharkawy:2026mev,silvergo,SilverGo2,Sutton:2018rl}, the pretraining task is a perturbation of the live RL training, and the dataset is composed in part with many probes computed during previous RL training runs. Hence, it is appropriate to first describe our setup for reinforcement learning in a live environment, as it drives both. See Figure \ref{fig:pretrain_workflow} for a diagram of our training workflow. 

The reward is designed to teach the agent to map viable regions thoroughly rather than any single dense region. Each probe earns the sum of four reward terms. A \emph{viability} bonus $R_V$ rewards any allowed point. A \emph{boundary} bonus $R_B$ rewards probes that sit at the viable/non-viable boundary\footnote{Detected when a probe has a near neighbor of the opposite viability within a small (scaled by parameter-space dimension) parameter-space radius.}. We include this to encourage the boundary where a model's experimental limits are sharply defined. A \emph{novelty} bonus rewards probes whose observable signature measured over the continuous components of a point's signature is far away from all other sampled points\footnote{The signature is taken from \textsc{micrOMEGAs} and includes normalized relic density, muon flux, gamma ray spectrum, etc. Distance is then computed in standardized signature space given by $\lVert \hat{z}_t-\hat{z}_s\rVert^2,~
\hat{z}_t = \frac{\sigma^{\text{obs}}_t - \mu}{\sigma}
$ and the minimum distance over all previous $\hat z_s$ is used for the reward given the turns $\hat{z}_t$.}. Finally, a \emph{testability} bonus $R_T$ rewards viable probes that are in principle testable by experiments given in Appendix \ref{app:future-experiments}. The per-probe reward is the sum,
\begin{equation}
R = w_VR_{\rm viable} + w_BR_{\rm boundary} + w_NR_{\rm novelty} + w_TR_{\rm test}, \qquad (w_V, w_B, w_N, w_T) = (3.0,0.1,0.5,0.2),
\end{equation}
and the turn reward $r_t$ is the sum of $R$ over all $N$ probes.

We use this reward to train the value and policy head via the standard PPO clip and value loss \cite{DBLP:journals/corr/SchulmanWDRK17}. Crucially, both rewards and value are computed each turn and not aggregated over the episode. Thus, the value head outputs every turn rather than assigning a single end-of-episode return to every action. This matters because the policy must change significantly within an episode, and this must be learned independently. An early-turn broad policy must not be punished for a poor late-turn policy (or vice versa). 

\subsection{Offline Pretraining}
\label{sec:pretraining}

Due to our computational pipeline illustrated in Figure \ref{fig:pipeline-fig} taking up to a few seconds per parameter evaluation, training the model from scratch with APPO alone is intractably slow as model size scales. We address this the way modern language models do. We use a cheap supervised pretraining stage over a large offline dataset, followed by RL
fine-tuning \cite{brown2020languagemodelsfewshotlearners,ouyang2022traininglanguagemodelsfollow}. Just as an LLM is pretrained by next-token prediction on a large amount of imperfect text, and RL is reserved for the final alignment of behavior, we pretrain with next-token prediction on a large corpus of augmented episode logs of many earlier RL training runs. This is also similar in spirit to \textsc{AlphaGo}, which includes offline pretraining on historical human moves.\footnote{\textsc{AlphaZero} later outperformed \textsc{AlphaGo}, partly by removing this human bias. Since our pretraining dataset is itself RL-generated and we can compute rewards at each turn, it is unclear whether a purely self-play-driven approach, if it were computationally feasible, would offer any advantage here.}

Each record contains the model
specification, every probed point $\theta$ with its signature and
viability flags, the per-turn rewards, and the episode's final outcome. Crucially, the data does \emph{not} need to come from a good policy. A mediocre agent that probes mostly non-viable
points still generates ground-truth labels about where viability lives, just as blog post text need not be written by experts to teach an LLM the structure of language. The dataset is composed of roughly 10,000 distinct Lagrangians, with over 100M unique parameter probes evaluated, and contains over a billion output tokens (probe $\times$ Lagrangian parameter dimension $d$) to fit. We take each distinct Lagrangian and corresponding episode and vary which viability is on to increase the number of episodes the model sees. See Table \ref{tab:pretrain_dataset} for more details. The differential evolution-style path is disabled during pretraining.

\begin{table}[h]
\centering
\caption{Pretraining dataset summary. Board counts include
cut augmentation (i.e., we include all combinations of cuts that can be on or off per Lagrangian). The relic-density allowed width $\tau$ provides another continuous direction of augmentation. The number of probes evaluated and output tokens are independent of augmentation (i.e., we did in fact evaluate 100M probes). $d$ here is the Lagrangian parameter dimension. }
\label{tab:pretrain_dataset}
\begin{tabular}{lr}
\hline
Quantity & Count \\
\hline
Distinct Lagrangians                  & $10,843$ \\
Boards (Lagrangians + viability criteria, computed with cut augmentation)        & $\approx 5.6\times 10^{7}$ \\
Probes evaluated                      & $111,046,650$ \\
Output tokens (probes $\times$ parameter dimension $d$)   & $1,308,007,956$ \\
\hline
\end{tabular}
\end{table}

To facilitate pretraining, we introduce two auxiliary heads to the model illustrated in Figure \ref{fig:model_arch}. Pretraining minimizes a weighted sum of five terms, where each training sample is one episode and turn pair drawn from the dataset. The total loss is,
\begin{equation}\mathcal{L} =\lambda_{\rm BC}\mathcal{L}_{\rm BC}+ \lambda_{V}\mathcal{L}_{V}+ \lambda_{\rm sum}\mathcal{L}_{\rm sum}+ \lambda_{\rm aux}\mathcal{L}_{\rm aux}+ \lambda_{\rm div}\mathcal{L}_{\rm div},\label{eq:pretrain_total}
\end{equation}
with $(\lambda_{\rm BC},\lambda_{V},\lambda_{\rm sum},\lambda_{\rm aux},\lambda_{\rm div}) = (0.05,0.3,0.3,0.3,0.1)$.  $\mathcal{L}_{\rm BC}$ maximize the likelihood of a policy head $\pi_k(\boldsymbol\theta|s)$ giving viable probes found in that turn, and can then be written as,
\begin{equation}
\mathcal{L}_{\rm BC}
= -\frac{1}{d}\Big\langle
\log\pi_k(\boldsymbol\theta|s) \Big\rangle_{k, \boldsymbol\theta\in V_k}
\label{eq:pretrain_bc}
\end{equation}
averaged over all heads and viable probes $V_k$. Computing loss only on the viable subset of mediocre RL policies creates a policy better than the one that generated them.  $\mathcal{L}_{V}$ trains the value head to regress to the reward from each episode. That is, training it to output the same value estimate it must later give in PPO RL, 
\begin{equation}
\mathcal{L}_{V} = ||V_t-V_t^{\rm true}||^2
\label{eq:pretrain_value}
\end{equation}
with $V_t^{\rm true}$ being the reward the episode received.

For $\mathcal{L}_{\rm sum}$ a temporary head is attached to the model in Figure \ref{fig:model_arch} that reads the encoded tokens and predicts a vector $\mathbf y\in\mathbb R^n$ of end-of-episode statistics (viable count and fraction, region count, \textsc{SModelS} exclusion rate and mean $r_{\max}$, HL-LHC-testable count, out-of-box fraction, total return). The loss is then simply,
\begin{equation}
\mathcal{L}_{\rm sum} = \tfrac{1}{n}
\lVert y^*-\mathbf y\rVert^2.
\label{eq:pretrain_sum}
\end{equation}
Next, the abundant probes are used to train a second temporary head. The model outputs a viability label $\ell$ that is used in a cross-entropy loss term against the true viability label in which the rare viable class is up-weighted, giving  $\mathcal{L}_{\rm aux}$. Finally, to prevent mode collapse, we use a pairwise KL repulsion between heads, given by $\mathcal{L}_{\rm div}$.

\subsection{Training the LLaM}
\label{sec:Training}
Training proceeds in three phases, the first is offline pretraining, the second is a small warm-up RL run, followed by a multi-node RL finetuning. Pretraining is done with 4x4 A100 GPU nodes for LLaM-small or 16x4 A100 GPU nodes for LLaM-medium. We aim to pass the pretraining dataset summarized in Table \ref{tab:pretrain_dataset} $\sim4$ times for each model size, which resulted in $\sim72$ hours of training for LLaM-small and $\sim60$ hours of training for LLaM-medium. 

RL training is done with asynchronous PPO (APPO). A single learner node holds the current model, and all other nodes are set to ``actors" who run search episodes. After each complete episode by any given actor, the learner receives one completed search, performs one PPO update, and returns updated weights to that actor. The actors each take the latest weights from the learner node, sample a Lagrangian, play one episode, and send the search results back. We do not wait for all actors to return their results to update the model, so fast actors are not stopped by slow ones. We find this to be essential because episode timing can range from minutes to hours, depending on the cost of the sampled model's physics evaluation. 

The use of APPO also motivates the two RL-training phases. We first train with 16 CPU-nodes, critically minimizing how stale a network's weights are for any given actor, until RL has stabilized\footnote{Each actor takes the latest model weights from the learner node at the beginning of an episode and returns search results after it is done. Given there are many actors and given the asynchronous nature of APPO, the model weights on the learner node may have updated before the search results of the actor are returned. That is, the weights an actor acts with can become stale (not up to date with the learner), and how stale they are is directly related to the number of actor nodes.}. We then scale training to 128 CPU-nodes for the second RL phase. Each CPU-node is comprised of 128 threads, and we run both phases for 24 hours each for LLaM-small and 48 hours each for LLaM-medium. For both model sizes and both phases, RL optimization uses AdamW \cite{loshchilov2019decoupledweightdecayregularization} at learning rate $1e-4$ with a short linear warmup and gradient-norm clipping at $0.5$. We describe training and model hyperparameters in more detail in Appendix \ref{app:modeltraininghparams}.

\section{Turning Degeneracies into Questions with Decision Trees}
\label{sec:DecisionTree}
\begin{figure}[t]
    \centering
    \includegraphics[width=.9\linewidth]{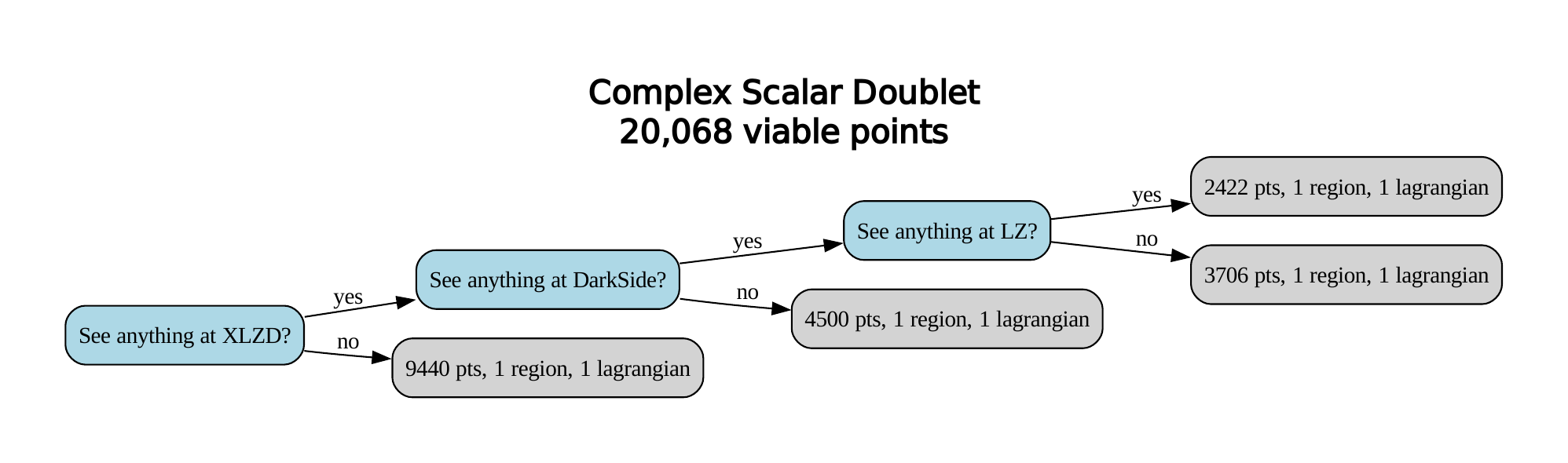}
    \vspace{1em}
    \includegraphics[width=.65\linewidth]{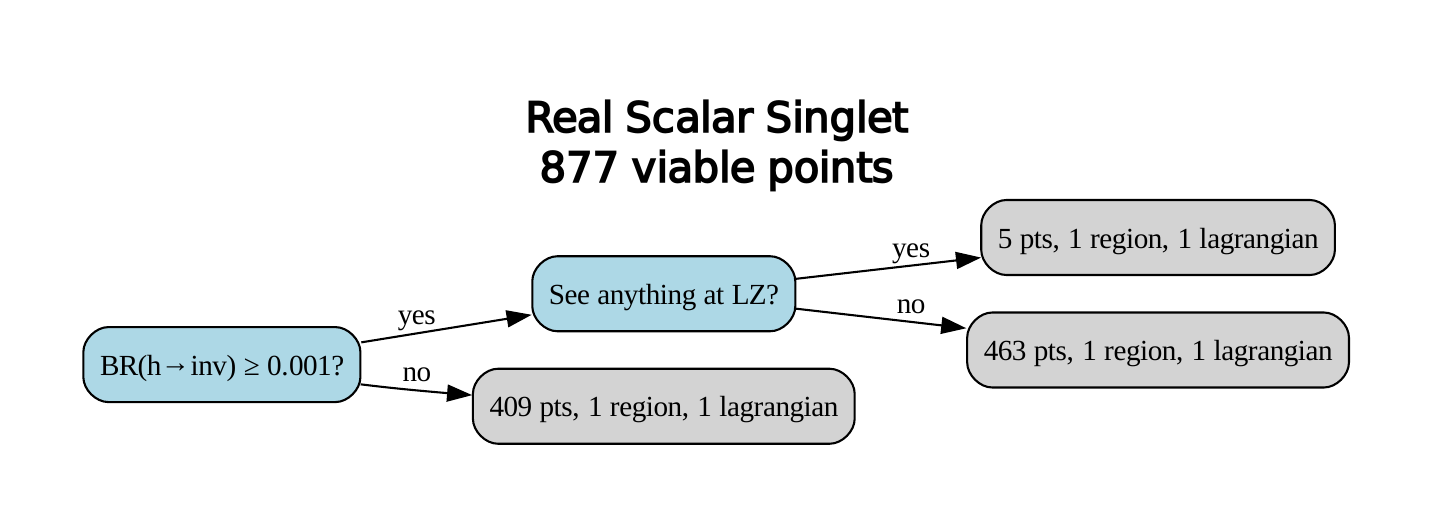}
    \caption{Two (very) small example decision trees, for (upper) a Lagrangian composed of a Complex Scalar Doublet and (lower) a Lagrangian composed of a Real Scalar Singlet. Here, viable means passes \emph{all} experimental constraints in Section \ref{sec:problem}.
    }
    \label{fig:example-dt}
\end{figure}

After training, the LLaM agent can return a set of viable points for a Lagrangian in our search space. We now aim to turn this set of points, whether from one Lagrangian or the union across many, into a decision tree composed of observables computed with trusted pheno tools whose resulting degenerate leaves become questions for a set of LLM-agents during a second pass over the tree. 

We construct this tree, for a set of viable points, by chaining nodes that represent a specific observable with a binary outcome (whether or not that signature is observed given an individual viable point). The resulting leaf nodes of the tree then represent a specific combination of signatures that define an experimental category. We order the tree by which signature splits the most nodes, giving a rough accounting of which experiments are maximally discriminating for that set of viable points. Figure \ref{fig:example-dt} shows two smaller examples.

\subsection{Constructing the Tree}
\label{sec:analytic-tree}
For each Lagrangian, we first scan with the LLaM, then we combine all viable probes from all of the LLaM search episodes. Each saved probe contains predicted observables and expected experimental outcomes for the set of future experiments described in Appendix \ref{app:future-experiments}. We then attempt to build a tree that separates these viable points based on this saved information. A simple binary flag per experiment discards most of the information an experiment offers. We thus instead split the tree with bins of predicted signal strength for each signature as a fraction of any given experiment's projected limit, split by channel when applicable. That is, for each observable $O_{\mathrm{pred}}$, we compute,
\begin{equation}
    \mu_e = \log_{10}\big(O_{\mathrm{pred}} / O_{\mathrm{lim}}\big)
\end{equation}
and bin into a discrete set. Bin widths are determined by the expected systematic error of each observable, details of which can be found in Appendix \ref{app:tree-bins}. The set of observables we consider is given in Appendix \ref{app:future-experiments}.

Two probes with identical binned observables across all features are
experimentally indistinguishable. We thus call their shared signatures a \emph{signature class}. The tree is grown organically. We start with all points at a starting node, and we find the binary split that maximizes the information gain, see Appendix \ref{app:infogain}.
We repeat on each following node until every leaf holds a single signature class. Crucially, this process does the experiment ranking for us. The observable that best separates all viable points rises to the root, and depth in the tree can be interpreted as the ability to split the Lagrangian we consider. We can of course compute the union of viable points in some set of Lagrangians and construct a tree over that set. In that case, the tree orders observables by their ability to distinguish points \emph{across} a set of models. 

We also cluster the probes in parameter space with \textsc{DBSCAN} \cite{3001460.3001507} to so called viable regions, see Appendix \ref{app:regions}. We report in the leaves the number of viable probes along with the number of regions that all share the signature class they define, see Figure \ref{fig:example-dt}.

\subsection{LLMs for Splitting Indistinguishable Leaves}
\label{sec:LLM-tree}
A leaf that holds more than one parameter-space region is degenerate under the set of experiments we consider. We hand these leaves to a set of LLM-agents and ask them to attempt to split the degeneracy per leaf. The agents are asked to extend the tree with new binary splits, until either all degeneracies have been split, a max LLM-depth is reached (we use a max LLM-depth of 4), or a proposed node is judged to be practically impossible to split. For adding binary splits, the agent has three choices for types of splits:
\begin{enumerate}
    \item A \textbf{Literature Split}: A published result for an observable whose current limits split the degenerate regions. For this kind of split, the LLM must cite the paper that is used, its reasoning for why this works, and a status \textsc{``Splits!"} or \textsc{``No Split"} that reports if the split was successful.
    \item A \textbf{Literature Projection}: An experiment or limit that is planned or described in the literature whose projected sensitivity splits the regions. For this split, the LLM must cite the most relevant papers for this projection, must report its reasoning for this split, an approximate sensitivity improvement factor to closest already ran experiment, and must report a qualitative judgment on the feasibility of this split as a choice from five categories: \textsc{Reanalysis}, \textsc{Possible}, \textsc{Next Generation}, \textsc{Speculative}, and \textsc{Impossible}.
    \item A \textbf{Novel Observable}: An experiment or limit that has not been proposed in the literature that may be capable of splitting the regions. The LLM agent must report its reasoning for this split, the paper that is closest to its idea, and must report a qualitative judgment on the feasibility of this split as a choice from four categories: \textsc{Possible}, \textsc{New Experiment}, \textsc{Speculative}, and \textsc{Impossible}.
\end{enumerate}

We enforce ordering rules for the proposal of these node choices. The first attempted split must be of type Literature Split to ensure the LLM searches what has already been done, and only after finding \textsc{``No Split"} is it permitted to propose other types. The agent then searches for \textbf{Literature Projection} nodes, and moves on to \textbf{Novel Observable} only if none are found. We also allow the LLM to propose alternatives. If a \textbf{Literature Projection} is qualitatively judged to be \textsc{Next Generation} or worse, the agent can propose a \textbf{Novel Alternative}, which is a signature with the same rules as the \textbf{Novel Observable} but acting as an alternative to the \textbf{Literature Projection} and must have the same or better feasibility qualitative judgment.

We create the full tree $N_{\rm repeat} = 5$ times (that is, we run $5\times N_{\rm {Degenerate-Leaf}}$ calls)  and then pass all trees through an aggregation agent, which chooses the best combination of observables from all runs. For all calls, tool use is enabled and set to web search and arXiv retrieval, which the agent must use to verify every
reference it cites. We keep the full reasoning trace for each node. All agent calls in this work are done with \textsc{Fable-5.1}-based agents with the exception of one of the repeated ensembling passes where we used \textsc{Opus-5} for diversity.

\section{Results}
\label{sec:results}
Here we present the results of our \haithem{} framework in two parts. In Section \ref{sec:LLaMResults} we benchmark the Large Lagrangian Model on its ability to find viable regions as compared to a baseline on a held-out set of Lagrangians. In comparison to the baseline, we find that the LLaM finds more and more diverse viable regions, and we also observe minimal collapse in the LLaM search ability as parameter space dimension grows. We then follow this in Section \ref{sec:DT-res} with a test of decision tree methodology. We scan through the space of all single-scalar-multiplet dark matter models our space allows with our LLaM and use the regions they find to build a decision tree over the set, with observables computed with trusted tools and LLM-agent-proposed nodes. We then scrutinize the tree and discuss interesting LLM proposals.

\begin{figure}
    \centering
    \includegraphics[width=0.7\linewidth]{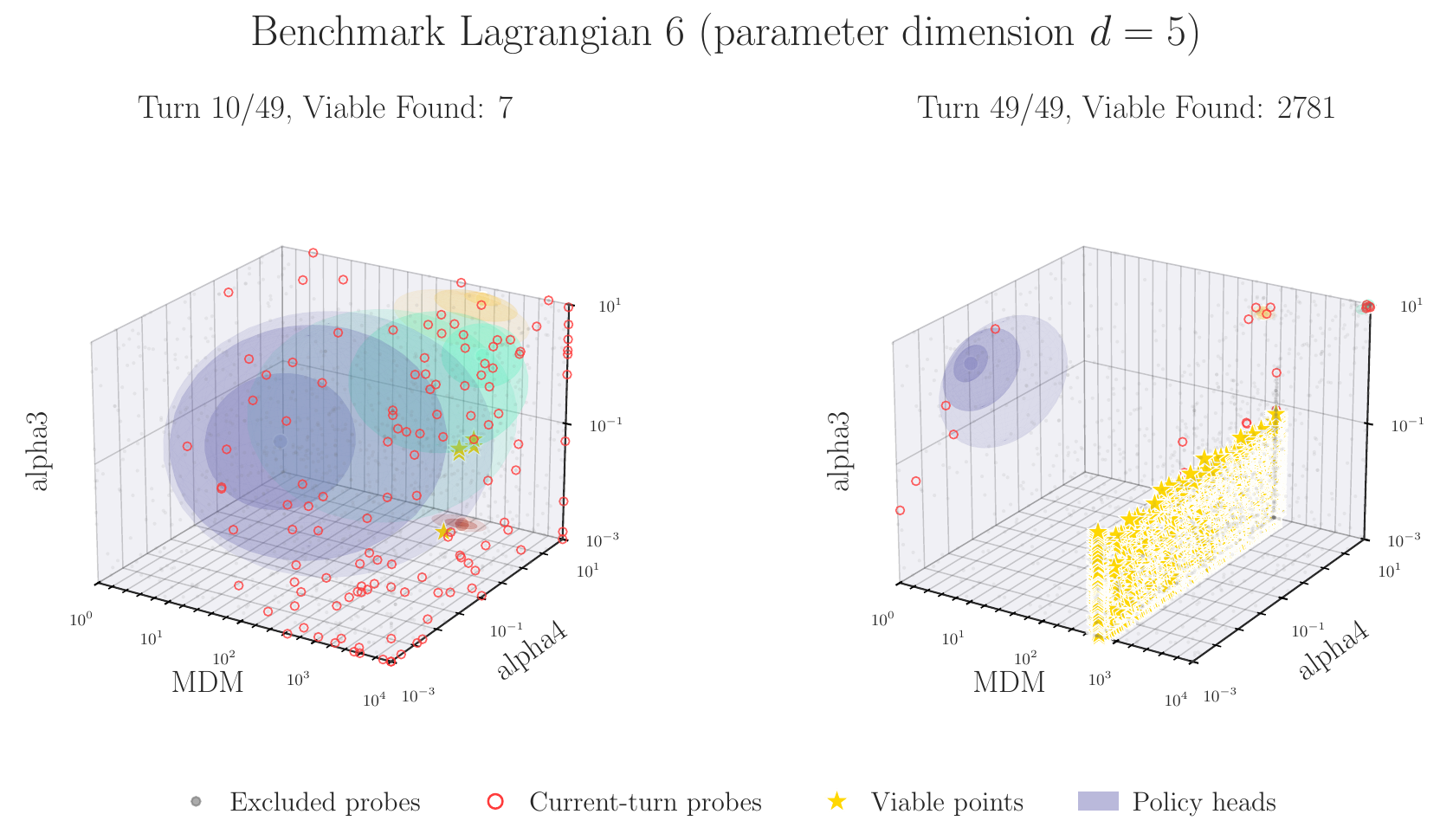}
    \includegraphics[width=0.7\linewidth]{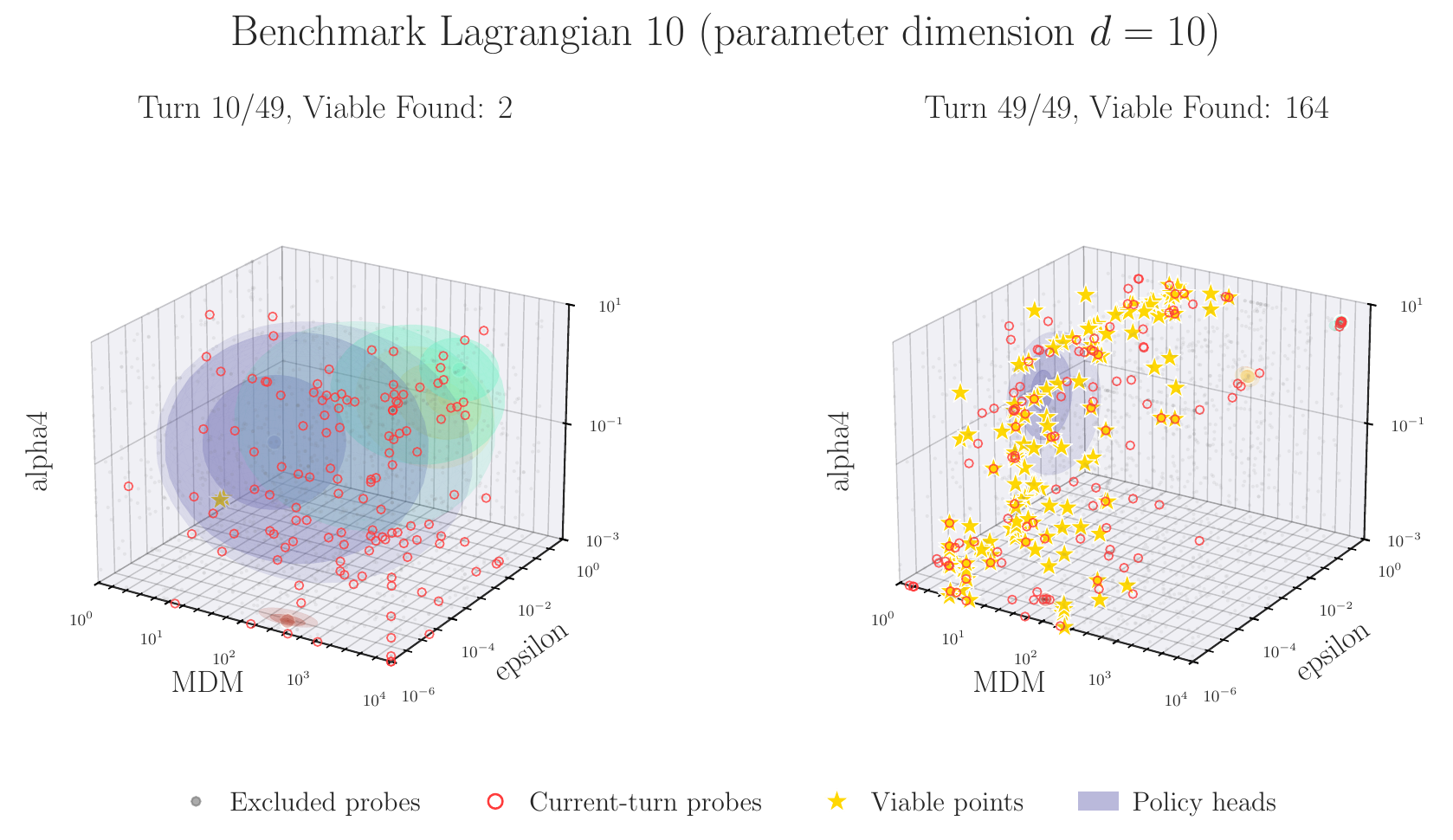}
    \includegraphics[width=0.7\linewidth]{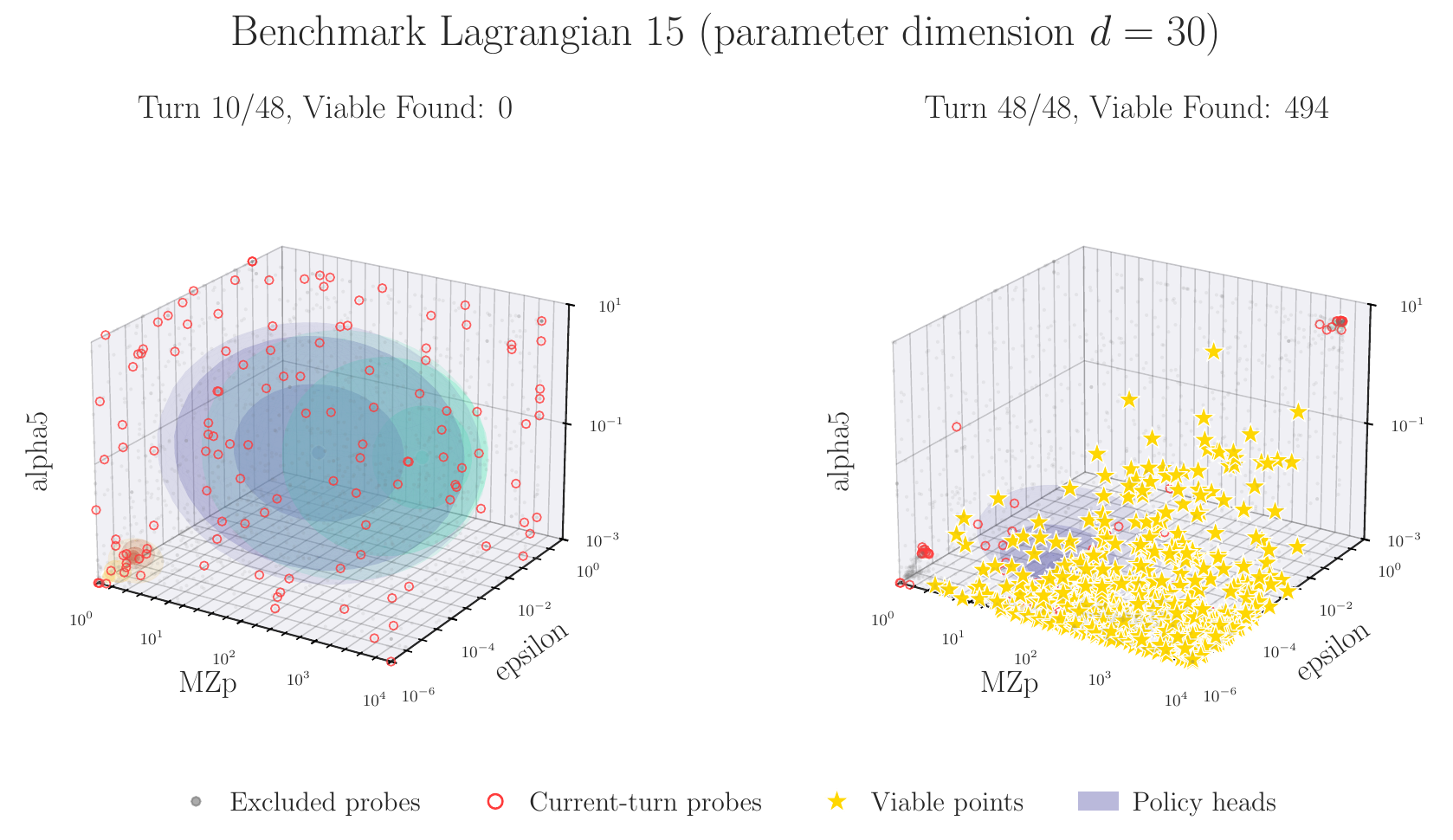}
    \caption{Example LLaM-small-RL search episodes at budget $B=50$ for
    three benchmark Lagrangians (see numbering of Appendix \ref{app:lag-bench}), shown early in the episode (left) and at the final turn (right). Grey points are excluded probes, red dots are current turn's probes, gold stars are the cumulative viable points found, and the shaded ellipses are the policy's beta distribution heads. The
    search acts in the full $d$-dimensional parameter space. We display the three parameters along which its viable points spread the most. \textbf{Policy heads need not be in the right location when Differential Evolution is chosen by the LLaM}. More examples can be found in Appendix \ref{app:lag-bench}, and interactive visualizations can be found online.}
    \label{fig:3d-episodes}
\end{figure}
\subsection{Benchmarking the Large Lagrangian Model}
\label{sec:LLaMResults}

To measure how well a search algorithm finds viable regions, we devise a
benchmark composed of 21 representative Lagrangians with all viability cuts turned on and with the relic density requirmetns at its tightest setting $\tau=1$. The LLaMs have not seen in training the configurations we test, and the specific Lagrangians are described and numbered in Appendix \ref{app:lag-bench}, along with per-Lagrangian results. In the benchmark, each policy plays a search episode for each Lagrangian with some budget of turns $B$ where every turn it proposes $N_{\rm probe}=128$ parameter points. We vary the search budget $B\in\{5,10,25,50\}$ turns (640 to 6,400 probes), and every combination of Lagrangian, budget, and policy is repeated three times to estimate the variance.

We report five metrics. $N_v$ is the total number of viable points found and
$\mathcal{L}_v$ is the number of Lagrangians where at least one viable point
was found. To measure diversity of parameter points found, we compute $R_{100}$, the number of regions found via \textsc{DBSCAN} per 100 viable points\footnote{Computed as is described in Appendix \ref{app:regions}.} \cite{3001460.3001507}. A policy that re-samples the same regions scores low, one that keeps discovering separated regions scores high, up to a maximum of 100. Next, $N_\sigma$ counts the distinct experimental-signature classes among the viable points, that is the number of leaf nodes the decision tree is composed of. This also acts as a measure of exploration, where sampling points with the same experimental signatures is not desired. Finally, $W$ is the so-called win rate and counts the number of Lagrangians on which a given policy was the one that found the most viable points.
\begin{table}[h]
    \centering
    \caption{Search benchmark composed of 21 representative Lagrangians. Here, summed over all Lagrangians, all budgets, and all three repeats. $N_v$ is the total number of viable points found, $\mathcal{L}_v$ the number of Lagrangians where at least one viable point was found, $R_{100}$ is the number of regions found via \textsc{DBSCAN} per 100 viable points (a measure of how diverse the viable points are) \cite{3001460.3001507}, $N_\sigma$ the distinct
    signature classes discovered, and $W$ is the win rate, i.e., the number of Lagrangians on which the policy found the most viable points. See Appendix \ref{app:lag-bench} for information on the Lagrangians used along with per-Lagrangian results. Bold marks the best performance per column.}
    \label{tab:policy-benchmark-combined}
\begin{tabular}{lccccc}
        \toprule
        Search Method & $N_v$ & $\mathcal{L}_v$ & $R_{100}$ & $N_\sigma$ & $W$ \\
        \midrule
        LLaM-small (pretrained)
            & $6,105$ & $9/21$ & \textbf{1.5} & $14$ & $0$ \\
        LLaM-small (RL)
            & $20,744$ & \textbf{14/21} & $0.9$ & \textbf{29} & $3$ \\
        \midrule
        LLaM-medium (pretrained)
            & $3,007$ & $13/21$ & $0.8$ & $17$ & $0$ \\
        LLaM-medium (RL)
            & \textbf{51,802} & $13/21$ & $0.5$ & $23$ & \textbf{12} \\
        \midrule
        Differential Evolution
            & $8,591$ & $11/21$ & $0.1$ & $24$ & $0$ \\
        \bottomrule
    \end{tabular}
\end{table}

\paragraph{Differential Evolution baseline.}
As a baseline, we use Differential Evolution (DE) \cite{Storn:1997uea}, a population-based,
gradient-free global optimizer, similar to the current state-of-the-art searchers such as \textsc{Diver} \cite{GAMBITDarkMatterWorkgroup:2017fax,GAMBITDarkMatterWorkgroup:2017fax,Martinez:2017lzg}. We do not use \textsc{Diver} itself due to our non-standard binary definition of viability. The comparison is also natural. Like the LLaM search `game', DE has a notion of turns, proposes a population of 128 points, and adapts from feedback alone. Specifically, we use \textsc{SciPy}'s \texttt{differential evolution}, operating in the same normalized $[0,1]^d$ parameter space as the LLaM policies \cite{Virtanen:2019joe,Storn:1997uea}. The population size equals the 128 probes per turn, one DE generation is evaluated per turn, and the number of generations equals the turn budget $B$. We minimize an objective that encourages viability, see Appendix \ref{app:diffevol}.

\begin{figure}[h]
    \centering
    \includegraphics[width=0.7\linewidth]{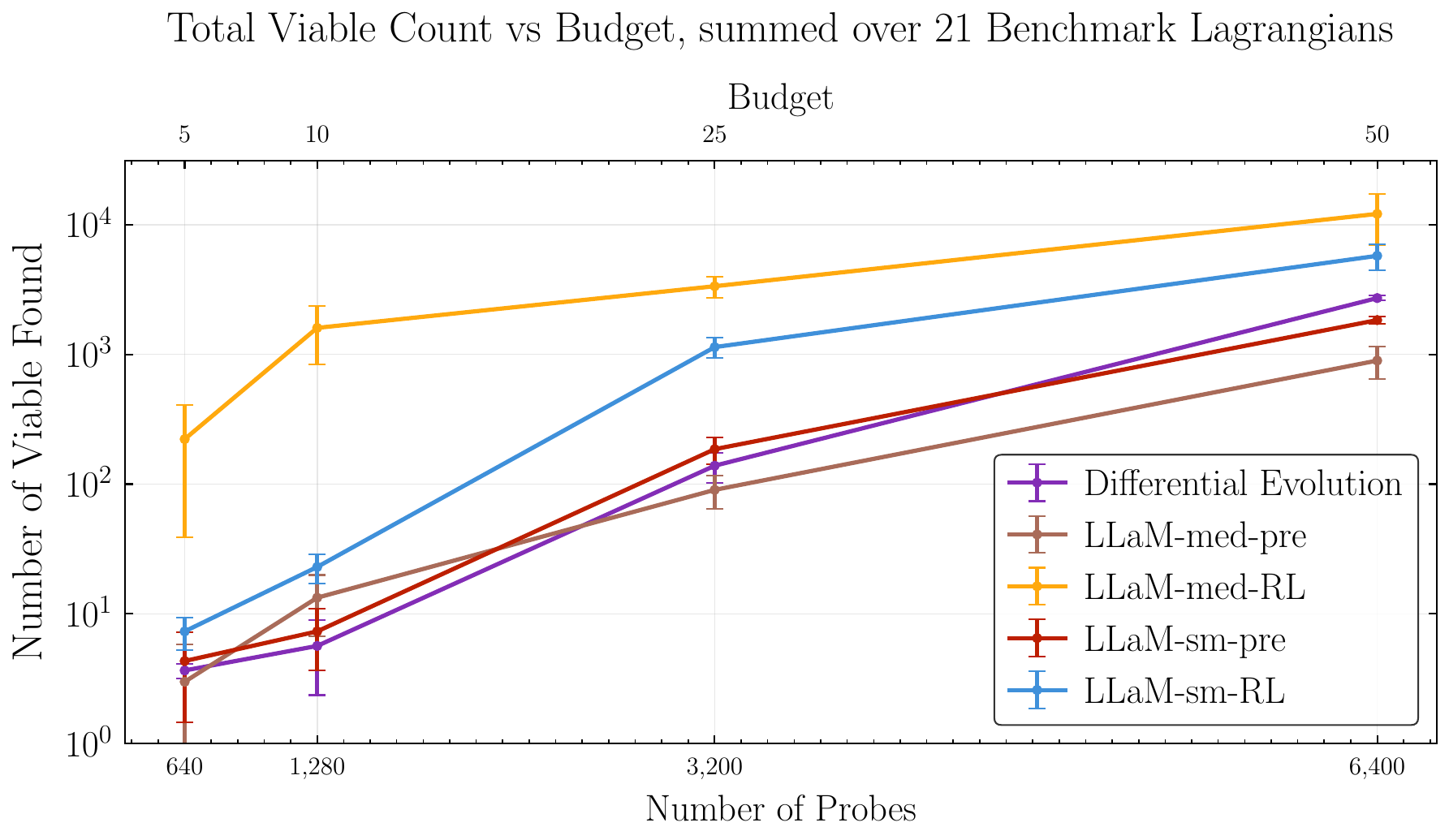}
    \caption{A Benchmark of search performance. Total number of viable points found as a function of search
    budget, summed over the 21 benchmark Lagrangians.}
    \label{fig:viable-vs-budget}
\end{figure}
\begin{table}[h]
    \centering
    \caption{Search benchmark composed of 21 representative Lagrangians,
    resolved by search budget: $B=5$, $10$, $25$, and $50$ turns
    (that is 640, 1280, 3200, and 6400 probes respectively), with mean and error
    computed over three repeats. Columns are as in
    Table \ref{tab:policy-benchmark-combined}. Bold marks the best
    performance per column within each budget block. See Appendix \ref{app:lag-bench} for per-Lagrangian results}
    \label{tab:policy-benchmark-budgets}
    \scriptsize
\begin{tabular}{llccccc}
        \toprule
        Budget & Search Method & $N_v$ & $\mathcal{L}_v$ & $R_{100}$ & $N_\sigma$ & $W$ \\
        \midrule
        \multirow{5}{*}{$B=5$}
        & LLaM-small (pretrained)
            & $4.3 \pm 2.9$ & $2.7 \pm 1.7$ & $0.0 \pm 0.0$ & $1.7 \pm 0.5$ & $1.3 \pm 1.2$ \\
        & LLaM-small (RL)
            & $7.3 \pm 2.1$ & $\mathbf{3.0 \pm 0.0}$ & $0.0 \pm 0.0$ & $1.7 \pm 0.5$ & $\mathbf{2.0 \pm 0.8}$ \\
        & LLaM-medium (pretrained)
            & $3.0 \pm 2.8$ & $2.0 \pm 1.4$ & $0.0 \pm 0.0$ & $2.0 \pm 1.4$ & $0.3 \pm 0.5$ \\
        & LLaM-medium (RL)
            & $\mathbf{222 \pm 183}$ & $2.3 \pm 0.9$ & $\mathbf{1.8 \pm 1.7}$ & $\mathbf{3.7 \pm 1.9}$ & $\mathbf{2.0 \pm 1.4}$ \\
        & Differential Evolution
            & $3.7 \pm 0.5$ & $\mathbf{3.0 \pm 0.8}$ & $0.0 \pm 0.0$ & $2.3 \pm 0.9$ & $1.0 \pm 0.8$ \\
        \midrule
        \multirow{5}{*}{$B=10$}
        & LLaM-small (pretrained)
            & $7.3 \pm 3.7$ & $3.3 \pm 1.2$ & $0.0 \pm 0.0$ & $3.0 \pm 0.8$ & $0.7 \pm 0.5$ \\
        & LLaM-small (RL)
            & $23.0 \pm 5.9$ & $4.0 \pm 0.0$ & $\mathbf{3.6 \pm 2.8}$ & $3.7 \pm 1.2$ & $1.3 \pm 0.5$ \\
        & LLaM-medium (pretrained)
            & $13.3 \pm 6.6$ & $4.0 \pm 0.8$ & $0.0 \pm 0.0$ & $3.3 \pm 1.2$ & $0.7 \pm 0.5$ \\
        & LLaM-medium (RL)
            & $\mathbf{1600 \pm 769}$ & $\mathbf{5.0 \pm 1.4}$ & $1.2 \pm 0.4$ & $\mathbf{5.3 \pm 3.7}$ & $\mathbf{4.7 \pm 1.2}$ \\
        & Differential Evolution
            & $5.7 \pm 3.3$ & $2.3 \pm 0.5$ & $3.3 \pm 4.7$ & $3.3 \pm 1.7$ & $0.3 \pm 0.5$ \\
        \midrule
        \multirow{5}{*}{$B=25$}
        & LLaM-small (pretrained)
            & $186 \pm 44$ & $5.3 \pm 0.5$ & $1.3 \pm 0.4$ & $5.7 \pm 0.5$ & $0.0 \pm 0.0$ \\
        & LLaM-small (RL)
            & $1137 \pm 202$ & $6.0 \pm 0.8$ & $\mathbf{2.0 \pm 0.5}$ & $6.7 \pm 1.2$ & $1.7 \pm 0.5$ \\
        & LLaM-medium (pretrained)
            & $90 \pm 26$ & $5.0 \pm 0.8$ & $1.9 \pm 0.8$ & $7.0 \pm 1.6$ & $0.7 \pm 0.5$ \\
        & LLaM-medium (RL)
            & $\mathbf{3351 \pm 636}$ & $5.7 \pm 0.5$ & $1.0 \pm 0.1$ & $\mathbf{8.7 \pm 4.5}$ & $\mathbf{4.7 \pm 0.5}$ \\
        & Differential Evolution
            & $139 \pm 36$ & $\mathbf{6.7 \pm 0.5}$ & $0.8 \pm 0.2$ & $8.0 \pm 1.6$ & $2.0 \pm 0.8$ \\
        \midrule
        \multirow{5}{*}{$B=50$}
        & LLaM-small (pretrained)
            & $1837 \pm 126$ & $6.3 \pm 0.9$ & $\mathbf{1.5 \pm 0.2}$ & $10.0 \pm 0.0$ & $0.3 \pm 0.5$ \\
        & LLaM-small (RL)
            & $5747 \pm 1302$ & $9.0 \pm 0.8$ & $0.7 \pm 0.1$ & $\mathbf{19.3 \pm 1.2}$ & $4.0 \pm 2.2$ \\
        & LLaM-medium (pretrained)
            & $896 \pm 249$ & $8.7 \pm 1.2$ & $0.7 \pm 0.3$ & $11.0 \pm 2.4$ & $1.3 \pm 0.9$ \\
        & LLaM-medium (RL)
            & $\mathbf{12094 \pm 5094}$ & $8.3 \pm 1.2$ & $0.3 \pm 0.0$ & $11.7 \pm 2.1$ & $\mathbf{6.0 \pm 0.8}$ \\
        & Differential Evolution
            & $2716 \pm 119$ & $\mathbf{9.7 \pm 0.5}$ & $0.1 \pm 0.0$ & $15.3 \pm 3.4$ & $2.0 \pm 0.0$ \\
        \bottomrule
    \end{tabular}
\end{table}

Figure \ref{fig:3d-episodes} shows what successful episodes look like for the LLaM on
three Lagrangians.  As a case study, we consider the complex scalar doublet ($20,068$ points) and the top plot in Figure \ref{fig:3d-episodes}. Here, mass-dependent gauge interactions with the SM $Z/W$ produce the right relic density, and thus, the Higgs portal coupling only needs to stay under bounds direct detection places upon it to achieve viability. We thus expect a thin band of viability in the mass of the dark matter. We see this exact structure in the top of Figure~\ref{fig:3d-episodes}. Even more reassuringly, we observe this band at $m_\chi\approx 563 \rm~ GeV$ in close agreement with Table 1 of \cite{Cirelli:2005uq}. Finding viability (at the right parameter values) where we expect is a good test of the entire pipeline, from operator construction through \textsc{micrOMEGAs} to the LLaM search. It is also clear in that figure that our LLaM has little trouble finding thin regions.
 
\begin{figure}
    \centering
    \includegraphics[width=0.75\linewidth]{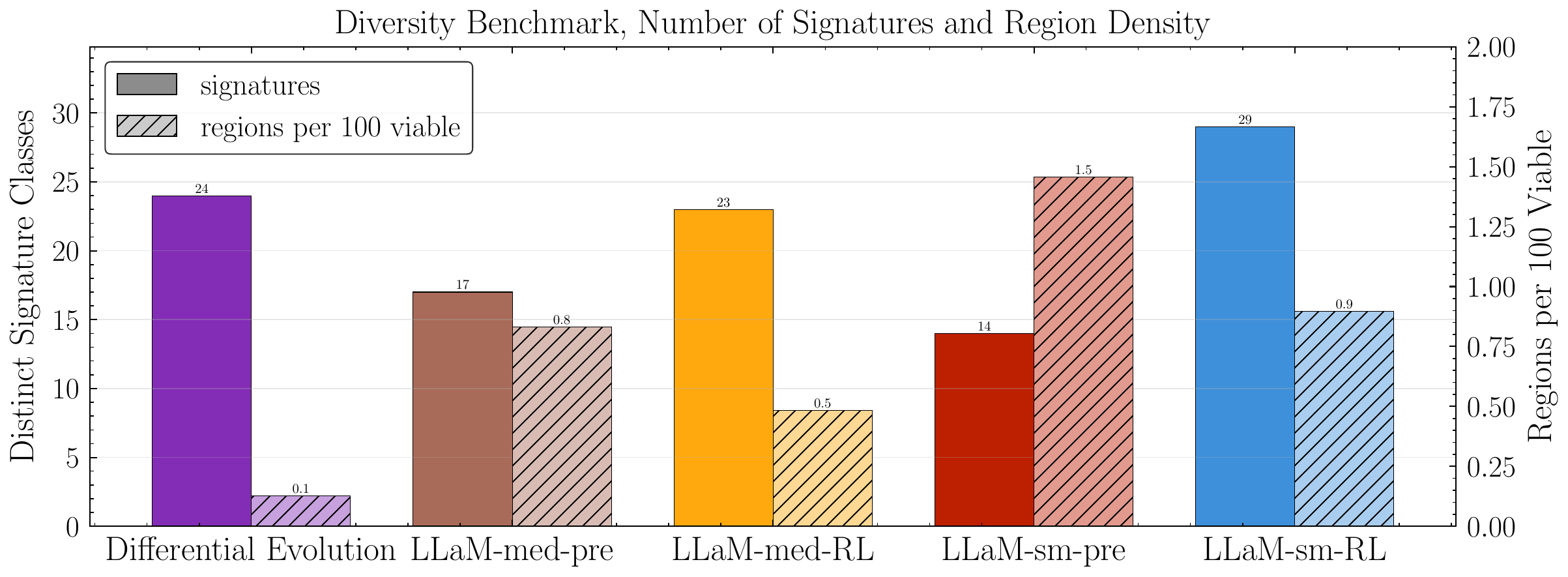}
    \caption{Diversity benchmark. Distinct experimental-signature classes discovered (that is, the number of decision-tree leaf nodes; solid, left axis) and \textsc{DBSCAN}-found regions per 100 viable points (striped, right axis) for each search method \cite{3001460.3001507}.}
    
    \label{fig:diversity}
\end{figure}

\begin{figure}
    \centering
    \includegraphics[width=0.6\linewidth]{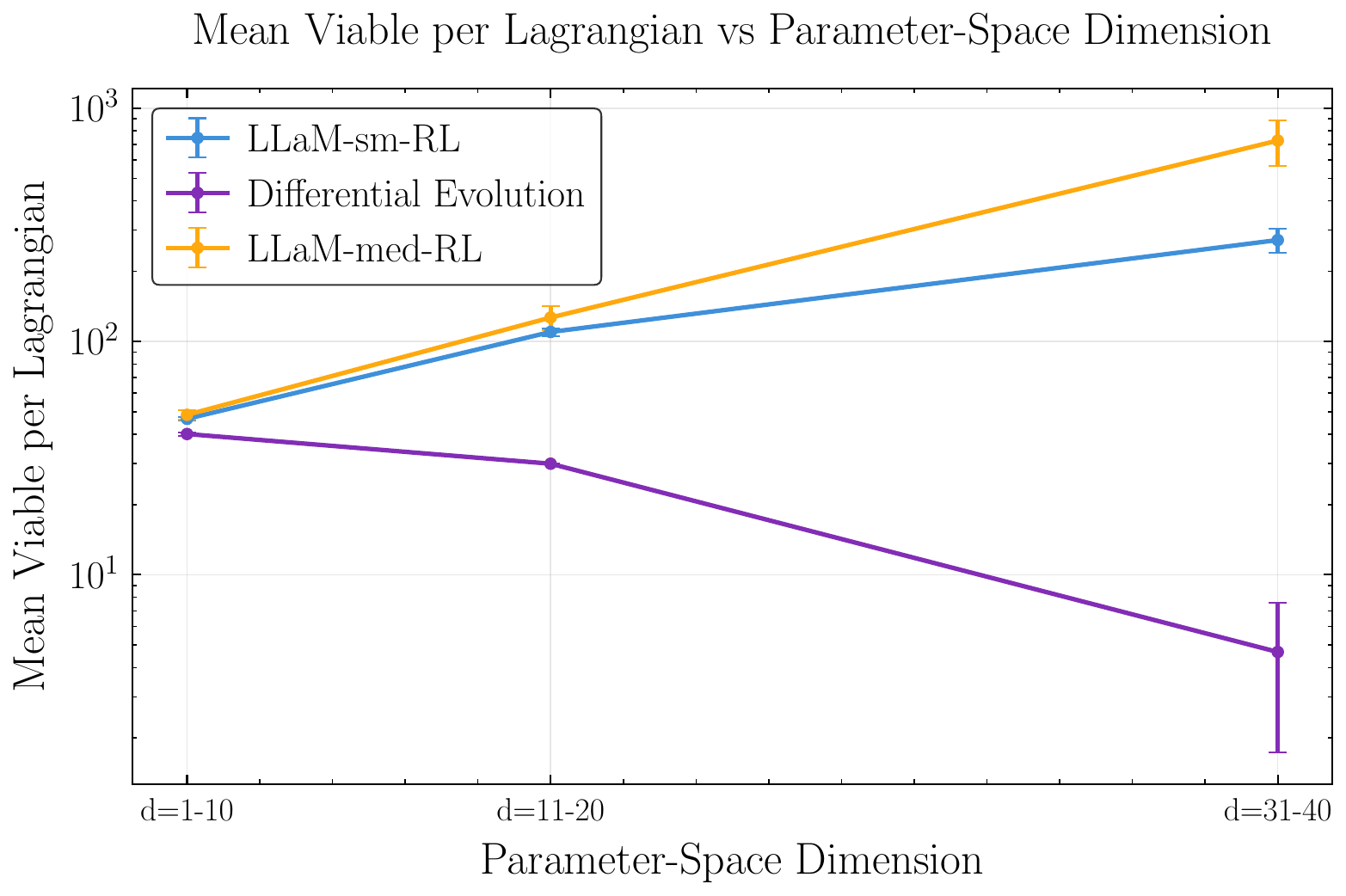}
    \caption{Mean number of viable points per benchmark Lagrangian as a function of the Lagrangian parameter-space dimension $d$. The LLaM becomes \emph{more} effective as the dimension grows, while differential evolution collapses, as expected from the curse of dimensionality. We only plot bins of Lagrangian parameter-space dimension $d$ for which we have sufficient statistics.}
    \label{fig:param-dim}
\end{figure}

Table \ref{tab:policy-benchmark-combined} summarizes the benchmark by summing over all three repeats and budget sizes. LLaM-small (RL) finds $N_v = 20,744$ viable
points, $3.4\times$ its pretrained starting point and $2.4\times$ DE, finds
viable points on the most Lagrangians ($14/21$), discovers the most
signature classes ($N_\sigma=29$), and wins the head-to-head comparison on
3 of the 21 Lagrangians. LLaM-medium (RL) finds $N_v = 51,802$  viable
points, $17.2\times$ its pretrained starting point, $2.5\times$ LLaM-small, and $6.02\times$ DE, finds
viable points on 13/21 Lagrangians, discovers $N_\sigma=23$
signature classes and wins the most head-to-head comparison on
12 of the 21 Lagrangians. 

Crucially, the LLaM's return of viable points does not come at the price
of diversity, meaning there is no major exploration-exploitation tradeoff with respect to DE. We summarize a diversity benchmark in Figure \ref{fig:diversity}. LLaM-small and LLaM-medium find $R_{100}\sim 1$ 5-10 times DE's $.1$. That is, LLaMs find 5-10 times as many viable regions in parameter space instead of sampling the same regions more than the more exploitative DE.

The same holds in signature space when looking at LLaM-small, where the LLaM-small (RL) model discovers $N_\sigma=29$ distinct signature classes (i.e., 29 leaves on the decision tree of Section \ref{sec:DecisionTree} over all 21 Lagrangians) against DE's $24$. The surprising caveat is LLaM-medium. Although we see a significant model scaling benefit in terms of the number of viable points found, that model scaling seems to result in a more exploitative rather than exploratory model. The region density $R_{100}$ drops between the model sizes, there is slightly worse Lagrangian coverage by the medium model, and, most drastically, the number of signatures found is significantly weaker for LLaM-medium. We leave diagnosing and addressing this collapse for future work.

Table~\ref{tab:policy-benchmark-budgets} shows the benchmark by budget, and Figure \ref{fig:viable-vs-budget} plots the curve of viable points as a function of budget. Three conclusions stand out. First, there is a clear threshold effect for LLaM-small and DE. At $B=5$ and $B=10$ (640 or 1280 probes), LLaM-small and DE find only a handful of viable points across all 21 Lagrangians. That is, below roughly a thousand probes there is simply not enough feedback to locate the narrow viable regions for those two policies. Second, once we are past this threshold, the policies quickly differentiate. At $B=25$, the RL model takes off ($1137 \pm 202$ viable points versus ${\sim}150$ for the DE
baseline, an $8\times$ difference), and at $B=50$ it reaches $5747 \pm 1302$, $\sim2.5\times$ ahead. Finally, we see a clear model-size benefit in terms of budget size, as seen in Figure \ref{fig:viable-vs-budget}. The medium model is dramatically better at low budgets, potentially indicating more knowledge of the physics that governs viability, leading to more efficient probe usage. Given that B=5 is 10x cheaper than B=50, we may be able to leverage larger models to search through 10x more Lagrangians with the same compute in the future.

A large difference appears when the results are plotted as a function of Lagrangian parameter dimension (see Figure~\ref{fig:param-dim}). Differential Evolution behaves
as expected given the curse of dimensionality. The mean number of viable probes found per Lagrangian falls from ${\sim}40$ at $d\le 10$ to near zero by $d\sim 35$. In contrast to standard intuition, the LLaM moves in the opposite direction. A possible explanation is that, having learned which parameters matter for a given Lagrangian (mass hierarchies, portal couplings, mixing angles), the learned policy effectively searches a low-dimensional manifold, so extra dimensions only add more space for viable regions. This trend also must be taken in context with the diversity benchmark in Figure \ref{fig:diversity}, we do not beat the curse of dimensionality simply by exploiting found regions, as the LLaM \emph{also} finds more diverse sets. 

\subsection{A Decision Tree over all Single-Scalar-Multiplet Lagrangians}
\label{sec:DT-res}
\begin{figure}
    \centering
    \includegraphics[width=0.9\linewidth]{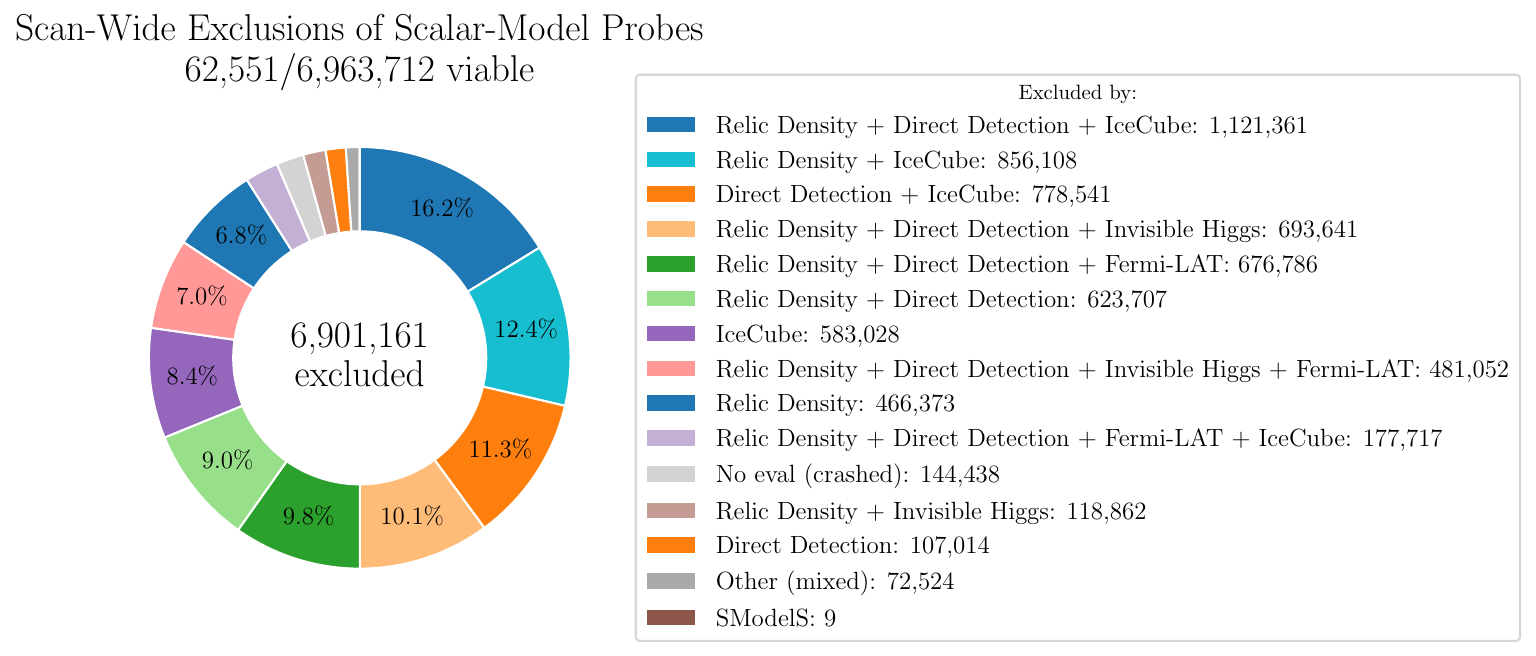}
    \caption{Pie chart of exclusions. Each point that is excluded by some set of constraints is tallied, and we compute the ratio of points that were excluded by that combination from the total exclusion count.}
    \label{fig:scan_exclusions}
\end{figure}

\begin{table}[b]
    \centering
    \caption{All 11 distinct Lagrangians in our single-dark-multiplet scan, split by spin and ordered by parameter-space dimension $d$. $N_v$ is the total number of viable points found. $N_\sigma$ is the number of leaves of each model's signature tree plotted in Appendix \ref{app:scan-trees}. $N_R$ is the total number of separated parameter-space regions (computed with DBSCAN as in Section \ref{sec:DecisionTree})\cite{3001460.3001507}. Models with $Y=0$ and Dark-U(1) are identical under a field redefinition between $Q=\pm1$ Dark-U(1) field charge.\footnote{Due to the self-conjugate condition, real scalars cannot carry hypercharge, and thus their doublets (which require $Y=\pm\frac{1}{2}$ for electric neutrality) are excluded. For the same reason, real scalars cannot carry a charge under a Dark-U(1) and thus those models are excluded.} \footnote{Since our search space allows for different $\mathbb{Z}_n$ charges for the different components in a complex scalar field, for each complex scalar model, we have two variants. One in which the complex component has a Higgs portal and one in which it is secluded. See Section \ref{sec:problem}.}}
    \label{tab:one-field-scan}
    \setlength{\tabcolsep}{4pt}
    \begin{tabular}{lcccc}
    \toprule
    Lagrangian & $d$ & $N_v$ & $N_\sigma$ & $N_R$ \\
    \midrule
    Real Scalar Singlet & $3$ & $877$ & $3$ & $3$ \\
    Real Scalar Triplet & $3$ & $0$ & $0$ & $0$ \\
    Complex Scalar Doublet & $5$ & $20,068$ & $4$ & $4$ \\
    Complex Scalar Doublet + dark $U(1)^{+}$ & $7$ & $0$ & $0$ & $0$ \\
    Complex Scalar Doublet + dark $U(1)^{-}$ & $7$ & $24$ & $4$ & $4$ \\
    Complex Scalar Singlet + dark $U(1)^{\pm}$ (secluded) & $8$ & $1,752$ & $18$ & $19$ \\
    Complex Scalar Triplet & $8$ & $0$ & $0$ & $0$ \\
    Complex Scalar Triplet + dark $U(1)^{\pm}$ & $8$ & $0$ & $0$ & $0$ \\
    Complex Scalar Singlet + dark $U(1)^{\pm}$ & $10$ & $32,346$ & $32$ & $38$ \\
    Complex Scalar Singlet (secluded) & $11$ & $0$ & $0$ & $0$ \\
    Complex Scalar Singlet & $17$ & $7,484$ & $5$ & $8$ \\
    \bottomrule
\end{tabular}

\end{table}
To validate the efficacy of our decision trees (coupled to our LLaMs), we restrict our search space to a set of Lagrangians that we can easily exhaust combinatorially. The natural choice is the subset of the space defined in Section \ref{sec:problem} that contains only a single dark scalar multiplet. That is, the subset of DM Lagrangians in our space that contain only one scalar field with one copy. The resulting set contains 11 Lagrangians, to which we apply both LLaM-small (RL) and LLaM-medium (RL) and sum their outcomes. Figure \ref{fig:scan_exclusions} contains a pie chart summarizing exclusions across the scan. The Lagrangians that are scanned, along with the number of viable points found per Lagrangian are summarized in Table \ref{tab:one-field-scan}.

\begin{sidewaysfigure}[p]
    \centering
    \includegraphics[width=\linewidth]{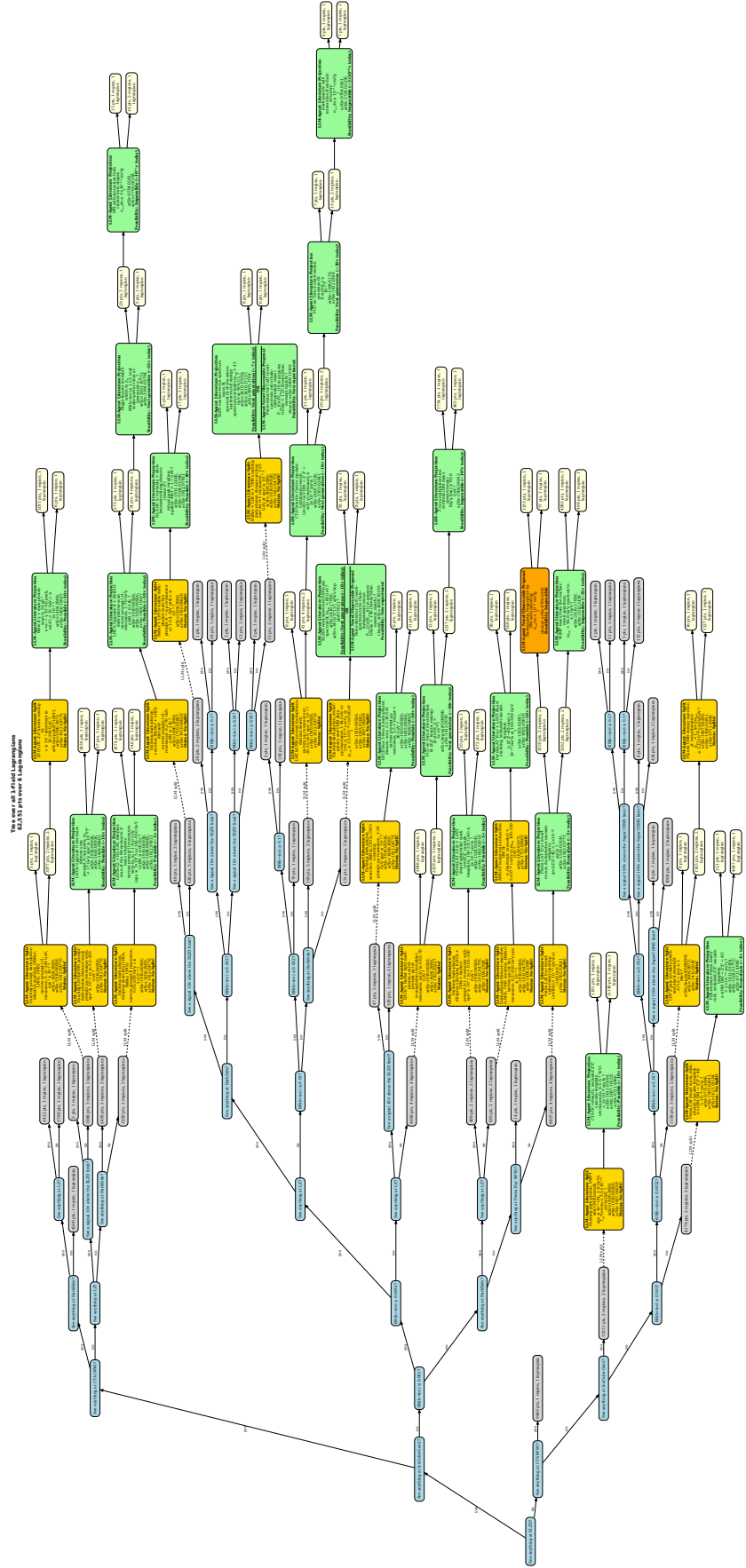}
    \caption{Decision Tree over all single-scalar-multiplet Lagrangians our space allows. The LLM-agent cites \cite{BaBar:2014zli,LHCb:2019vmc,CMS:2019buh,LHCb:2017trq,BaBar:2017tiz,Ilten:2016tkc,Belle-II:2018jsg,LHCb:2018roe,Autran:2015mfa,ATLAS:2019lng,CidVidal:2018eel,ALEPH:2005ab,Hook:2010tw,Curtin:2014cca,Babu:1997st,TLEPDesignStudyWorkingGroup:2013myl,Cuoco:2016eej,Reinert:2017aga,Fermi-LAT:2015att,Fermi-LAT:2016afa,Fermi-LAT:2016uux,Fermi-LAT:2015kyq,CTA:2020qlo,Elor:2015tva,Pospelov:2007mp,Planck:2018vyg,Slatyer:2015jla,LZ:2024zvo,XLZD:2024nsu,Pato:2010zk,Green:2008rd,Theodosopoulos:2026eym,Green:2007rb,Randall:2008ppe,Tulin:2017ara,ATLAS:2023tkt,deBlas:2019rxi}. Interactive visualizations that show reasoning traces can be found online.}
    \label{fig:global_tree}
\end{sidewaysfigure}

Given the outcome of the LLaM search, we take the union over all viable regions found over all Lagrangians scanned in Table \ref{tab:one-field-scan}, and build a combined non-LLM tree as described in Section \ref{sec:analytic-tree}. We then use a set of LLM-agents to split degeneracies as described in Section \ref{sec:LLM-tree}. The (large) tree with LLM-agent nodes appended is plotted in Figure \ref{fig:global_tree}. We also do the same independently per-Lagrangian and report those trees in Appendix \ref{app:scan-trees}. LLM-proposed nodes are colored based on the category of split and appended to deterministic nodes.

Table \ref{tab:llm-proposal-types} summarizes the types of observables the LLM-agent uses to split nodes across all calls. We see a focus on collider signatures in the Literature split stage, focusing on dark-photon searches, and we then see a shift away from collider and toward astrophysical signatures as we move to the Novel Observable category. Many of the ``Novel Observables" proposed experiments are for probing dark-matter self-interaction well beyond what is realistic and thus rated impossible (we generally agree with the LLM judgment). This is the case for the one Novel Observable node in Figure \ref{fig:global_tree}, and many in the $N_{\rm repeat}=5$ repeats. This also explains the large drop in the fraction of ``Novel Observable" nodes from the total over all repeats to the final aggregated tree. The aggregation LLM-agent call (likely rightly) chooses alternatives over the impossible ``Novel Observables".
\begin{table}[h]
\centering
\caption{Types of measurements proposed by the LLM agent over global and model trees for all $N_{\rm repeat}=5$ repeats. $n$ is the number in that category. The bottom row is the total per split-type for the final aggregated Trees in Figure \ref{fig:global_tree} and in Appendix \ref{app:scan-trees}. Percentages for categories are percentages over category type. Percentages for Totals are percentages over LLM split type.}
\label{tab:llm-proposal-types}
\begin{tabular}{l cc @{\hspace{1.2em}} cc @{\hspace{1.2em}} cc @{\hspace{1.2em}} cc}
\hline
& \multicolumn{2}{c}{Literature Splits} & \multicolumn{2}{c}{Literature projections}
& \multicolumn{2}{c}{Novel Observables} & \multicolumn{2}{c}{Novel Alternatives} \\
\cmidrule(lr){2-3} \cmidrule(lr){4-5} \cmidrule(lr){6-7} \cmidrule(lr){8-9}
Category & $n$ & Percentage & $n$ & Percentage & $n$ & Percentage & $n$ & Percentage \\
\hline
Collider signatures      & 94 & $73.4\%$ & 50 & $42.0\%$ &  5 & $13.9\%$ & 0 & $0.0\%$  \\
Astrophysical            & 24 & $18.8\%$ & 53 & $44.5\%$ & 29 & $80.6\%$ & 4 & $80.0\%$ \\
Dedicated DM experiments &  2 & $1.6\%$  & 15 & $12.6\%$ &  2 & $5.6\%$  & 1 & $20.0\%$ \\
Cosmology                &  8 & $6.2\%$  &  1 & $0.8\%$  &  0 & $0.0\%$  & 0 & $0.0\%$  \\
\hline
\hline
Total                    & 128 & $44.4\%$ & 119 & $41.3\%$ & 36 & $12.5\%$ & 5 & $1.7\%$ \\
Final Tree after Aggregation   &  27 & $45.0\%$ &  28 & $46.7\%$ &  3 & $5.0\%$  & 2 & $3.3\%$ \\
\hline
\end{tabular}
\end{table}

\subsubsection{Interesting Observable Proposals}
A few LLM-agent nodes (labeled as either novel alternative or literature projection) provide observables that may highlight the efficacy of the decision-tree method for inspiring underexplored LLM signatures. They further emphasize the value of LLMs' ability to combine ideas across the literature in a context-aware fashion.

One motivating example is a ``Novel Alternative" shown in two nodes in Figure \ref{fig:global_tree}, titled \emph{``Paleo-detector Ca/O recoil-spectrum ratio"}. In both cases, the agent was in need of a precise measurement of the dark matter mass $m_\chi$ to split the regions that were otherwise nearly identical. The agent proposes using XLZD's nuclear-recoil spectrum (i.e., the spectrum of the energy deposition seen by the yet-to-be-built XLZD \cite{XLZD:2024nsu}) as the $m_\chi$ probe and rates the observable as ``Next-Generation". Given the rules stated in Section \ref{sec:DecisionTree}, it then also looks for a novel alternative that is similar or less qualitatively difficult.

For this alternative, the agent recognizes that \cite{Drees:2008bv}'s description of the expected maximum kinetic energy a dark matter particle deposits in a direct detection detector is a good probe for $m_\chi$ and given by,
\begin{equation}
E_{\rm max} = \frac{v^2_{\rm esc}}{\alpha^2}, \quad \alpha^2 \equiv \frac{m_N}{2\mu_\chi^2 }
\end{equation}
where $v_{\rm esc}$ is the galactic escape velocity of the dark matter particles, $m_N$ is the mass of the nucleus of the detector, and $\mu_\chi\equiv \frac{m_\chi m_N}{m_\chi+m_N}$ is the reduced mass of the dark matter candidate \cite{Drees:2008bv}. Thus, a statistically robust measurement of $E_{\rm max}$ may be a precise probe of the dark matter mass if $v_{\rm esc}$ is known.

The challenge is that the exposure needed for measuring such an observable is very large, requiring many dark matter detections near $v_{\rm esc}$ despite the significantly decreased frequency of particles at such a velocity, given a Boltzmann-like distribution. As a solution, the agent reaches for paleo-detectors, minerals found deep underground, protected from cosmic rays and high temperatures for order giga-years. Disturbances of the crystal lattice of these detectors contain permanent records of nuclear recoil events throughout their lifetime from many sources of background along with potential dark matter interactions \cite{Baum:2018tfw,Baum:2023cct,Baum:2021jak,Edwards:2018hcf,Drukier:2018pdy}.

However, by the nature of their long exposure taking place throughout our galaxy's assembly history, paleo-detectors introduce much more uncertainty in $v^2_{\rm esc}$. As a solution, the agent proposes the use of the ratio of two direct detection targets,
\begin{equation}
    \frac{E_{\rm max}^{N_1}}{E_{\rm max}^{N_2}} = \frac{\alpha^2_{N_2}}{\alpha^2_{N_1}} = \frac{\frac{\mu^2_{N_1}}{m_{N_1}}}{\frac{\mu^2_{N_2}}{m_{N_2}}}= \frac{\mu^2_{N_1}m_{N_2}}{\mu^2_{N_2}m_{N_1}}\equiv f(m_{N_1},m_{N_2},m_\chi)
\end{equation}
which removes $v^2_{\rm esc}$ dependence entirely. It further reasons that identifying a single mineral containing two sufficiently sensitive target nuclei would be ideal for controlling backgrounds,\footnote{For example, neutrons from radioactive decays of surrounding elements, where the background rate may be shared between atoms in the same crystal.} and identifies calcium and oxygen in gypsum as such a pair, inspired by recent work in \cite{Theodosopoulos:2026eym} (though they do not consider this observable). The agent also reports a more statistically robust ``moment-matching" (and likely more practical) observable taken from \cite{Drees:2008bv} and applied to paleo-detectors that expands the observable beyond just the $E_{\rm max}$ spectrum end point.

The agent does not describe how to do the measurement (e.g., what form of microscopy, at what resolution, measuring what properties of what tracks), nor does it describe how best to identify which nucleus (Ca or O) formed a specific measured track. There is work proposing secondary track properties as a method for per-event nucleus discrimination \cite{Hedges:2026pgf,Calabrese-Day:2026soq}, although an in-depth study is required to understand if this is practical and, if not, if there is a realistic version of this observable. It is thus difficult to claim the practicality of this observable, but its proposal highlights the utility of the decision-tree formulation for finding unexplored combinations of ideas.

We note a few more interesting examples not of novelty but of application of known observables to our context. One node in Figure \ref{fig:global_tree} and one in Appendix \ref{app:scan-trees} note a ratio of spin-independent cross section to the Higgs-to-invisible branching ratio,
\begin{equation}
    \frac{\sigma_{SI}}{\mathrm Br(h \to inv)}
\end{equation}
as an observable for counting the number of Higgs-to-invisible degrees of freedom. In the global tree in Figure \ref{fig:global_tree}, the agent proposes this observable when attempting to split a complex scalar singlet from a real scalar singlet that share all bins of observables we compute. This relation has been previously noted in \cite{Djouadi:2012zc,Baek:2014jga} for distinguishing dark matter spin and testing the Higgs-to-invisible mediator structure. Its application here is a small and natural extension. 

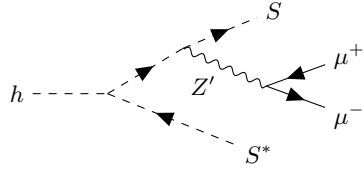
\begin{figure}[h]
\centering
\begin{tikzpicture}
\begin{feynman}
  \vertex (h)   at (-1.6, 0)   {\(h\)};
  \vertex (v1)  at (-0.4, 0);
  \vertex (v2)  at ( 0.6, 0.6);
  \vertex (S)   at ( 1.8, 1.1) {\(S\)};
  \vertex (Sb)  at ( 1.6,-0.8) {\(S^*\)};
  \vertex (v3)  at ( 1.7, 0.1);
  \vertex (mup) at ( 2.8, 0.5) {\(\mu^+\)};
  \vertex (mum) at ( 2.8,-0.3) {\(\mu^-\)};
  \diagram*{
    (h)  -- [scalar] (v1),
    (v1) -- [charged scalar] (v2),
    (v2) -- [charged scalar] (S),
    (v1) -- [anti charged scalar] (Sb),
    (v2) -- [photon, edge label'=\(\Zp\)] (v3),
    (v3) -- [fermion] (mum),
    (mup) -- [fermion] (v3),
  };
\end{feynman}
\end{tikzpicture}
\caption{Diagram of Dark Bremsstrahlung possible with the scalar + dark U(1) models we consider. This process results in a dimuon + MET final state near the Higgs mass.}
\label{fig:mono-zprime}
\end{figure}

Another example is given in Figure \ref{fig:global_tree} as a dark bremsstrahlung search (illustrated in Figure \ref{fig:mono-zprime}). The agent is given two regions, both from the same Lagrangian containing a dark-$U(1)'$. The regions only differ in the range of self-coupling parameters and in the mass range of the $Z'$ mediator. The agent chooses to separate the regions by proposing an observable that measures $M_{Z'}$. Due to the regions sharing a similar range of the kinetic-mixing parameter $\epsilon$ and due to that range being very small $\epsilon\sim10^{-6}$, the agent decides that whatever observable it chooses needs a $Z'$ production mechanism that is independent of $\epsilon$. Its choice for this is a (dark-U(1)-charged) scalar produced via a Higgs decay that then radiates the desired $Z'$, all without kinetic mixing. 

This is then formalized as a specific collider dark bremsstrahlung search: using the HL-LHC to search for dimuon + MET events tagged as from a Higgs decay. The reconstructed dimuon invariant mass in such a search is the $\epsilon$-independent measurement of $M_{Z'}$ needed to split the regions it was given\footnote{Unlike dark Drell-Yan, the $Z'$ is produced by the scalar coupling depending only on $g_{Z'}$. The $\epsilon$ independence of the dark bremsstrahlung process follows from this and the cancellation of $\epsilon$ factors in the branching ratio $\rm BR(Z'\rightarrow\mu\mu)$ when $M_{Z'}<2m_\chi$.}. Other papers have noted this final state \cite{Dawson:2025dmi,Aguilar-Saavedra:2022xrb}, though in a different Higgs-decay context (i.e., not in the context of dark bremsstrahlung or dark scalars), and \cite{Autran:2015mfa} notes a similar search independent of a Higgs decay (direct $qq\rightarrow Z'\rightarrow \rm inv+\mu\mu$). Further, the bump hunt on $m_{\mu\mu}$ in Higgs-tagged events with MET, to the best of our knowledge, has also not been done, where the closest is \cite{ATLAS:2019lng} which does not filter events as coming from a Higgs decay. 

Imposing a Higgs-tag on a final state with MET is the primary challenge that would need to be overcome to make such an analysis feasible. The agent notes the Higgs vector-boson-fusion (VBF) channel as the solution \cite{ATLAS:2022yvh,Cahn:1986zv,Kleiss:1987cj}, standard in invisible-Higgs searches. The VBF channel's relatively small contribution to the total Higgs cross section, in addition to the suppressed rate of the bremsstrahlung search as compared to just Higgs-to-invisible ($h\rightarrow SS^*$), leads to the HL-LHC requirement.\footnote{Such a search acts as a measurement for $M_{Z'}$ (and $g_{Z'}$, see above footnote) and also may be an interesting channel to search before dark-matter discovery. The inclusion of a $Z'$ makes the channel subdominant to Higgs-to-invisible. However, the requirement of a dimuon in the final state cuts the main backgrounds of VBF-based Higgs-to-invisible searches such as $pp\rightarrow Z+jj\rightarrow\nu\bar\nu + jj$ \cite{ATLAS:2022yvh}. A detailed analysis is required to study the feasibility of such a search, given that new backgrounds may also be introduced.}
\section{Conclusion and Outlook}
\label{sec:conclusion}

We introduced \haithem{} (\textbf{H}idden-sector \textbf{AI} for \textbf{T}estable \textbf{H}ypothesis \textbf{E}nu\textbf{M}eration), a framework for searching for underexplored model signatures with LLM-empowered decision trees. In order to build this tree, we must have a fast method for scanning through an arbitrary Lagrangian's high-dimensional parameter space that is not biased towards prior literature, to find diverse sets of viable regions. For this, we develop a Large Lagrangian Model (LLaM), an autoregressive transformer trained at scale with a dataset composed of $\sim1~\rm billion$ output tokens. The LLaM learns to play a Battleship-style game to search the parameter space and is trained with RL, similar to \textsc{AlphaGo} in the use of pretraining and RL, and similar to \textsc{AlphaZero} in the lack of human data in training \cite{silvergo,SilverGo2,doi:10.1126/science.aar6404}.

In order for the LLaM to function and train, we need an automated method to generate Lagrangians from allowed symmetries and field content, and an automated method to take those Lagrangians along with proposed parameter points and compute observables and compare those observables to known bounds. We implement this with a pipeline that incorporates \textsc{Sym2IntPy}, \textsc{FlavorBuilder} \cite{Baretz:2025zsv}, \textsc{micrOMEGAs} \cite{Alguero:2023zol}, \textsc{SModelS} \cite{Alguero:2020grj}, and several other physics tools, and is stably scalable to hundreds of compute nodes. This pipeline is intended as a rigorous proof of concept; additional constraints, tools, and NLO corrections could be incorporated to produce more detailed phenomenology.

We compare our LLaM to a differential evolution baseline and find that on a benchmark of 21 Lagrangians, LLaM-medium finds over 5x, and LLaM-small over 2x, the number of viable points found by the baseline. Both also find 5-10x the number of distinct and disconnected viable regions per 100 viable points, and both either match or significantly exceed the number of experimental signature classes discovered by differential evolution over the benchmark. That is, according to our metrics, the models both explore and exploit the space more effectively than the baseline. 

We also find that the LLaM is able to handle high-dimensional parameter spaces significantly better than the baseline, which may be because the LLaM has implicitly learned a lower-dimensional subspace of parameter space. 

The number of viable points found by the LLaMs increases with parameter space dimension, while the number found by differential evolution collapses, as expected given the curse of dimensionality. Further, by training two model sizes of the LLaM, we observed model scaling benefits in the number of viable points found, especially when only a few probes are allowed. With only 640 probes per Lagrangian, LLaM-medium finds $\sim$60x the number of viable points of differential evolution and $\sim$30x the number of LLaM-small.

We then scan, with our LLaM, the space of all one-scalar-multiplet Lagrangians we consider and build per-Lagrangian decision trees, along with one tree across the whole scan. The LLaM finds sufficient parameter-space diversity to produce topologically nontrivial trees, and it can identify extremely thin regions of viability. We find that LLM-agents are capable of proposing discriminating and physically reasonable combinations of observables that further break decision-tree degeneracies. A few examples of such previously unexplored combinations of known signatures are also discussed, and the closest known literature is credited as inspiration for the idea.

\subsection{Outlook}
This work opens the door to large-scale searches for unexplored observables across different classes of dark matter and broader BSM Lagrangians. Since combinatorics grows quickly, such scans would require a large quantity of compute and potentially intelligent Lagrangian sampling strategies. The payoff, however, may be substantial. Such searches may reveal the value of different future and proposed experiments in distinguishing between diverse dark matter and BSM models, and may help the community make better-informed multi-decade scientific investments. Extensions to larger classes of Lagrangians would require the inclusion of additional computable probes and a software suite that goes beyond \textsc{micrOMEGAs} and \textsc{SModelS}~\cite{Alguero:2020grj,Alguero:2023zol}. This work further demonstrates that the observables we choose to include in the non-LLM section of tree could be informed by the LLM suggestions reported here and beyond. It is also worth noting that while LLMs are used to generate new ideas, our framework explicitly attempts to credit the related literature that inspired the LLM's suggestion.

While this work shows how knowledge and tools from across fundamental physics can be assimilated, its reliable application requires coordination among physicists from different disciplines. It is therefore natural to build  on existing communities at the intersection of particle phenomenology, experimental particle physics, and cosmology, such as those behind GAMBIT, SModelS, and micrOMEGAs. 

\section*{Code Availability, Dataset, and Interactive Visualizations}
\label{sec:code-avail}
The code and dataset used for this study are to be released soon. A blog post on the method with interactive visualizations of the LLaM episodes and decision trees can be found at \url{https://ibrahimelsharkawy.com/blog/bsm-rl-search.html}.

\vspace{-1em}
\section*{Acknowledgments}

AG thanks Genevieve Belanger, their expertise in dark matter phenomenology was critical in the early stages of this project. We also thank Sean Benevedes, Yoni Kahn, Sabine Kraml, Max Fieg, Jake Rudolph and Daniel Whiteson for comments on the manuscript that helped improve clarity and scientific relevance of the paper. This research used computing resources allocated to AG’s NESAP `AI Physicist' project at the National Energy Research Scientific Computing Center (NERSC), a U.S. Department of Energy Office of Science User Facility located at Lawrence Berkeley National Laboratory.

I.E. is supported in part by the Connaught International Scholarship at the University of Toronto and the Natural Sciences and Engineering Research Council of Canada (NSERC) Canada Graduate Research Scholarship. I.E. thanks Marat Freytsis and Anthropic for access to Claude and Claude Code, which assisted with aspects of building the model and running the LLM agents used in this work. The work of VKP was supported in part by the U.S. National Science Foundation under Grant PHY-221028.

\bibliography{bib}
\bibliographystyle{apsrev4-2}
\clearpage

\appendix
\section{On the Physics Setting and Computational Pipeline}
\label{app:physsetting}

\subsection{Experimental Bounds}
\label{app:interp}

All experimental bounds are given to our pipeline as sets of $\{(m_i, L_i)\}$ stored in csv-files sourced from the literature, where $L_i$ is the limit on the relevant observable at dark matter mass $m_i$ \cite{IceCube:2025fcu,IceCube:2016dgk,Alguero:2023zol,IceCube-Gen2:2020qha,Boddy:2024tiu,Fermi-LAT:2015att,CTA:2020qlo,PICO:2019vsc,DARWIN:2016hyl,LZ:2022lsv,DarkSide-20k:2017zyg,SuperCDMS:2016wui}. Every curve is interpolated in log-log. A model point with predicted observable $\mathcal{O}$ is then compared to the curve with $r = \mathcal{O}/ L(m_{\rm DM})$. Here $r>1$ means the prediction lies above the experimental limit.
\begin{figure}
    \centering
    \includegraphics[width=\linewidth]{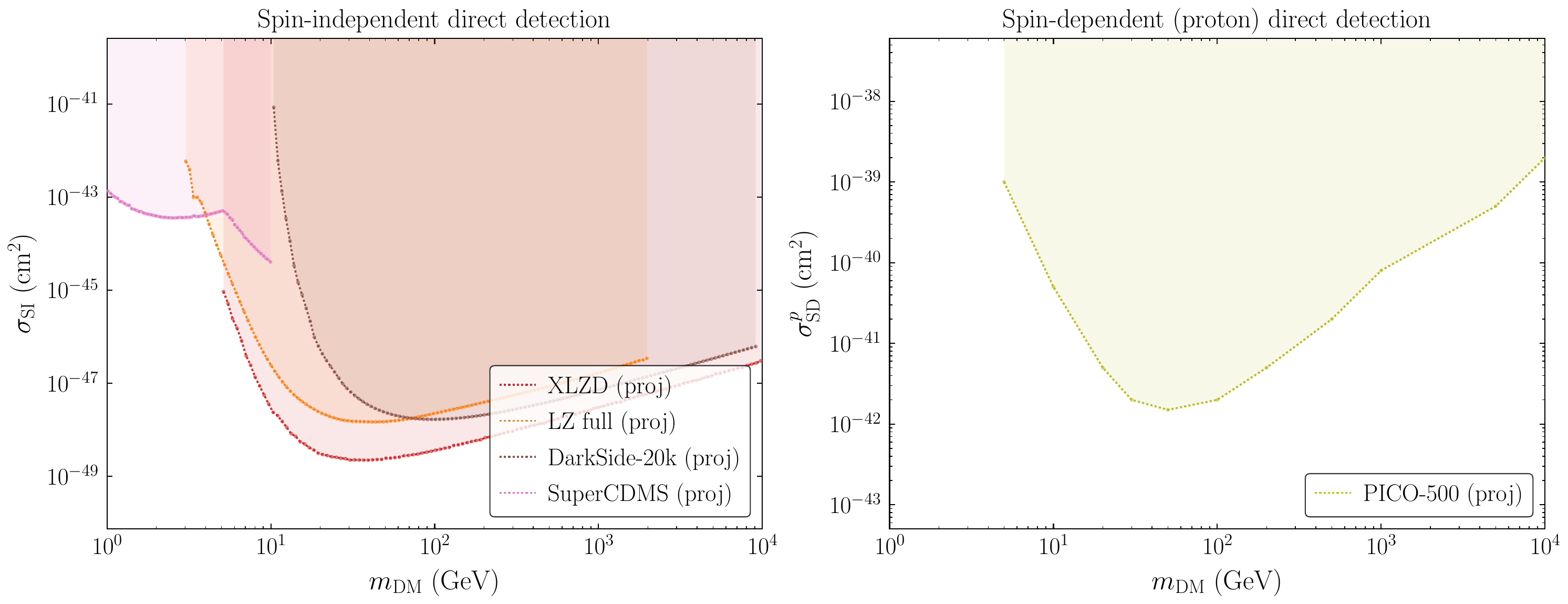}
    \caption{Direct-detection Observables $\sigma_{\rm SI}$ and $\sigma^p_{\rm SD}$ bounds interpolated from \cite{DARWIN:2016hyl,LZ:2022lsv,DarkSide-20k:2017zyg,PICO:2019vsc,SuperCDMS:2016wui}. SuperCDMS combines the  Si-HV, Si-iZIP, Ge-HV, and Ge-iZIP constraints.}
    \label{fig:SI-Bounds}
\end{figure}

For the direct-detection observables $\sigma_{\rm SI}$ and $\sigma^p_{\rm SD}$, the bound $\sigma_{\rm bound}$ depends on the dark matter mass. Each experiment provides one curve that we interpolate. The model cross-section, computed by \textsc{micrOMEGAs}, is converted from pb to cm$^2$, and compared directly with $r = \sigma_{\rm model}(m_{\rm DM}) / \sigma_{\rm bound}(m_{\rm DM})$. These curves are shown in Figure \ref{fig:SI-Bounds}.

\begin{figure}
    \centering
    \includegraphics[width=\linewidth]{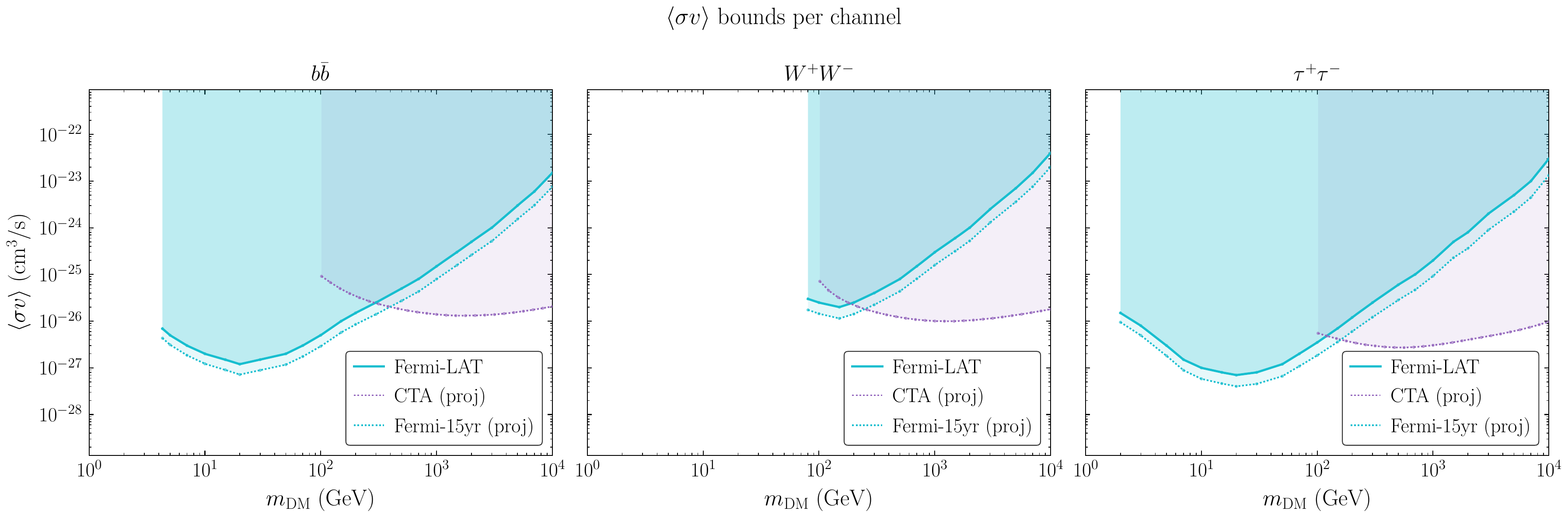}
    \caption{$\langle\sigma v\rangle$ bounds per channel interpolated from \cite{CTA:2020qlo,Fermi-LAT:2016afa,Fermi-LAT:2015att}. The WW channel is non-constraining below $\sim m_W$, similarly for other channels.}
    \label{fig:indirect-bound}
\end{figure}

For the indirect observables, like $\langle\sigma v\rangle$ (CTA, Fermi-LAT)
and the solar upward-muon flux $\mu_{\rm solar}$ (IceCube), the bound depends on the dark matter mass \emph{and} the annihilation final state. For CTA and Fermi-LAT, which probe $\langle\sigma v\rangle$, we compare channel by channel. \textsc{micrOMEGAs} returns
the annihilation cross-section for each individual final states $c$, so each contribution $\langle\sigma v\rangle_c$ can be compared against the limit
curve $L_c(m_{\rm DM})$ for that same channel. 

\begin{figure}
    \centering
    \includegraphics[width=0.5\linewidth]{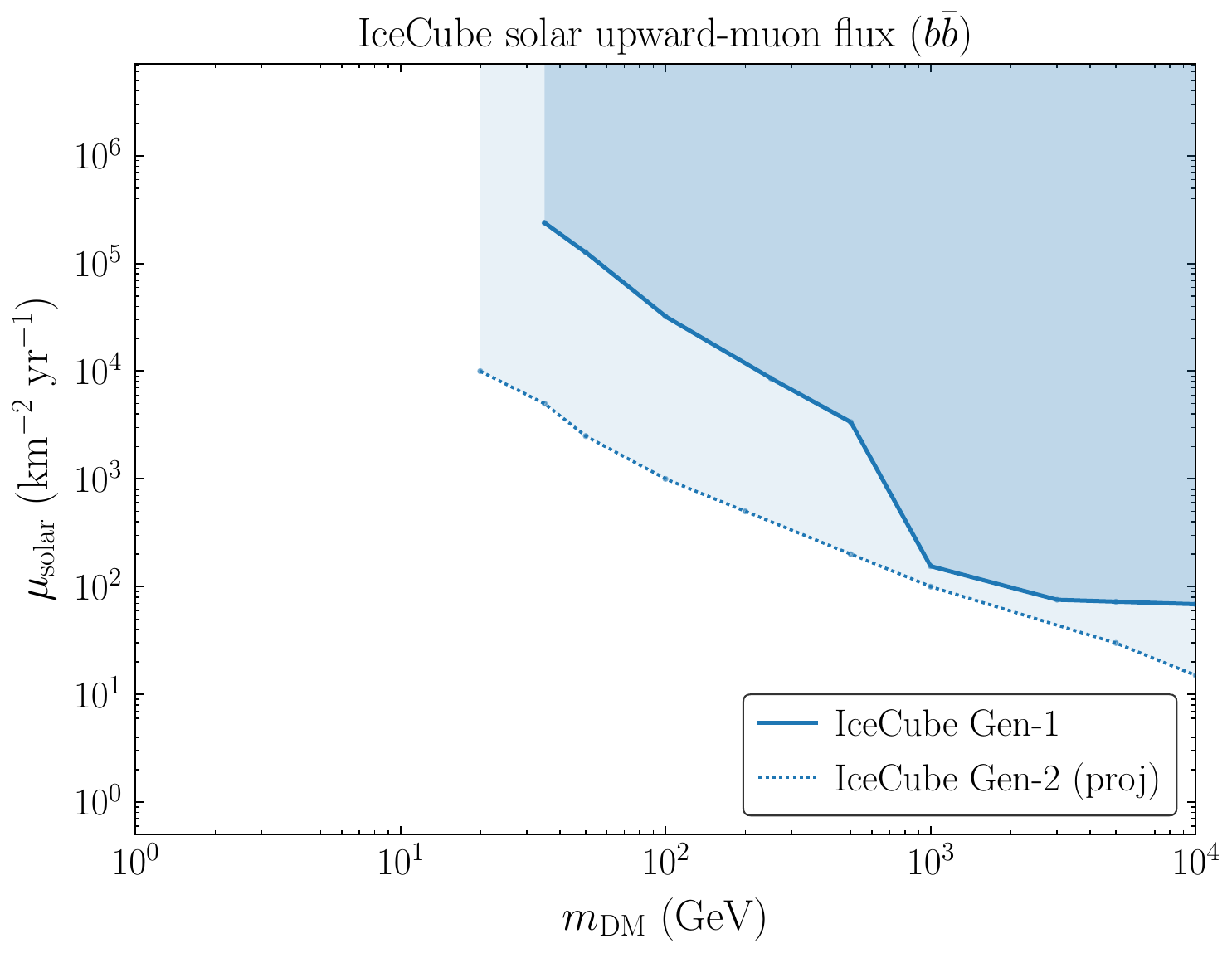}
    \caption{IceCube $b\bar b$ channel bound on $\mu_{\rm solar}$ interpolated from \cite{IceCube:2016dgk,IceCube-Gen2:2020qha}}
    \label{fig:icecube-bound}
\end{figure}
For IceCube the situation differs. \textsc{micrOMEGAs} outputs the upward-muon flux summed over all annihilation channels. That is the predicted flux $\mu_{\rm solar}$ is a single number and cannot (as far as we are aware) be decomposed to match a per-channel curve. We must therefore compare against one curve, and we choose the $b\bar b$ bound, which is the least constraining bound \cite{IceCube:2016dgk,IceCube-Gen2:2020qha,IceCube:2025fcu}. Since $L_{b\bar b}(m_{\rm DM})\ge L_{c}(m_{\rm DM})$, where $L$ is upper flux limit, for any final state, this choice can only under-constrain a model, never over-constrain it. We adopt the same conservative $b\bar b$ treatment for both the current (Gen-1) and projected (Gen-2) IceCube bounds. This bound can be seen in Figure \ref{fig:icecube-bound}.

Finally, for the relic abundance cut we introduce a parameter $\tau$ that parametrizes how wide our cut on relic density is, given by the correspondence show in Table \ref{tab:relicwidth}.

Table \ref{tab:relicwidth}. 
\begin{table}[h]
\centering
\caption{Example Relic-band $\tau$ and the acceptance range it
maps to.\footnote{Given by $\Omega_{\pm}(\tau)=\sqrt{\Omega_{\rm lo}\Omega_{\rm hi}}\cdot\left(\frac{\Omega_{\rm hi}}{\Omega_{\rm lo}}\right)^{\pm\tau/2}$ with $\Omega_{\rm lo}=0.118$, $\Omega_{\rm hi}=0.126$}}
\label{tab:relicwidth}
\begin{tabular}{cc}
\hline
$\tau$ & $\Omega h^2$ range \\
\hline
1  & $[0.118,\ 0.126]$ \\
10 & $[0.088,\ 0.169]$ \\
50 & $[0.024,\ 0.629]$ \\
\hline
\end{tabular}
\end{table}

\subsection{Invisible Higgs Branching Ratio}
\label{app:invHiggsanddrell}

For dark matter lighter than half the Higgs mass, the Higgs can decay
invisibly, $h\to\chi\chi$. We read the decay width directly from the
\textsc{micrOMEGAs} decay table, which is constructed for each model. For multi-component models, the Higgs decay table already contains a channel for each
dark sector ($h\to\chi_1\chi_1$, $h\to\chi_2\chi_2$, $\dots$). We define the total invisible width as the sum of the partial widths of all channels whose final-state particles are dark matter candidates,
\begin{equation}
  \Gamma_{\rm inv} = \sum_{c \in \{h\to\chi_i\chi_i\}} \Gamma(h\to c),
  \label{eq:gamma-inv}
\end{equation}

We use the SM width in the denominator to compute the branching ratio,
\begin{equation}
  \mathrm{BR}(h\to\mathrm{inv})
  = \frac{\Gamma_{\rm inv}}
         {\Gamma_h^{\rm SM} + \Gamma_{\rm inv}},
  \qquad \Gamma_h^{\rm SM} = 4.07~\mathrm{MeV}.
  \label{eq:br-inv}
\end{equation}
For $m_{\rm DM} \ge m_h/2$ the decay is kinematically closed, no invisible channel appears in the decay table, and $\mathrm{BR}(h\to\mathrm{inv}) = 0$. A point is rejected as non-viable when $\mathrm{BR}(h\to\mathrm{inv}) > 0.11$, the
current LHC bound \cite{LHCHiggsCrossSectionWorkingGroup:2016ypw,ATLAS:2023tkt,ATLAS:2020kdi}.

\subsection{\textsc{SModelS} Large Hadron Collider Bounds}
\label{app:smodels}
For every parameter point not excluded by other bounds, we attempt to rule out the proposed parameter point with LHC bounds using \textsc{SModelS} \cite{Alguero:2020grj}. For a given Lagrangian and parameter point, we first compute approximate production cross sections for a variety of dark-matter production channels as we describe below in Sec \ref{app:smodelscross} and export this along with dark-matter decay branching ratios in an SLHA file. \textsc{SModelS} then takes the SLHA file and computes the expected cross section of many possible final state topologies. \textsc{SModelS} then compares each topology's predicted cross section with the corresponding upper-limit in an official database of ATLAS and CMS results. The output is the ratio $r$ for every matched search. We use the most constraining value,
$r_{\max}$, and a parameter-point with $r_{\max}>1$ is LHC-excluded. The same setup, but where we rescale limits given the HL-LHC luminosity, provides the HL-LHC topology-reach testability flag of Appendix ~\ref{app:future-experiments}.

\subsubsection{Computing Approximate Cross Sections for SModelS}
\label{app:smodelscross}
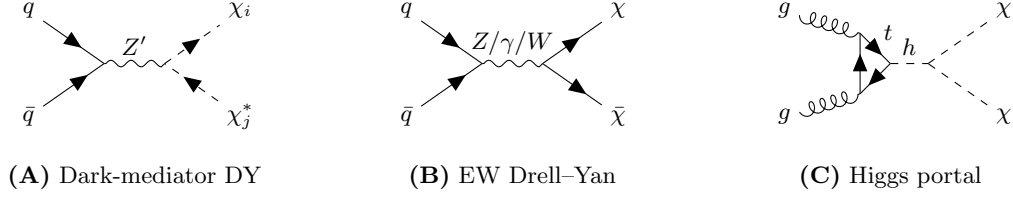
\begin{figure}[h]
\centering
\begin{tikzpicture}
\begin{scope}[shift={(0,0)}]
\begin{feynman}
  \vertex (qin)  at (-1.4, 0.7) {\(q\)};
  \vertex (qbin) at (-1.4,-0.7) {\(\bar q\)};
  \vertex (v1)   at (-0.4, 0);
  \vertex (v2)   at ( 0.4, 0);
  \vertex (chi)  at ( 1.4, 0.7) {\(\chi_i\)};
  \vertex (chib) at ( 1.4,-0.7) {\(\chi_j^*\)};
  \diagram*{
    (qin)  -- [fermion]      (v1),
    (v1)   -- [anti fermion] (qbin),
    (v1)   -- [photon, edge label=\(\Zp\)] (v2),
    (v2)   -- [charged scalar] (chi),
    (chib) -- [charged scalar] (v2),
  };
\end{feynman}
\node at (0,-1.5) {\textbf{(A)} Dark-mediator DY};
\end{scope}
\begin{scope}[shift={(5.0,0)}]
\begin{feynman}
  \vertex (qin)  at (-1.4, 0.7) {\(q\)};
  \vertex (qbin) at (-1.4,-0.7) {\(\bar q\)};
  \vertex (v1)   at (-0.4, 0);
  \vertex (v2)   at ( 0.4, 0);
  \vertex (chi)  at ( 1.4, 0.7) {\(\chi\)};
  \vertex (chib) at ( 1.4,-0.7) {\(\bar\chi\)};
  \diagram*{
    (qin)  -- [fermion]      (v1),
    (v1)   -- [anti fermion] (qbin),
    (v1)   -- [boson, edge label=\(Z/\gamma/W\)] (v2),
    (v2)   -- [fermion]      (chi),
    (chib) -- [anti fermion] (v2),
  };
\end{feynman}
\node at (0,-1.5) {\textbf{(B)} EW Drell--Yan};
\end{scope}
\begin{scope}[shift={(10.0,0)}]
\begin{feynman}
  \vertex (gin1) at (-1.4, 0.7) {\(g\)};
  \vertex (gin2) at (-1.4,-0.7) {\(g\)};
  \vertex (t1)   at (-0.4, 0.4);
  \vertex (t2)   at (-0.4,-0.4);
  \vertex (t3)   at ( 0.0, 0);
  \vertex (h)    at ( 0.5, 0);
  \vertex (chi)  at ( 1.5, 0.7) {\(\chi\)};
  \vertex (chib) at ( 1.5,-0.7) {\(\chi\)};
  \diagram*{
    (gin1) -- [gluon] (t1),
    (gin2) -- [gluon] (t2),
    (t1)   -- [fermion, edge label=\(t\)] (t3),
    (t3)   -- [fermion] (t2),
    (t2)   -- [fermion] (t1),
    (t3)   -- [scalar, edge label=\(h\)] (h),
    (h)    -- [scalar] (chi),
    (chib) -- [scalar] (h),
  };
\end{feynman}
\node at (0,-1.5) {\textbf{(C)} Higgs portal};
\end{scope}
\end{tikzpicture}
\caption{The three production channel classes. \textbf{(A)} $\Zp$-mediated
Drell--Yan for DM charged under $\Up$. \textbf{(B)} Electroweak Drell--Yan via
SM gauge bosons for DM in non-singlet $\SUL$ reps. \textbf{(C)} Higgs-portal
production (here gluon fusion through a top loop) for SM-singlet DM with no $\Up$.}
\label{fig:channels}
\end{figure}

For each model in the search space of Section~\ref{sec:problem}, we compute an approximate DM production cross section in the LHC. Depending on the quantum numbers of the dark matter field, this process is mediated by one or more of three distinct channels, shown in Figure~\ref{fig:channels}.

Channel~A in Figure~\ref{fig:channels} is the $\Zp$-mediated Drell--Yan process
$q\bar q \to \Zp{}^* \to \chi_i\chi_j^*$. For scalar dark matter its partonic
cross section is
\begin{equation}
\hat\sigma_A^{\rm scalar}(\shat; m_i, m_j)
 = \frac{g_q^2g_\chi^2}{144\pi}
   \frac{\shat \beta_{ij}^{3}(\shat)}
        {(\shat - \MZp^2)^2 + \MZp^2\GZp^2},
\label{eq:A_scalar}
\end{equation}
while for Dirac fermion dark matter it is
\begin{equation}
\hat\sigma_A^{\rm fermion}(\shat)
 = \frac{g_q^2g_\chi^2}{12\pi N_c}
   \frac{\shat \beta_\chi(1 + 2m_\chi^2/\shat)}
        {(\shat - \MZp^2)^2 + \MZp^2\GZp^2},
\label{eq:A_fermion}
\end{equation}
where $g_q$ and $g_\chi$ are the $\Zp$ couplings to the quark and to the dark
matter, $\MZp$ and $\GZp$ are the mediator mass and total width, with $\GZp$
computed per point as a sum over all kinematically open decay channels, $N_c = 3$,
and $\beta_{ij}(\shat) = \sqrt{1 - 2(m_i^2 + m_j^2)/\shat + (m_i^2 - m_j^2)^2/\shat^2}$
reduces to $\beta_\chi = \sqrt{1 - 4m_\chi^2/\shat}$ for the diagonal pair
$m_i = m_j = m_\chi$.

Channel~B in Figure~\ref{fig:channels} is electroweak Drell--Yan through the SM
$\gamma$ and $Z$ (and $W^\pm$ for charged-current pairs), available when the
dark matter sits in a non-singlet $\SUL$ multiplet $(n, Y)$. For scalar dark
matter,
\begin{equation}
\hat\sigma_B^{\rm scalar}
 = \frac{\shat\beta_\chi^3}{144\pi}
   \left|\frac{g_q^\gamma g_\chi^{\gamma}}{D_\gamma(\shat)}
       + \frac{g_q^Z g_\chi^{Z}}{D_Z(\shat)}\right|^2,
\label{eq:B_scalar}
\end{equation}
and for Dirac fermion dark matter,
\begin{equation}
\hat\sigma_B^{\rm fermion}
 = \frac{\shat\beta_\chi}{12\pi N_c}
   \left|\frac{g_q^\gamma g_\chi^{\gamma}}{D_\gamma}
       + \frac{g_q^Z g_\chi^{V}}{D_Z}\right|^2
   \left(1 + \frac{2m_\chi^2}{\shat}\right)
 + \frac{\shat\beta_\chi^3}{12\pi N_c}
   \left|\frac{g_q^Z g_\chi^{A}}{D_Z}\right|^2,
\label{eq:B_fermion}
\end{equation}
where $D_\gamma(\shat) = \shat$ and $D_Z(\shat) = \shat - m_Z^2 + i m_Z\Gamma_Z$
are the photon and $Z$ propagator denominators, and $g_\chi^{V,A}$ are the
vector and axial $Z$--dark-matter couplings. All Dirac multiplets we consider are vector-like. All couplings are fixed by gauge
invariance once $(n, Y)$ is specified, so Channel~B carries no free
parameters beyond the dark matter mass. The Majorana case in our space does not couple to the $Z$. 

Channel~C in Figure~\ref{fig:channels} is Higgs-portal production via gluon
fusion, $gg \to h^* \to \chi\chi$, through a top-quark loop. Under the
renormalizable restriction it exists only for scalar dark matter, with
partonic cross section
\begin{equation}
\hat\sigma_C^{\rm scalar}
 = \frac{\alpha_s^2 g_{h\chi\chi}^2}{576\pi v^2 \shat}
   \frac{|\mathcal{F}_{1/2}(\tau_t)|^2\beta_\chi}
        {(\shat - m_h^2)^2 + m_h^2\Gamma_h^2}\frac{1}{S},
\label{eq:C_scalar}
\end{equation}
where $g_{h\chi\chi} = 2\lambda_{h\chi} v$ for a real scalar and
$\lambda_{h\chi} v$ for a complex scalar, $\mathcal{F}_{1/2}(\tau_t)$ is the
top-loop form factor with $\tau_t = \shat/(4m_t^2)$, $v$ is the Higgs vacuum
expectation value, and $S = 2$ ($S = 1$) is the identical-particle factor for
a real (complex) scalar. There is no fermionic counterpart: the analogous
$|H|^2\bar\chi\chi$ coupling is dimension-5 and excluded by the renormalizable
restriction. 

Moreover, when a dark matter field carries both an $\SUL$ charge and a $\Up$ charge, the
$Z$ and $\Zp$ amplitudes contribute coherently to the same final state and
interfere. The squared matrix element is then
\begin{equation}
\left|\mathcal{M}_Z + \mathcal{M}_{\Zp}\right|^2
  = \left|\mathcal{M}_Z\right|^2
  + \left|\mathcal{M}_{\Zp}\right|^2
  + 2\mathrm{Re}\left(\mathcal{M}_Z^{*}\mathcal{M}_{\Zp}\right),
\label{eq:ABinterference}
\end{equation}
so that the total cross section is not the incoherent sum of the Channel~A and
Channel~B rates computed separately. The interference term must be retained.

Table~\ref{tab:primitives} lists every dark matter candidate in the search space of Section~\ref{sec:problem}, together with the dominant LHC production channel (Figure ~\ref{fig:channels}) for each.

\begin{table}[h]
\centering
\small
\begin{tabular}{c l l l l}
\toprule
\# & Lorentz & $(\SUL, Y)$ & $\Up$? & Channel(s) \\
\midrule
1 & complex scalar  & singlet, $Y{=}0$ & yes & A \\
2 & Dirac fermion   & singlet, $Y{=}0$ & yes & A \\
3 & real scalar     & singlet, $Y{=}0$ & no  & C \\
4 & complex scalar  & singlet, $Y{=}0$ & no  & C \\
5 & Majorana / Dirac fermion & singlet, $Y{=}0$ & no & --- \\
\midrule
6 & complex scalar & doublet, $Y{=}\tfrac{1}{2}$ & no  & B \\
7 & Dirac fermion  & doublet, $Y{=}\tfrac{1}{2}$ & no  & B \\
\midrule
8 & real scalar    & triplet, $Y{=}0$ & no & B \\
9 & Majorana       & triplet, $Y{=}0$ & no & B \\
10 & Dirac          & triplet, $Y{=}0$ & no & B \\
\midrule
11 & any of 6-10, \emph{plus} $Q^\prime \neq 0$ & --- & yes & A+B w/ interference \\
12 & multi-component $\chi_i \chi_j$, $i\neq j$ & --- & yes & A, off-diagonal \\
\bottomrule
\end{tabular}
\caption{Dominant production DM channel in the LHC in the search space of Section \ref{sec:problem}.}
\label{tab:primitives}
\end{table}

The partonic cross sections of Channels~A, B, and C give $\hat\sigma(\shat)$
for a fixed parton--parton energy $\sqrt{\shat}$, whereas the LHC collides
protons. To obtain the hadronic cross section we convolve each partonic cross
section $\hat\sigma_X$ ($X \in \{A, B, C\}$) with the corresponding parton
luminosity,
\begin{equation}
\sigma(pp \to \chi\chi)
 = \int d\tau dy
   \mathcal{L}(\tau, y; \mu_F)
   \hat\sigma_X(\shat = \tau s),
\label{eq:convolution}
\end{equation}
where $\sqrt{s}$ is the proton--proton centre-of-mass energy,
$\tau = x_1 x_2 = \shat / s$ and $y$ are the momentum fraction product and
rapidity of the produced system ($x_{1,2} = \sqrt{\tau}e^{\pm y}$), and
$\mathcal{L}(\tau, y; \mu_F)$ is the parton luminosity built from the parton
distribution functions evaluated at factorization scale $\mu_F = \sqrt{\shat}$.
The integration runs from the production threshold
$\tau_{\min} = (m_i + m_j)^2 / s$ up to $\tau = 1$. The same integration
machinery serves all three channels; only the initial state and the integrand
change. Channels~A and B ($\hat\sigma_A$, $\hat\sigma_B$) are quark-initiated,
so $\mathcal{L}$ is the $q\bar q$ luminosity summed over the active light
flavours, with the appropriate quark couplings inserted per flavour, while
Channel~C ($\hat\sigma_C$) is gluon-initiated and uses the $gg$ luminosity. The parton luminosities are built from parton distribution functions provided
by \textsc{LHAPDF}~\cite{Buckley:2014ana}, which supplies the proton's PDFs
$f_a(x, \mu_F)$ for each parton flavour $a$ at momentum fraction $x$ and
factorization scale $\mu_F$. We use the leading-order
\texttt{NNPDF31\_lo\_as\_0118} set~\cite{NNPDF:2017mvq}, whose perturbative
order matches that of the partonic cross sections computed above. The resulting cross section is passed to SModelS.

\subsection{EWSB, Mass Mixing, and Dark-U(1) Kinetic Mixing}
\label{app:massmix}

When two or more dark-sector components carry identical post-EWSB conserved quantum numbers, the gauge-basis fields are not mass eigenstates. The physical states and the physical couplings follow from diagonalizing the mass matrix, whose eigenvalues are the physical masses and whose eigenvectors make up the mixing matrix. \textsc{micrOMEGAs} requires these physical quantities to compute any observable, so the diagonalization must happen for every proposed parameter point \cite{Alguero:2023zol}.

Our pipeline detects how mixing must be done at model-build time. First, the $\mathrm{SU}(2)_L$ multiplets
are expanded into components, and second, a VEV-substitution is done for terms containing Higgs \footnote{Specifically, we take the Feynman gauge at tree level, setting the VEV to $v \approx 243$ GeV. A limitation in our work is that we do not impose a vacuum-stability requirement for parameter points. Our $\mathbb{Z}_n$ symmetry and positive coupling implies a local vacuum minimum, but not necessary a global minimum.} After which, all components of \emph{equal} electric charge, spin, and $\mathbb{Z}_n$ charge are grouped into a mixing block. The pipeline then diagonalizes each block numerically. For Scalars,
\begin{equation}
M^2 = O\mathrm{diag}(m_1^2,\dots,m_n^2)O^{\top},
\label{eq:massdiag}
\end{equation}
where $M^2$ is the $n\times n$ mass matrix squared in the gauge-basis. $O$ is the orthogonal mixing matrix whose columns are the eigenvectors of $M^2$. Thus, the eigenvalues $m_i^2$ are the squared masses of the physical states, and $O$ relates the two bases,
\begin{equation}
\phi_a^{\rm gauge} = \sum_{i} O_{ai}\phi_i^{\rm mass}.
\label{eq:massrotation}
\end{equation}
Our method of diagonalization is dependent on the block's spin \footnote{Given the symmetric mass matrix of a Majorana fermion, we use Takagi, for Dirac fermions we use the more general singular value decomposition.}, and the resulting masses and mixing matrices are injected into \textsc{CalcHep}.

\subsubsection{Kinetic Mixing}
When a model contains a dark $\mathrm{U}(1)'$, we do not handle the diagonalization of the neutral gauge sector numerically as we did above. We canonically normalize with a standard $Z'$ kinetic term field redefinition given in Eq 2.3 in \cite{Curtin:2014cca}, transform to the post-EWSB Standard Model basis via a Weinberg rotation $\{W_{1,2,3},B\}\to\{W^{\pm},Z,A\}$ 
\cite{Peskin:1995ev}, and then diagonalize the remaining non-diagonal mass matrix of $Z$ and $Z'$ as in Eq 2.6 \cite{Curtin:2014cca}. The result is that a (mostly) dark eigenstate couples to the SM photon, and we thus get one operator per Standard Model charged fermion,
\begin{equation}
\mathcal{L}_{Z'\bar f f} = \epsilon c_WeQ_fZ'_\mu\bar f\gamma^\mu f.
\end{equation}
This coupling is the SM portal that leads to dilepton and direct-detection signal in the $\mathrm{U}(1)'$ models\footnote{Our kinetic term is given by $\frac{\epsilon }{2}B_{\mu\nu} Z'^{\mu\nu}$, rather than $\frac{\epsilon}{c_W} B_{\mu\nu} Z'^{\mu\nu}$ as in \cite{Curtin:2014cca}. That is, we do not absorb $c_W \equiv \cos\theta_W$ into our definition of $\epsilon$.}.

The SM-photon stays massless throughout as $U(1)_{\rm EM}$ stays unbroken. The failure of this approximation at the Z pole $M_{Z'}\approx m_Z$ is a limitation of our pipeline. The window of non-validity of the approximation is $\epsilon$ dependent and only a few-GeV at its widest. A post-hoc scan over all viable points finds fewer than $0.1\%$ points landing near it.

\subsection{Limiting the Copy Coupling Combinatorics}
\label{app:darkmfv}

A field with $N$ copies is expanded into $N$ fields with identical quantum numbers and independent masses. Left unrestricted, the combinatorics explode. The number of (renormalizable) couplings of a Lagrangian grows as $\sim n^4$ in the number of fields, causing these models to quickly surpass the limit of what we can sample (our hard cap in training was $d=128$). To remedy this, we impose a discrete symmetry on the copied sector such that the copies of a given field are interchangeable everywhere except in their masses. That is, each copy retains an independent mass, but every copy carries the same self-coupling, every pair of copies carries the same cross-coupling (i.e. one $\beta$ for all $s_i^2 s_j^2$, $i\neq j$), and couplings to fields outside the copy set do not depend on
which copy is being acted on. 

\subsection{The Computational Pipeline}
\label{app:comp-pipeline}

Below we describe our computation pipeline summarized in Figure \ref{fig:pipeline-fig}.
\subsubsection{From Symmetries to Lagrangian }

The Mathematica package \textsc{Sym2Int}, presented in \cite{Fonseca:2017lem}, generates the list of Lorentz- and gauge-invariant operators that can appear in the Lagrangian of a given model, taking as input the internal symmetries and the field content. We have translated this package into Python, which we call \textsc{Sym2IntPy}. The gauge symmetry can be any product of simple compact Lie groups and $\U{1}$ factors, and for each field the user specifies its representation under the Lorentz group and under the gauge group, together with the number of generations. \textsc{Sym2IntPy} returns all invariant operators up to a mass dimension chosen by the user (we keep all 4-field operators), including those with covariant derivatives and gauge field-strength tensors. Redundant operators are eliminated using the equations of motion and integration by parts.

From the list of invariant operators obtained with \textsc{Sym2IntPy}, the \textsc{FlavorBuilder} package builds the explicit, fully expanded Lagrangian. Its input is the gauge group used by \textsc{Sym2IntPy} together with any discrete flavor group that further constrains the allowed terms. \textsc{FlavorBuilder} then expands the operators in both their gauge and flavor indices. To do this, the user specifies the names of the component fields, and the expansion is carried out using the corresponding gauge-group generators. The information on the discrete flavor group needed for the flavor-index expansion is obtained from the computer algebra system GAP~\cite{GAP4} through the \textsc{FlavorBuilder} subpackage \textsc{PyDiscrete}, a Python translation of \textsc{Discrete}~\cite{Holthausen:2011vd}.

\subsubsection{From Lagrangian to Observable}
After generating the \textsc{SymPy} symbolic Lagrangian, we normalize our dark-U(1) kinetic terms if present, and we construct post-EWSB mass matrices (see Appendix \ref{app:massmix}). The result is saved as a \textsc{LanHEP} model file (the standard format needed for \textsc{CalcHep} and \textsc{micrOMEGAs} \cite{Semenov:2008jy,Alguero:2023zol,Belyaev:2012qa}), where Lagrangian parameters (masses and couplings) are taken as the file's inputs. \textsc{LanHEP} then compiles this file into \textsc{CalcHep} model files (particles, vertices, and parameter tables), and we parse the resulting free parameters to generate a \texttt{main.c} executable for \textsc{micrOMEGAs}. In this executable, mass matrices are diagonalized numerically at evaluation time, once per proposed gauge-basis parameter point.

Each parameter point is evaluated by running that executable, and we thus build each model's \texttt{main.c} once per Lagrangian (not per parameter point). Given \texttt{main.c}, we compute (via micrOMEGAs) the relic abundance, the spin-independent and spin-dependent direct-detection cross-sections, the present-day annihilation spectra, and indirect-detection fluxes, and the Higgs decay table, from which we can compute the invisible branching ratio $\mathrm{BR}(h\to\mathrm{inv})$ (see Appendix \ref{app:invHiggsanddrell}). 

For points that survive these micrOMEGAs-level cuts \emph{and} contain electroweak-charged multiplets, we also apply LHC limits through \textsc{SModelS}. For each DM field, we numerically compute the expected $pp\rightarrow\chi_i\chi_i$ cross-section (to good approximation, see App.~\ref{app:smodels} for more detail). We save the computed cross section and the decay table for each $\chi_i$ from \textsc{micrOMEGAs} into an SLHA file, which is passed to \textsc{SModelS}.\textsc{SModelS} then decomposes this input into a set of expected final state cross-sections, compares with LHC-bounds for those, and returns the maximal ratio $r_{\max}$ of all predicted rates to all LHC-observed $95\%$ CL limits.

As we described in Sec \ref{sec:problem}, a parameter point is declared viable when it passes every active cut of Table \ref{tab:viability} for the episode's configuration (see Section \ref{sec:ExperimentalConstraints}). The same observables are compared against the projected sensitivities of future experiments to assign a \emph{testability} score. The viability flag and testability score are returned for every point an RL agent proposes.

\subsection{Allowed Quantum Numbers, Parameter Ranges, and Representative Examples}
Given a Lagrangian with $d$ gauge-basis parameters, the agent proposes a point in $[0,1]^d$, which is mapped to physical Lagrangian parameters, given the type-dependent parameter ranges given in Table \ref{tab:params}. Moreover, the ranges in Table \ref{tab:params} are the maximum ranges. At the start of each episode, a tighter sub-range is sampled independently for each parameter type, with a minimum width of 0.2 of the full logarithmic range described in Table \ref{tab:params}, and the model is conditioned on the sampled sub-ranges. This teaches a single agent to scan both broad and zoomed-in slices of parameter space. Table \ref{tab:examples} contains small representative dark matter Lagrangian examples from our space, and Table \ref{tab:reps} counts allowed quantum numbers.

\begin{table}[h]
\centering
\caption{Representative scalar dark-matter Lagrangians from the agent's
search space. $H$ is the Higgs doublet, $s$ is a real scalar, $S$ is a complex
scalar, and $Z'$ is the $U(1)'$ gauge boson with charge $q$. Kinetic and mass
terms are not shown except for $U(1)'$. $\alpha_i$ are coupling parameters the agent learns to scan through and $g_{Z'}$, $\epsilon$ the dark $U(1)'$'s coupling and kinetic mixing terms.}
\label{tab:examples}
\begin{tabular}{cll}
\hline
\# & Model & $\in\mathcal{L}_{\rm BSM}$ \\
\hline
1 & $\mathbb{Z}_2$ real scalar singlet
  & $-\alpha_1 H^\dagger H s^2 - \alpha_2 s^4$ \\
2 & $\mathbb{Z}_2$ complex scalar singlet
  & $-\alpha_1 H^\dagger H|S|^2 - \alpha_2 |S|^4
     - \left(\alpha_3 H^\dagger H S^2 + \alpha_4 S^4 + \text{h.c.}\right)$ \\
3 & $\mathbb{Z}_2$ complex scalar singlet (secluded)
  & $-\alpha_1 H^\dagger H s_1^2 - \alpha_2 s_1^4 - \alpha_3 s_1^2 s_2^2 - \alpha_4 s_2^4$ \\
4 & $\mathbb{Z}_2$ complex scalar singlet $+$ dark $U(1)'$
  & $-\alpha_1 H^\dagger H|S|^2 - \alpha_2|S|^4
     + |(\partial_\mu - i g_{Z'} q Z'_\mu)S|^2
     - \tfrac{\epsilon}{2} B_{\mu\nu}Z'^{\mu\nu}
     + \tfrac12 M_{Z'}^2 Z'_\mu Z'^\mu$ \\
\hline
\end{tabular}
\end{table}

\begin{table}[h]
\centering
\caption{Spin and quantum numbers for a single dark multiplet we consider. Electrical neutrality requires $Y=0$ for singlets and $Y=\pm\frac{1}{2}$ for doublets and we restrict to  $Y=0$ for triplets. Real scalars or Majorana fermions cannot carry hypercharge, and similarly cannot carry a dark U(1)' charge. \cite{Peskin:1995ev}. }
\label{tab:reps}
\begin{tabular}{lccccc}
\hline
$\mathrm{SU}(2)_L$ & $Y$ & Real scalar & Complex scalar & Majorana & Dirac \\
\hline
Singlet & $0$            & \checkmark & \checkmark & \checkmark & \checkmark \\
Doublet & $\tfrac12$     & $\times$         & \checkmark & $\times$         & \checkmark \\
Triplet & $0$            & \checkmark & \checkmark & \checkmark & \checkmark \\
\hline
\end{tabular}
\end{table}
\begin{table}[h]
\centering
\caption{Continuous parameters sampled by the agent and their full ranges. Each episode further restricts every type to a randomly drawn sub-range of the full range below.\footnote{Our EFT scale range is  $\Lambda\in[10^{-1},10^{3}]$ GeV for the two non-renormizable operators we consider. No points are accepted that do not satisfy $m_{\rm DM}\leq \Lambda$. This is not relevant for our scan of scalar models who contain only renormalizable operators.}}
\label{tab:params}
\begin{tabular}{lll}
\hline
Type & Examples & Maximum Allowed Range \\
\hline
DM or mediator mass & $M_{\rm DM}$, $M_{Z'}$        & 1 GeV to 10 TeV \\
Dimensionless Coupling           & $a_2$, $\lambda_i$, $g_{Z'}$  & $10^{-2}$ to $4\pi$\\
Kinetic Mixing Parameter         & $\epsilon$  & $10^{-6}$ to $10^{-1}$ \\
\hline
\end{tabular}
\end{table}
\subsection{Future Experiments}
\label{app:future-experiments}

Twelve experiments are chosen as observables for the non-LLM decision tree and the testability reward. They are the next-generation direct detection (XLZD, full-exposure LZ, DarkSide-20k, SuperCDMS, PICO-500), indirect detection (CTA, Fermi-LAT 15-yr, IceCube-Gen2), and projected invisible-Higgs measurements (HL-LHC and FCC-ee) and an HL-LHC pair production testability flag obtained by rescaling the \textsc{SModelS} $r_{\rm max}$ of the viability pipeline to the projected HL-LHC exposure \cite{DARWIN:2016hyl,XLZD:2024nsu,LZ:2018qzl,DarkSide-20k:2017zyg,PICO:2019vsc,CTA:2020qlo,Fermi-LAT:2015att,IceCube-Gen2:2020qha,Cepeda:2019klc,FCC:2025lpp,ATLAS:2019erb,ATLAS:2018tvr,SuperCDMS:2016wui,Fermi-LAT:2016afa}. Together they cover the reach of many upcoming experiments, and they reuse exactly the signatures \textsc{micrOMEGAs} already computes for the viability, so no additional computation is needed. See Table \ref{tab:future} for a summary.

\begin{table}
\centering
\caption{Projected experiments scored as testability targets. Bounds are interpolated log-log in $m_{\rm DM}$ when needed. HL-LHC and FCC-ee Invisible Higgs thresholds are flat in $m_{\rm DM}$. The HL-LHC collider testability flag is computed by rescaling the \textsc{SModelS} $r_{\rm max}$ to projected HL-LHC luminosity. How we interpolate bounds from literature values is given in App. \ref{app:interp}.}
\label{tab:future}
\begin{tabular}{lllll}
\hline
Experiment & Observable & Threshold & Software \\
\hline
XLZD        & $\sigma_{\rm SI}$           &  bound on $\sigma_{\rm SI}$ as a function of $m_{\rm DM}$ \cite{XLZD:2024nsu} & \textsc{micrOMEGAs} \\
LZ full exposure   & $\sigma_{\rm SI}$           & bound on $\sigma_{\rm SI}$ as a function of $m_{\rm DM}$ \cite{LZ:2018qzl} & \textsc{micrOMEGAs} \\
DarkSide-20k       & $\sigma_{\rm SI}$           & bound on $\sigma_{\rm SI}$ as a function of $m_{\rm DM}$ \cite{DarkSide-20k:2017zyg} & \textsc{micrOMEGAs} \\
SuperCDMS      & $\sigma_{\rm SI}$           & bound on $\sigma_{\rm SI}$ as a function of $m_{\rm DM}\leq10$ GeV \cite{SuperCDMS:2016wui} & \textsc{micrOMEGAs} \\
PICO-500      & $\sigma^p_{\rm SD}$         & bound on $\sigma^p_{\rm SD}$ as a function of $m_{\rm DM}$ \cite{PICO:2019rsv} & \textsc{micrOMEGAs} \\
CTA                & $\langle\sigma v\rangle_{c^\star}$ &  annihilation-channel dependent bound  \cite{CTA:2020qlo} & \textsc{micrOMEGAs} \\
Fermi-LAT 15-yr  & $\langle\sigma v\rangle_{c^\star}$ &  annihilation-channel dependent bound \cite{Fermi-LAT:2016afa} & \textsc{micrOMEGAs} \\
IceCube-Gen2            & $\mu_{\rm solar}$ & $b\bar b$ bound as a function of $m_{\rm DM}$ \cite{IceCube-Gen2:2020qha} & \textsc{micrOMEGAs} \\
HL-LHC inv Higgs      & $\mathrm{BR}(h\to\mathrm{inv})$  & $>0.025$ \cite{Cepeda:2019klc} & \textsc{micrOMEGAs}, see App. \ref{app:invHiggsanddrell} \\
FCC-ee inv Higgs     & $\mathrm{BR}(h\to\mathrm{inv})$ & $>0.003$ \cite{FCC:2025lpp} & \textsc{micrOMEGAs}, see App. \ref{app:invHiggsanddrell} \\
HL-LHC pair production    & $r_{\rm max}^{\rm HL}$ & luminosity-rescaled $r_{\rm max}^{\rm HL}>1$ & \textsc{SModelS}, see App. \ref{app:smodels}\\
\hline
\end{tabular}
\end{table}

\section{Large Lagrangian Model Architecture and Training}

\subsection{Input Tokens}
Below we describe each input and then describe the model's body and output heads in detail.
\begin{figure}
    \centering
    \includegraphics[width=0.65\linewidth]{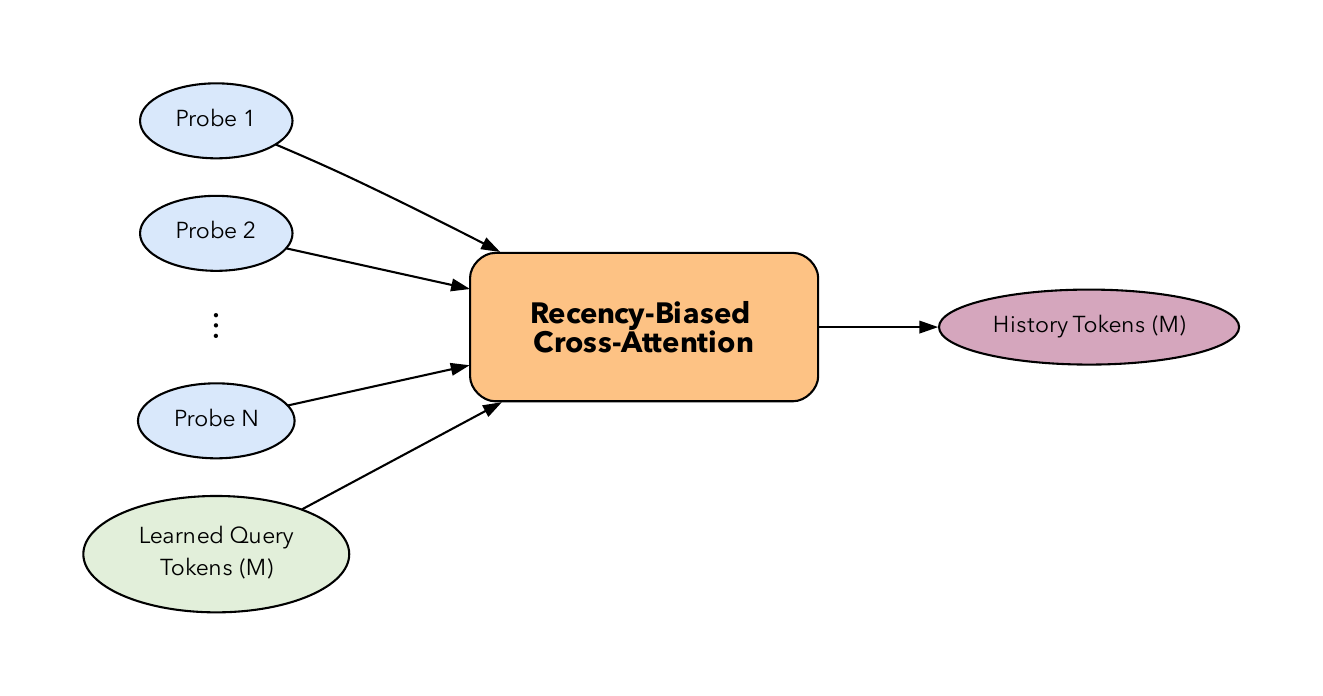}
    \caption{The probe-history encoder. $M$ learned queries
    cross-attend, with a learned recency bias, over the embeddings of
    every probe taken so far, and the $M$ outputs join the input token set as
    history tokens.}
    \label{fig:hist_and_ar}
\end{figure}
\begin{enumerate}
    \item \textbf{Lagrangian Input}. Given the set of possible choices in
Section ~\ref{sec:problem}, we map each randomly sampled Lagrangian to a set of tokens. Each dark-sector
field is represented a one-hot vector encoding its configuration (spin, $SU(2)$ rep, etc.) which is then embedded via an MLP into a \emph{field token}, one per field. The remaining global content (field count, $\mathbb{Z}_N$, and a binary $U(1)'$ indicator) is embedded into a single \emph{global Lagrangian token}. See Figure \ref{fig:Lagrangian_tokens} for the encoding of an example Lagrangian.

\item \textbf{Viability Configuration Input.} The quantities defining ``viable'' in a given episode are
inputs, so a single network can be trained on different experimental constraint configurations. Each experimental cut (direct detection, invisible Higgs width, LHC with \textsc{SModelS}, etc.) contributes one \emph{cut token} carrying its identity and if it is active in a given episode.

Further, the allowed scan range of a parameter also changes per episode and we thus introduce a \emph{range token} to inform the network. A \emph{budget token} encodes the episode's turn budget (in our case, randomly sampled from $T\in[5,50]$), a \emph{current-turn-token} encodes the current turn, a \emph{probes-used-token} encodes the number of probes used for that current turn, and a \emph{relic-width token} encodes the width of the allowed relic-density range $\tau$. All of these are randomized per episode
during training.

\item \textbf{State and Region Input.} Episode-level summaries are encoded through two types of tokens built each turn. \emph{State tokens} summarize the search so far. One token
carries the viable and non-viable fractions of all probes to date, and one carries the previous turn's \textsc{SModelS} results (fraction evaluated, fraction excluded, mean $r_{\max}$).
\emph{Region tokens} describe up to eight discovered \emph{regions},
each embedding the region's signature centroid, size, and boundary
fraction. Here, a region is defined as a connected component of viable
points in parameter space with the same testability signature.

\item \textbf{History Tokens} give the model access to every individual probe of the episode. Each past probe is embedded from its signature (viability and testability flags), tagged with the identity of the policy head that proposed it and an absolute-turn positional encoding. A set of $M=64$ learned queries then
cross-attends over all probe embeddings (see Figure ~\ref{fig:hist_and_ar}). Similar to \cite{Elsharkawy:2026kwp,Mikuni:2025ocp,Bhimji:2025isp}, this attention is subject to a learned
\emph{recency bias}. An integer that describes a probe's age in turns is added to the cross-attention logits, that lets the encoder re-weight probes by how recently they were taken. The $M$ query outputs join all other tokens as history tokens. This running, attention summarized, map of where viable territory has
been found is what lets the agent's next proposals depend on what previous turns revealed.

\item \textbf{Autoregressive Chain and the Value  Output Tokens}  Since parameter values are sampled autoregressively (i.e. model outputs a distribution of each parameter one at a time and ordered, see Figure \ref{fig:ar_samp}), we must embed the value sampled for the previous parameters (and use a learned \textsc{bos} token for the first). We thus feed an autoregressive \emph{chain token}. These chain tokens attend to the full token set and, causally, to their own token (i.e. to previous chain tokens), while context tokens never attend back. That is, the state encoding is computed once per turn and is independent of the actions sampled from it. These tokens are used to sample the proposed probe outputs. 

Finally, the \emph{value token}, akin to a cls token \cite{devlin2019bertpretrainingdeepbidirectional} in language models, is a single learned vector. It attends to all context tokens (but, like them, never to the autoregressive chain), and its post-body embedding is read by the value head to produce the per-turn value (needed for PPO).

\end{enumerate}
\subsection{Concentration Schedule.}
\begin{figure}[t]
    \centering
    \includegraphics[width=.7\linewidth]{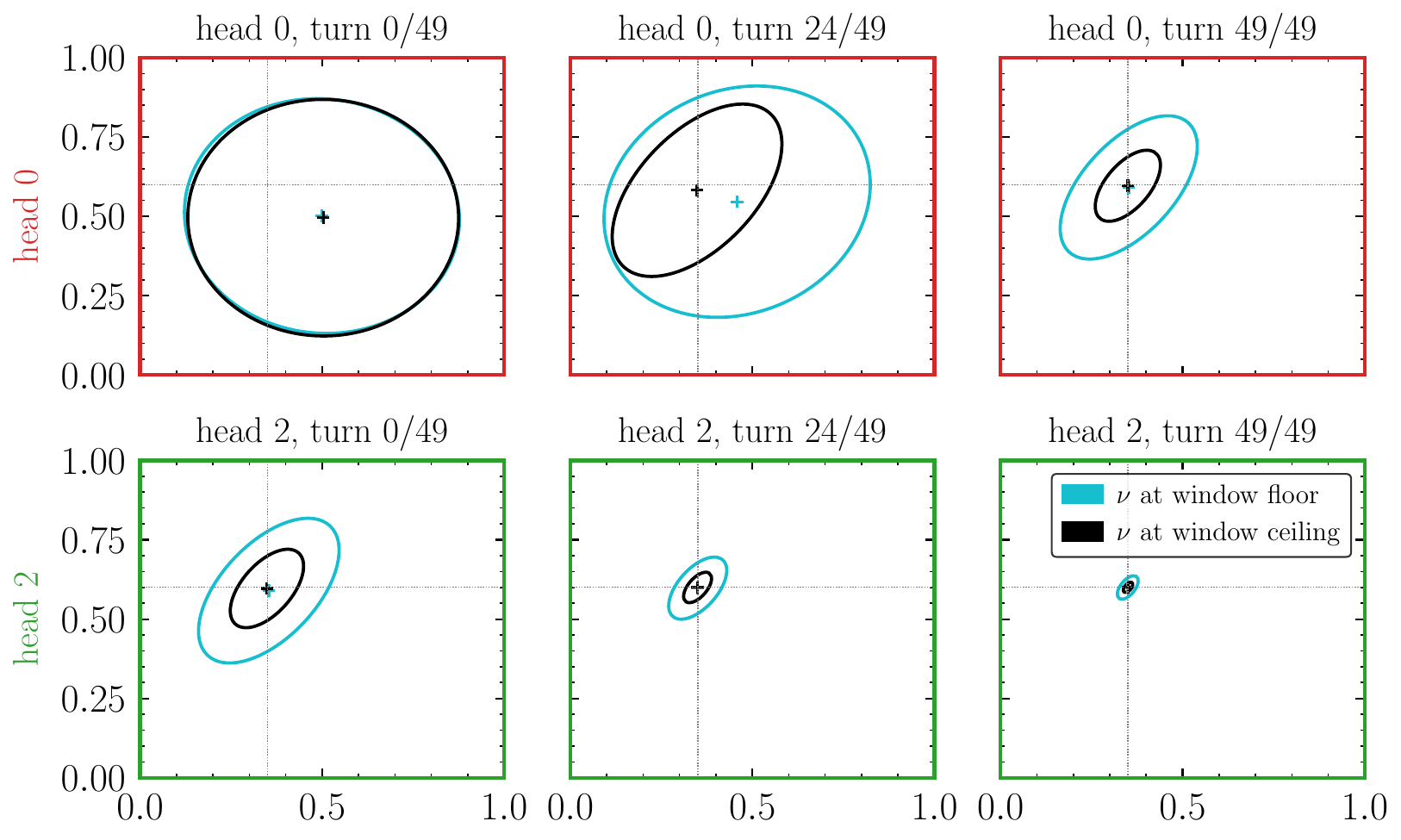}
\includegraphics[width=0.7\linewidth]{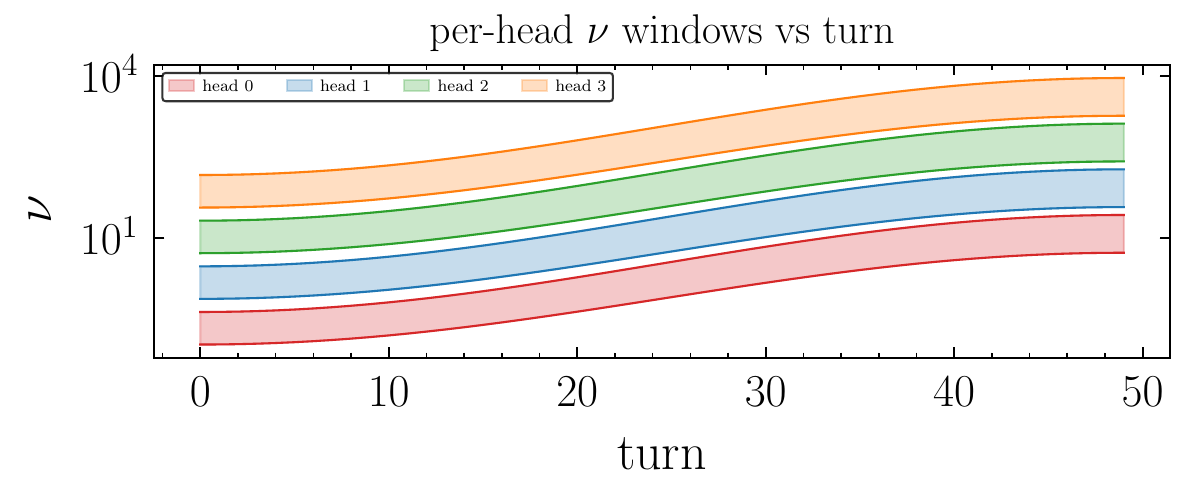}
    \caption{(\emph{Upper}) Effect of the $\nu$ window on a two-parameter toy parameter space.
    $1\sigma$ contours of the broadest (cyan) and sharpest (black)
    densities each window permits, for a broad and a sharp head at the
    start, middle and end of an episode. (\emph{Lower}) Per-head concentration windows over an episode with $T=50$ turns per episode. Each head's hard $\nu$ window slides from
    broad (exploration) to sharp (exploitation), with a stagger for a division of labor across heads.}
    \label{fig:nu_ellipses}
\end{figure}
We found it essential to control the explore-exploit choice of the beta distribution by hand rather than leaving it to be learned. We impose a hard, non-learnable window on each head's concentration (roughly the inverse of the variance of the output Beta distribution), scheduled over the episode. For a given turn $t$ out of a $T$ turn episode, we define a schedule,
\begin{equation}
    \rho(t)=\tfrac12\left(1-\cos\frac{\pi t}{T-1}\right)
\end{equation}
and then restrict the allowed concentration to a turn-dependent window given by $\nu \in \left[\nu_{\rm lo}(t),\nu_{\rm hi}(t)\right]$ and a schedule for the first policy heads window given by, 
\begin{align}
\log \nu_{\rm lo}(t) &= [1-\rho(t)]\log 2 + \rho(t)\log 100,
\label{eq:nuwindow_lo}\\
\log \nu_{\rm hi}(t) &= [1-\rho(t)]\log 8 + \rho(t)\log 500.
\label{eq:nuwindow_hi}
\end{align}
That is, the $\nu \in \left[\nu_{\rm lo}(t),\nu_{\rm hi}(t)\right]$ band slides for the first policy head from $[2,8]$ at turn 0 (broad, exploration is forced) to $[100,500]$ at the final turn (sharp, exploitation is forced). The network chooses the means and where in the window each $\nu$ sits, but the broad-to-sharp structure cannot be violated. See Figure \ref{fig:nu_ellipses} for a representation of this schedule. 

The four heads are also varied along this schedule. That is, the $\nu \in \left[\nu_{\rm lo}(t),\nu_{\rm hi}(t)\right]$ band varies between heads in a controlled manner to hard-wire a division of labor from a broad scouting head to a fine-resolution head. The $N=128$ probes of a turn are split evenly across heads, so each head's joint distribution is
sampled $N_{\rm probes}/H=32$ times, and each sample is mapped to physical Lagrangian
parameters by the rules of Sec.~\ref{sec:problem}. Also see Figure \ref{fig:nu_ellipses} for a representation of this spread. 

\subsection{Differential Evolution}
\label{app:diffevol}
For both differential evolution as a baseline and the differential evolution path in the model, the minimized objective for a probe $\theta$ is
\begin{equation}
    f(\theta) = \tfrac{1}{2} \Omega^{2}(\theta)
+ \tfrac{1}{2} \sum_{e} \mu_e(\theta)/\sigma^{2}
- 10^{3}\cdot (\texttt{is $\theta$ viable})
\end{equation}
where $\Omega(\theta)$ is the deviation of the probe's relic density from the experimental value, $\mu_e$ is the size by which the point exceeds the experimental bound $e$ summed over all experiments, $\sigma = 0.5$, and the last term adds $-10^{3}$ only for viable points so that viability dominates the objective.

\subsection{Model and Training Hyperparameters}
\label{app:modeltraininghparams}
Table \ref{tab:hparams_shared} contains shared Hyperparameters for both sizes for the LLaM. Table \ref{tab:hparams_sizes} contain hyperparameter distinct to each model size.

\begin{table}[h]
\centering
\small
\begin{tabular}{l cc}
\toprule
 & \textbf{small} & \textbf{medium} \\
\midrule
Token hidden dimension $d_{\rm tok}$ & 256  & 512 \\
Body depth                           & 4    & 12 \\
Body attention heads                 & 8    & 16 \\
MLP width                            & $4d_{\rm tok}=1024$ & $4d_{\rm tok}=2048$ \\
\midrule
Total parameters                     & $4.8$M & $44.0$M \\
\bottomrule
\end{tabular}
\caption{LLaM-small and LLaM-medium hyperparameters. Parameters shared between the two sizes are given in Table~\ref{tab:hparams_shared}.}
\label{tab:hparams_sizes}
\end{table}
\begin{table}[h]
\centering
\scriptsize
\renewcommand{\arraystretch}{1.15}
\begin{tabular}{l c @{\hspace{1.5em}} l c}
\toprule
\multicolumn{4}{c}{\textbf{Shared hyperparameters}} \\
\midrule
\multicolumn{2}{l}{\emph{Architecture}} & \multicolumn{2}{l}{\emph{PPO and Training Parameters}} \\
Policy heads $H$                          & 4      & Learning rate (AdamW, RL and Pre)& $1\times10^{-4}$ \\
History query tokens $M$                  & 64     & LR warmup episodes for RL  & 1000 \\
Attention head dim                        & 32     & Clip $\epsilon$      & 0.1 \\
Max Lagrangian parameter dim $d_{\max}$   & 128    &                      & \\
Probes per turn $N$                       & 128    &                      & \\
Sampled Episode Budget                    & 5--50  &                      & \\
\midrule
\multicolumn{2}{l}{\emph{Beta-Distribution Parameters}} & \multicolumn{2}{l}{\emph{Differential-evolution head}} \\
$[\nu_{\rm lo},\nu_{\rm hi}]$ at turn 0        & $[2, 8]$     &  Top-$K$ for $\theta_b$ selection & 128 \\
$[\nu_{\rm lo},\nu_{\rm hi}]$ at turn $T=50$   & $[100, 500]$ & $F$            & $0.7 \pm 0.2$ \\
Beta Distribution Parameter Floors $\alpha,\beta \geq$ & 0.4  & $c$  & 0.7 \\
$\nu$ spread for Head $H\in[1,4]$              & $\times 7.0^{k-(H-1)/2}$ & & \\
\bottomrule
\end{tabular}
\caption{Hyperparameters shared between the two model sizes.}
\label{tab:hparams_shared}
\end{table}

\section{On Building the Tree and LLM-agent Prompts}
\label{app:tree-app}

\subsection{Deterministic Observable Bin Widths}
\label{app:tree-bins}
Bin widths are determined by the expected systematic error of each observable. High-error observables are binned into large bins, while less-error prone observables are binned much tighter (See Table~\ref{tab:tree-observables}). Indirect-detection signatures are annihilation channel dependent, and we thus compute and bin $\mu_e$ per channel. HL-LHC signatures can also be split by final-state topology, and here we choose to keep HL-LHC binary for each final-state for simplicity. See Table \ref{tab:tree-observables} for a summary of all observables we use to build our tree along with how much we bin $\mu_e$ by and any channel dependence. 

\begin{table}[h]
    \centering
    \caption{Observables that are used to split nodes in the decision tree. Each
    $\mu_e = \log_{10}(O_{\rm pred}/O_{\rm lim})$ is a continuous ratio of
    the experiment's projected limit in log space. We discretize this $\mu_e $ into fixed-width
    bins with overflow bins. All values below the lower overflow bin are merged into a single ``far-below-reach'' bin, all values above the upper overflow bin into a single
    ``far-above-reach'' bin.}
    \label{tab:tree-observables}
    \vspace{1em}
    \begin{tabular}{llcc}
        \toprule
        Observable & Channel Dependence & Bin Width in $\log_{10}$ & Overflow Bin Edges in $\log_{10}$\\
        \midrule
        $\sigma_{\rm SI}$ for XLZD, LZ, and DarkSide & N/A & 1 & $\pm 2$ \\
        $\sigma_{\rm SD}$ for PICO & N/A & 1 & $\pm 2$ \\
        $\langle\sigma v\rangle$ for CTA & channel ($b\bar b$, $WW$, $\tau\tau$) & 1 & $\pm 1.5$ \\
        $\langle\sigma v\rangle$ for Fermi-LAT (15 yr) & channel ($b\bar b$, $WW$, $\tau\tau$) & 1 & $\pm 1.5$ \\
        Upward muon flux for IceCube-Gen2 & N/A & 1 & $\pm 2$ \\
        BR($h\to$ inv): HL-LHC & N/A & 0.5 & $\pm 3$ \\
        Topology reach: HL-LHC & final state & N/A & N/A  \\
        \bottomrule
    \end{tabular}
\end{table}

\subsection{Choosing Splits That Maximize Information Gain}
\label{app:infogain}
The best split at a given node is the measurement that most unmixes the signature classes. This can be computed numerically with so-called information gain. For a set of viable points $S$ split into
$S_{\rm yes}$ and $S_{\rm no}$ we can compute $H(S) - \sum_{i \in \{\rm yes, no\}} \tfrac{|S_i|}{|S|} H(S_i)$, where $H(S) = -\sum_c p_c \log p_c$ and $p_c$ is the fraction of points in $S$ with signature class $c$. $H$ is largest when many classes are mixed together and zero when only one remains.



\subsection{Defining Regions}
\label{app:regions}
Regions are computed by clustering the viable points with \textsc{DBSCAN} in $\log\theta$, using a neighborhood radius $\epsilon = 0.8\sqrt{d/3}$ where d is the dimension of the parameter space, and with the minimum number of points being $5$ (that is, fewer points can never be called a region). We then define the region as a d-dimensional bounding box of the DBSCAN clustered parameter points. We include a filter where, if one region's parameter ranges are strictly contained in another, the regions are combined. 


\section{Benchmark and One Dark-Multiplet Scan Configuration and Detailed Results}
\subsection{The Benchmark}
\label{app:lag-bench}
Here we describe in detail each Lagrangian used in the 21-Lagrangian benchmark and give per-policy, per-budget, and per-Lagrangian results with the same metrics described in Section \ref{sec:LLaMResults}. Table \ref{tab:benchmark-classes} contains the 21-Lagrangian we use, numbered and ordered by their parameter space dimension $d$. We use this numbering to organize per-Lagrangian results in the following Tables. Table \ref{tab:policy-benchmark-per-Lagrangian} gives per-Lagrangian results summed over all budgets and repeats similar to Table \ref{tab:policy-benchmark-combined}. Tables \ref{tab:policy-benchmark-per-Lagrangian-b5} through \ref{tab:policy-benchmark-per-Lagrangian-b50} give per-Lagrangian results for each budget size. We plot 3D depiction of all LLaM-small search episodes that found viable points in Figure \ref{fig:3d-episodes-1} and Figure \ref{fig:3d-episodes-2}.

\begin{table}[h]
    \centering
    \caption{The 21 Lagrangian classes of the benchmark with $n_f$ number of fields,
    ordered by parameter-space dimension $d$.}
    \label{tab:benchmark-classes}
    \vspace{1em}
    \scriptsize
    \begin{tabular}{rlccc}
        \toprule
        \# & Field content & parameter dim $d$ & $n_f$ & U(1)$'$ \\
        \midrule
        1  & $\mathbb{Z}_3$: Real Scalar Singlet & 3 & 1 & -- \\
        2  & $\mathbb{Z}_3$: Majorana Singlet & 3 & 1 & -- \\
        3  & $\mathbb{Z}_3$: Majorana Triplet & 3 & 1 & -- \\
        4  & $\mathbb{Z}_3$: Dirac Doublet & 4 & 1 & -- \\
        5  & $\mathbb{Z}_3$: Dirac Triplet & 4 & 1 & -- \\
        6  & $\mathbb{Z}_3$: Complex Scalar Doublet & 5 & 1 & -- \\
        7  & $\mathbb{Z}_3$: $3\times$ Majorana Singlet & 5 & 1 & -- \\
        8  & $\mathbb{Z}_3$: Dirac Singlet & 6 & 1 & \checkmark \\
        9  & $\mathbb{Z}_3$: Majorana Singlet $+$ Dirac Doublet & 8 & 2 & -- \\
        10 & $\mathbb{Z}_3$: Complex Scalar Singlet & 10 & 1 & \checkmark \\
        11 & $\mathbb{Z}_3$: Complex Scalar Singlet & 17 & 1 & -- \\
        12 & $\mathbb{Z}_5$: Complex Scalar Singlet & 17 & 1 & -- \\
        \addlinespace
        13 & $\mathbb{Z}_4$: Dirac Doublet $+$ $3\times$ Complex Scalar Singlet $+$ $2\times$ Dirac Doublet & 27 & 5 & \checkmark \\
           & $\quad+$ Dirac Singlet $+$ Dirac Singlet & & & \\
        \addlinespace
        14 & $\mathbb{Z}_4$: Real Scalar Singlet $+$ Majorana Singlet $+$ Dirac Singlet & 28 & 3 & -- \\
        \addlinespace
        15 & $\mathbb{Z}_2$: Complex Scalar Doublet $+$ Dirac Singlet $+$ $3\times$ Dirac Singlet & 30 & 4 & \checkmark \\
           & $\quad+$ $2\times$ Dirac Doublet & & & \\
        \addlinespace
        16 & $\mathbb{Z}_4$: $2\times$ Dirac Doublet $+$ $2\times$ Complex Scalar Singlet $+$ $3\times$ Complex Scalar Doublet & 30 & 3 & \checkmark \\
        17 & $\mathbb{Z}_3$: Complex Scalar Singlet $+$ Dirac Singlet $+$ Real Scalar Singlet & 34 & 3 & \checkmark \\
        18 & $\mathbb{Z}_2$: Majorana Singlet $+$ Complex Scalar Singlet & 35 & 2 & -- \\
        19 & $\mathbb{Z}_2$: $3\times$ Majorana Singlet $+$ Majorana Singlet $+$ $2\times$ Complex Scalar Singlet & 94 & 3 & -- \\
        20 & $\mathbb{Z}_3$: $2\times$ Complex Scalar Singlet $+$ $3\times$ Majorana Singlet $+$ $3\times$ Complex Scalar Singlet & 173 & 3 & -- \\
        \addlinespace
        21 & $\mathbb{Z}_5$: $2\times$ Complex Scalar Singlet $+$ $3\times$ Complex Scalar Singlet $+$ Dirac Doublet & 218 & 4 & -- \\
           & $\quad+$ Complex Scalar Singlet & & & \\
        \bottomrule
    \end{tabular}
\end{table}
\begin{table}[h]
    \centering
    \caption{Per-Lagrangian benchmark results, summed over all budgets and three repeats, where row numbers refer to Table~\ref{tab:benchmark-classes}. Bold marks the policy with the most viable points on each Lagrangian (ties bolded jointly).}
    \label{tab:policy-benchmark-per-Lagrangian}
    \vspace{1em}
    \scriptsize
    \setlength{\tabcolsep}{2pt}
    \begin{tabular}{rccc ccc ccc ccc ccc}
        \toprule
        & \multicolumn{3}{c}{LLaM-small (pretrained)}
        & \multicolumn{3}{c}{LLaM-small (RL)}
        & \multicolumn{3}{c}{LLaM-medium (pretrained)}
        & \multicolumn{3}{c}{LLaM-medium (RL)}
        & \multicolumn{3}{c}{Differential Evolution} \\
        \cmidrule(lr){2-4} \cmidrule(lr){5-7} \cmidrule(lr){8-10} \cmidrule(lr){11-13} \cmidrule(lr){14-16}
        \# & $N_v$ & $R_{100}$ & $N_\sigma$
           & $N_v$ & $R_{100}$ & $N_\sigma$
           & $N_v$ & $R_{100}$ & $N_\sigma$
           & $N_v$ & $R_{100}$ & $N_\sigma$
           & $N_v$ & $R_{100}$ & $N_\sigma$ \\
        \midrule
        1 & $0$ & -- & $0$ & \textbf{418} & $0.2$ & $2$ & $1$ & $0.0$ & $1$ & $0$ & -- & $0$ & $0$ & -- & $0$ \\
        2 & $0$ & -- & $0$ & $0$ & -- & $0$ & $0$ & -- & $0$ & $0$ & -- & $0$ & $0$ & -- & $0$ \\
        3 & $0$ & -- & $0$ & $0$ & -- & $0$ & $0$ & -- & $0$ & $0$ & -- & $0$ & $0$ & -- & $0$ \\
        4 & $0$ & -- & $0$ & $0$ & -- & $0$ & $0$ & -- & $0$ & $0$ & -- & $0$ & $0$ & -- & $0$ \\
        5 & $0$ & -- & $0$ & $0$ & -- & $0$ & $0$ & -- & $0$ & $0$ & -- & $0$ & $0$ & -- & $0$ \\
        6 & $3414$ & $0.2$ & $4$ & $8102$ & $0.1$ & $4$ & $1876$ & $0.3$ & $4$ & \textbf{15574} & $0.1$ & $4$ & $6602$ & $0.1$ & $4$ \\
        7 & $0$ & -- & $0$ & $0$ & -- & $0$ & $0$ & -- & $0$ & \textbf{1920} & $0.1$ & $4$ & $0$ & -- & $0$ \\
        8 & $0$ & -- & $0$ & $0$ & -- & $0$ & $0$ & -- & $0$ & $0$ & -- & $0$ & $0$ & -- & $0$ \\
        9 & $0$ & -- & $0$ & $0$ & -- & $0$ & $0$ & -- & $0$ & $0$ & -- & $0$ & $0$ & -- & $0$ \\
        10 & $15$ & $0.0$ & $5$ & $357$ & $1.1$ & $21$ & $131$ & $0.8$ & $6$ & \textbf{1283} & $0.2$ & $8$ & $136$ & $1.5$ & $15$ \\
        11 & $0$ & -- & $0$ & \textbf{1146} & $0.1$ & $5$ & $5$ & $0.0$ & $2$ & $0$ & -- & $0$ & $0$ & -- & $0$ \\
        12 & $0$ & -- & $0$ & $1250$ & $0.1$ & $1$ & $5$ & $0.0$ & $1$ & \textbf{3928} & $0.0$ & $1$ & $1$ & $0.0$ & $1$ \\
        13 & $962$ & $3.6$ & $4$ & $1463$ & $3.4$ & $5$ & $398$ & $2.8$ & $5$ & \textbf{3849} & $3.7$ & $2$ & $65$ & $3.1$ & $3$ \\
        14 & $2$ & $0.0$ & $1$ & $2200$ & $0.0$ & $3$ & $0$ & -- & $0$ & \textbf{4230} & $0.0$ & $1$ & $7$ & $0.0$ & $4$ \\
        15 & $0$ & -- & $0$ & $1429$ & $0.7$ & $4$ & $51$ & $3.9$ & $4$ & \textbf{2590} & $0.3$ & $4$ & $11$ & $0.0$ & $4$ \\
        16 & $1028$ & $2.9$ & $5$ & $1815$ & $2.3$ & $7$ & $477$ & $0.6$ & $5$ & \textbf{3258} & $1.8$ & $4$ & $33$ & $0.0$ & $2$ \\
        17 & $12$ & $8.3$ & $5$ & $14$ & $0.0$ & $4$ & $8$ & $0.0$ & $4$ & \textbf{1388} & $0.2$ & $3$ & $12$ & $0.0$ & $8$ \\
        18 & $0$ & -- & $0$ & $5$ & $0.0$ & $1$ & $7$ & $0.0$ & $4$ & \textbf{1475} & $0.1$ & $4$ & $0$ & -- & $0$ \\
        19 & $1$ & $0.0$ & $1$ & $2$ & $0.0$ & $1$ & $1$ & $0.0$ & $1$ & \textbf{5687} & $0.0$ & $6$ & $14$ & $0.0$ & $1$ \\
        20 & $9$ & $0.0$ & $1$ & $4$ & $0.0$ & $1$ & $2$ & $0.0$ & $1$ & \textbf{4539} & $0.1$ & $1$ & $23$ & $0.0$ & $1$ \\
        21 & $662$ & $2.6$ & $1$ & \textbf{2539} & $2.8$ & $2$ & $45$ & $4.4$ & $1$ & $2081$ & $0.7$ & $6$ & $1687$ & $0.0$ & $1$ \\
        \bottomrule
    \end{tabular}
\end{table}

\begin{table}[h]
    \centering
    \caption{Per-Lagrangian benchmark results at budget $B=5$ (mean and error over three repeats), where row numbers refer to Table~\ref{tab:benchmark-classes}. Bold marks the policy with the most viable points (ties bolded jointly).}
    \label{tab:policy-benchmark-per-Lagrangian-b5}
    \vspace{1em}
    \scriptsize
    \setlength{\tabcolsep}{1pt}
    \begin{tabular}{rccc ccc ccc ccc ccc}
        \toprule
        & \multicolumn{3}{c}{LLaM-small (pretrained)}
        & \multicolumn{3}{c}{LLaM-small (RL)}
        & \multicolumn{3}{c}{LLaM-medium (pretrained)}
        & \multicolumn{3}{c}{LLaM-medium (RL)}
        & \multicolumn{3}{c}{Differential Evolution} \\
        \cmidrule(lr){2-4} \cmidrule(lr){5-7} \cmidrule(lr){8-10} \cmidrule(lr){11-13} \cmidrule(lr){14-16}
        \# & $N_v$ & $R_{100}$ & $N_\sigma$
           & $N_v$ & $R_{100}$ & $N_\sigma$
           & $N_v$ & $R_{100}$ & $N_\sigma$
           & $N_v$ & $R_{100}$ & $N_\sigma$
           & $N_v$ & $R_{100}$ & $N_\sigma$ \\
        \midrule
         1 & $0$ & -- & $0$ & $0$ & -- & $0$ & $0$ & -- & $0$ & $0$ & -- & $0$ & $0$ & -- & $0$ \\
        2 & $0$ & -- & $0$ & $0$ & -- & $0$ & $0$ & -- & $0$ & $0$ & -- & $0$ & $0$ & -- & $0$ \\
        3 & $0$ & -- & $0$ & $0$ & -- & $0$ & $0$ & -- & $0$ & $0$ & -- & $0$ & $0$ & -- & $0$ \\
        4 & $0$ & -- & $0$ & $0$ & -- & $0$ & $0$ & -- & $0$ & $0$ & -- & $0$ & $0$ & -- & $0$ \\
        5 & $0$ & -- & $0$ & $0$ & -- & $0$ & $0$ & -- & $0$ & $0$ & -- & $0$ & $0$ & -- & $0$ \\
        6 & $0$ & -- & $0$ & $0$ & -- & $0$ & $1.0 \pm 1.4$ & $0 \pm 0$ & $0.3 \pm 0.5$ & $\mathbf{109 \pm 150}$ & $8 \pm 8$ & $2.3 \pm 1.7$ & $0.7 \pm 0.9$ & $0 \pm 0$ & $0.7 \pm 0.9$ \\
        7 & $0$ & -- & $0$ & $0$ & -- & $0$ & $0$ & -- & $0$ & $0$ & -- & $0$ & $0$ & -- & $0$ \\
        8 & $0$ & -- & $0$ & $0$ & -- & $0$ & $0$ & -- & $0$ & $0$ & -- & $0$ & $0$ & -- & $0$ \\
        9 & $0$ & -- & $0$ & $0$ & -- & $0$ & $0$ & -- & $0$ & $0$ & -- & $0$ & $0$ & -- & $0$ \\
        10 & $0.3 \pm 0.5$ & $0 \pm 0$ & $0.3 \pm 0.5$ & $\mathbf{0.7 \pm 0.5}$ & $0 \pm 0$ & $0.7 \pm 0.5$ & $0$ & -- & $0$ & $0$ & -- & $0$ & $0.3 \pm 0.5$ & $0 \pm 0$ & $0.3 \pm 0.5$ \\
        11 & $0$ & -- & $0$ & $0$ & -- & $0$ & $0$ & -- & $0$ & $0$ & -- & $0$ & $0$ & -- & $0$ \\
        12 & $0$ & -- & $0$ & $0$ & -- & $0$ & $0$ & -- & $0$ & $0$ & -- & $0$ & $\mathbf{0.3 \pm 0.5}$ & $0 \pm 0$ & $0.3 \pm 0.5$ \\
        13 & $2.0 \pm 0.8$ & $0 \pm 0$ & $1.0 \pm 0.0$ & $2.7 \pm 1.2$ & $0 \pm 0$ & $1.0 \pm 0.0$ & $0.3 \pm 0.5$ & $0 \pm 0$ & $0.3 \pm 0.5$ & $\mathbf{74 \pm 63}$ & $4 \pm 0$ & $1.0 \pm 0.8$ & $1.3 \pm 0.5$ & $0 \pm 0$ & $1.0 \pm 0.0$ \\
        14 & $0$ & -- & $0$ & $0$ & -- & $0$ & $0$ & -- & $0$ & $0$ & -- & $0$ & $0$ & -- & $0$ \\
        15 & $0$ & -- & $0$ & $0$ & -- & $0$ & $0$ & -- & $0$ & $0$ & -- & $0$ & $0$ & -- & $0$ \\
        16 & $1.0 \pm 0.8$ & $0 \pm 0$ & $1.0 \pm 0.8$ & $1.0 \pm 1.4$ & $0 \pm 0$ & $0.3 \pm 0.5$ & $1.0 \pm 0.0$ & $0 \pm 0$ & $1.0 \pm 0.0$ & $\mathbf{39 \pm 27}$ & $2 \pm 2$ & $1.0 \pm 0.0$ & $0.3 \pm 0.5$ & $0 \pm 0$ & $0.3 \pm 0.5$ \\
        17 & $\mathbf{0.3 \pm 0.5}$ & $0 \pm 0$ & $0.3 \pm 0.5$ & $0$ & -- & $0$ & $0$ & -- & $0$ & $0$ & -- & $0$ & $0$ & -- & $0$ \\
        18 & $0$ & -- & $0$ & $0$ & -- & $0$ & $0$ & -- & $0$ & $0$ & -- & $0$ & $0$ & -- & $0$ \\
        19 & $0$ & -- & $0$ & $0$ & -- & $0$ & $0$ & -- & $0$ & $0$ & -- & $0$ & $0$ & -- & $0$ \\
        20 & $0$ & -- & $0$ & $\mathbf{0.3 \pm 0.5}$ & $0 \pm 0$ & $0.3 \pm 0.5$ & $0$ & -- & $0$ & $0$ & -- & $0$ & $0$ & -- & $0$ \\
        21 & $0.7 \pm 0.9$ & $0 \pm 0$ & $0.3 \pm 0.5$ & $\mathbf{2.7 \pm 2.1}$ & $0 \pm 0$ & $0.7 \pm 0.5$ & $0.7 \pm 0.9$ & $0 \pm 0$ & $0.3 \pm 0.5$ & $0$ & -- & $0$ & $0.7 \pm 0.5$ & $0 \pm 0$ & $0.7 \pm 0.5$ \\
        \bottomrule
    \end{tabular}
\end{table}

\begin{table}[h]
    \centering
    \caption{Per-Lagrangian benchmark results at budget $B=10$ (mean and error over three repeats), where row numbers refer to Table~\ref{tab:benchmark-classes}. Bold marks the policy with the most viable points (ties bolded jointly).}
    \label{tab:policy-benchmark-per-Lagrangian-b10}
    \vspace{1em}
    \scriptsize
    \setlength{\tabcolsep}{1pt}
    \begin{tabular}{rccc ccc ccc ccc ccc}
        \toprule
        & \multicolumn{3}{c}{LLaM-small (pretrained)}
        & \multicolumn{3}{c}{LLaM-small (RL)}
        & \multicolumn{3}{c}{LLaM-medium (pretrained)}
        & \multicolumn{3}{c}{LLaM-medium (RL)}
        & \multicolumn{3}{c}{Differential Evolution} \\
        \cmidrule(lr){2-4} \cmidrule(lr){5-7} \cmidrule(lr){8-10} \cmidrule(lr){11-13} \cmidrule(lr){14-16}
        \# & $N_v$ & $R_{100}$ & $N_\sigma$
           & $N_v$ & $R_{100}$ & $N_\sigma$
           & $N_v$ & $R_{100}$ & $N_\sigma$
           & $N_v$ & $R_{100}$ & $N_\sigma$
           & $N_v$ & $R_{100}$ & $N_\sigma$ \\
        \midrule
        1 & $0$ & -- & $0$ & $0$ & -- & $0$ & $0$ & -- & $0$ & $0$ & -- & $0$ & $0$ & -- & $0$ \\
        2 & $0$ & -- & $0$ & $0$ & -- & $0$ & $0$ & -- & $0$ & $0$ & -- & $0$ & $0$ & -- & $0$ \\
        3 & $0$ & -- & $0$ & $0$ & -- & $0$ & $0$ & -- & $0$ & $0$ & -- & $0$ & $0$ & -- & $0$ \\
        4 & $0$ & -- & $0$ & $0$ & -- & $0$ & $0$ & -- & $0$ & $0$ & -- & $0$ & $0$ & -- & $0$ \\
        5 & $0$ & -- & $0$ & $0$ & -- & $0$ & $0$ & -- & $0$ & $0$ & -- & $0$ & $0$ & -- & $0$ \\
        6 & $2.7 \pm 1.9$ & $0.0 \pm 0.0$ & $1.3 \pm 0.9$ & $3.3 \pm 2.6$ & $4.8 \pm 6.7$ & $1.7 \pm 0.9$ & $2.7 \pm 1.7$ & $0.0 \pm 0.0$ & $1.3 \pm 0.5$ & $\mathbf{322 \pm 228}$ & $0.2 \pm 0.0$ & $2.7 \pm 1.9$ & $4.0 \pm 3.7$ & $5.6 \pm 5.6$ & $2.3 \pm 1.7$ \\
        7 & $0$ & -- & $0$ & $0$ & -- & $0$ & $0$ & -- & $0$ & $\mathbf{78 \pm 110}$ & $0.4 \pm 0.0$ & $1.3 \pm 1.9$ & $0$ & -- & $0$ \\
        8 & $0$ & -- & $0$ & $0$ & -- & $0$ & $0$ & -- & $0$ & $0$ & -- & $0$ & $0$ & -- & $0$ \\
        9 & $0$ & -- & $0$ & $0$ & -- & $0$ & $0$ & -- & $0$ & $0$ & -- & $0$ & $0$ & -- & $0$ \\
        10 & $\mathbf{0.7 \pm 0.5}$ & $0.0 \pm 0.0$ & $0.7 \pm 0.5$ & $\mathbf{0.7 \pm 0.9}$ & $0.0 \pm 0.0$ & $0.7 \pm 0.9$ & $0$ & -- & $0$ & $0$ & -- & $0$ & $0.3 \pm 0.5$ & $0.0 \pm 0.0$ & $0.3 \pm 0.5$ \\
        11 & $0$ & -- & $0$ & $0$ & -- & $0$ & $0$ & -- & $0$ & $0$ & -- & $0$ & $0$ & -- & $0$ \\
        12 & $0$ & -- & $0$ & $0$ & -- & $0$ & $0$ & -- & $0$ & $0$ & -- & $0$ & $0$ & -- & $0$ \\
        13 & $1.3 \pm 1.9$ & $0.0 \pm 0.0$ & $0.3 \pm 0.5$ & $9.7 \pm 3.1$ & $0.0 \pm 0.0$ & $1.0 \pm 0.0$ & $5.0 \pm 5.0$ & $0.0 \pm 0.0$ & $1.0 \pm 0.0$ & $\mathbf{333 \pm 94}$ & $2.5 \pm 1.0$ & $1.0 \pm 0.0$ & $0.7 \pm 0.5$ & $0.0 \pm 0.0$ & $0.7 \pm 0.5$ \\
        14 & $0$ & -- & $0$ & $0$ & -- & $0$ & $0$ & -- & $0$ & $0$ & -- & $0$ & $0$ & -- & $0$ \\
        15 & $0$ & -- & $0$ & $0$ & -- & $0$ & $0.7 \pm 0.9$ & $0.0 \pm 0.0$ & $0.7 \pm 0.9$ & $\mathbf{127 \pm 180}$ & $0.5 \pm 0.0$ & $1.3 \pm 1.9$ & $0$ & -- & $0$ \\
        16 & $1.7 \pm 0.5$ & $0.0 \pm 0.0$ & $1.0 \pm 0.0$ & $2.7 \pm 0.9$ & $0.0 \pm 0.0$ & $1.7 \pm 0.5$ & $4.3 \pm 1.2$ & $0.0 \pm 0.0$ & $1.7 \pm 0.9$ & $\mathbf{150 \pm 127}$ & $3.0 \pm 0.2$ & $0.7 \pm 0.5$ & $0$ & -- & $0$ \\
        17 & $0$ & -- & $0$ & $0$ & -- & $0$ & $0$ & -- & $0$ & $\mathbf{153 \pm 216}$ & $0.2 \pm 0.0$ & $0.3 \pm 0.5$ & $0$ & -- & $0$ \\
        18 & $0$ & -- & $0$ & $0$ & -- & $0$ & $0$ & -- & $0$ & $0$ & -- & $0$ & $0$ & -- & $0$ \\
        19 & $0$ & -- & $0$ & $0$ & -- & $0$ & $0$ & -- & $0$ & $0$ & -- & $0$ & $0$ & -- & $0$ \\
        20 & $0.7 \pm 0.9$ & $0.0 \pm 0.0$ & $0.3 \pm 0.5$ & $0$ & -- & $0$ & $0.3 \pm 0.5$ & $0.0 \pm 0.0$ & $0.3 \pm 0.5$ & $\mathbf{245 \pm 296}$ & $0.7 \pm 0.6$ & $0.7 \pm 0.5$ & $0$ & -- & $0$ \\
        21 & $0.3 \pm 0.5$ & $0.0 \pm 0.0$ & $0.3 \pm 0.5$ & $6.7 \pm 8.1$ & $5.6 \pm 5.6$ & $0.7 \pm 0.5$ & $0.3 \pm 0.5$ & $0.0 \pm 0.0$ & $0.3 \pm 0.5$ & $\mathbf{192 \pm 148}$ & $0.5 \pm 0.5$ & $1.0 \pm 0.0$ & $0.7 \pm 0.5$ & $0.0 \pm 0.0$ & $0.7 \pm 0.5$ \\
        \bottomrule
    \end{tabular}
\end{table}

\begin{table}[h]
    \centering
    \caption{Per-Lagrangian benchmark results at budget $B=25$ (mean and error over three repeats), where row numbers refer to Table~\ref{tab:benchmark-classes}. Bold marks the policy with the most viable points (ties bolded jointly).}
    \label{tab:policy-benchmark-per-Lagrangian-b25}
    \vspace{1em}
    \scriptsize
    \setlength{\tabcolsep}{1pt}
    \begin{tabular}{rccc ccc ccc ccc ccc}
        \toprule
        & \multicolumn{3}{c}{LLaM-small (pretrained)}
        & \multicolumn{3}{c}{LLaM-small (RL)}
        & \multicolumn{3}{c}{LLaM-medium (pretrained)}
        & \multicolumn{3}{c}{LLaM-medium (RL)}
        & \multicolumn{3}{c}{Differential Evolution} \\
        \cmidrule(lr){2-4} \cmidrule(lr){5-7} \cmidrule(lr){8-10} \cmidrule(lr){11-13} \cmidrule(lr){14-16}
        \# & $N_v$ & $R_{100}$ & $N_\sigma$
           & $N_v$ & $R_{100}$ & $N_\sigma$
           & $N_v$ & $R_{100}$ & $N_\sigma$
           & $N_v$ & $R_{100}$ & $N_\sigma$
           & $N_v$ & $R_{100}$ & $N_\sigma$ \\
        \midrule
        1 & $0$ & -- & $0$ & $0$ & -- & $0$ & $0$ & -- & $0$ & $0$ & -- & $0$ & $0$ & -- & $0$ \\
        2 & $0$ & -- & $0$ & $0$ & -- & $0$ & $0$ & -- & $0$ & $0$ & -- & $0$ & $0$ & -- & $0$ \\
        3 & $0$ & -- & $0$ & $0$ & -- & $0$ & $0$ & -- & $0$ & $0$ & -- & $0$ & $0$ & -- & $0$ \\
        4 & $0$ & -- & $0$ & $0$ & -- & $0$ & $0$ & -- & $0$ & $0$ & -- & $0$ & $0$ & -- & $0$ \\
        5 & $0$ & -- & $0$ & $0$ & -- & $0$ & $0$ & -- & $0$ & $0$ & -- & $0$ & $0$ & -- & $0$ \\
        6 & $119 \pm 39$ & $0.9 \pm 0.3$ & $4.0 \pm 0.0$ & $387 \pm 117$ & $0.3 \pm 0.1$ & $4.0 \pm 0.0$ & $15 \pm 11$ & $11 \pm 5$ & $2.7 \pm 1.2$ & $\mathbf{867 \pm 668}$ & $0.1 \pm 0.0$ & $2.0 \pm 1.6$ & $89 \pm 46$ & $1.5 \pm 0.7$ & $4.0 \pm 0.0$ \\
        7 & $0$ & -- & $0$ & $0$ & -- & $0$ & $0$ & -- & $0$ & $0$ & -- & $0$ & $0$ & -- & $0$ \\
        8 & $0$ & -- & $0$ & $0$ & -- & $0$ & $0$ & -- & $0$ & $0$ & -- & $0$ & $0$ & -- & $0$ \\
        9 & $0$ & -- & $0$ & $0$ & -- & $0$ & $0$ & -- & $0$ & $0$ & -- & $0$ & $0$ & -- & $0$ \\
        10 & $0.3 \pm 0.5$ & $0.0 \pm 0.0$ & $0.3 \pm 0.5$ & $0.3 \pm 0.5$ & $0.0 \pm 0.0$ & $0.3 \pm 0.5$ & $1.3 \pm 1.9$ & $0.0 \pm 0.0$ & $0.3 \pm 0.5$ & $\mathbf{151 \pm 214}$ & $0.2 \pm 0.0$ & $2.0 \pm 2.8$ & $3.0 \pm 1.4$ & $0.0 \pm 0.0$ & $2.3 \pm 1.2$ \\
        11 & $0$ & -- & $0$ & $0$ & -- & $0$ & $0$ & -- & $0$ & $0$ & -- & $0$ & $0$ & -- & $0$ \\
        12 & $0$ & -- & $0$ & $0$ & -- & $0$ & $\mathbf{1.7 \pm 2.4}$ & $0.0 \pm 0.0$ & $0.3 \pm 0.5$ & $0$ & -- & $0$ & $0$ & -- & $0$ \\
        13 & $37 \pm 14$ & $0.6 \pm 0.9$ & $1.3 \pm 0.5$ & $157 \pm 71$ & $3.2 \pm 0.8$ & $1.7 \pm 0.5$ & $29 \pm 5$ & $0.0 \pm 0.0$ & $2.0 \pm 0.8$ & $\mathbf{448 \pm 65}$ & $4.9 \pm 1.6$ & $1.0 \pm 0.0$ & $3.0 \pm 0.8$ & $0.0 \pm 0.0$ & $1.3 \pm 0.5$ \\
        14 & $0$ & -- & $0$ & $0$ & -- & $0$ & $0$ & -- & $0$ & $0$ & -- & $0$ & $\mathbf{0.3 \pm 0.5}$ & $0.0 \pm 0.0$ & $0.3 \pm 0.5$ \\
        15 & $0$ & -- & $0$ & $1.0 \pm 1.4$ & $0.0 \pm 0.0$ & $0.3 \pm 0.5$ & $0$ & -- & $0$ & $\mathbf{294 \pm 208}$ & $0.3 \pm 0.1$ & $2.7 \pm 1.9$ & $0$ & -- & $0$ \\
        16 & $22 \pm 12$ & $0.9 \pm 1.2$ & $1.0 \pm 0.0$ & $150 \pm 81$ & $1.5 \pm 1.2$ & $1.3 \pm 0.5$ & $32 \pm 20$ & $0.0 \pm 0.0$ & $2.7 \pm 0.5$ & $\mathbf{436 \pm 60}$ & $1.3 \pm 0.8$ & $1.7 \pm 0.5$ & $2.7 \pm 1.2$ & $0.0 \pm 0.0$ & $1.0 \pm 0.0$ \\
        17 & $0$ & -- & $0$ & $1.0 \pm 1.4$ & $0.0 \pm 0.0$ & $0.7 \pm 0.9$ & $0$ & -- & $0$ & $\mathbf{169 \pm 239}$ & $0.2 \pm 0.0$ & $0.3 \pm 0.5$ & $0.7 \pm 0.9$ & $0.0 \pm 0.0$ & $0.3 \pm 0.5$ \\
        18 & $0$ & -- & $0$ & $0$ & -- & $0$ & $\mathbf{1.3 \pm 1.9}$ & $0.0 \pm 0.0$ & $0.7 \pm 0.9$ & $0$ & -- & $0$ & $0$ & -- & $0$ \\
        19 & $0$ & -- & $0$ & $\mathbf{0.7 \pm 0.9}$ & $0.0 \pm 0.0$ & $0.3 \pm 0.5$ & $0$ & -- & $0$ & $0$ & -- & $0$ & $0.3 \pm 0.5$ & $0.0 \pm 0.0$ & $0.3 \pm 0.5$ \\
        20 & $1.0 \pm 0.0$ & $0.0 \pm 0.0$ & $1.0 \pm 0.0$ & $1.0 \pm 0.8$ & $0.0 \pm 0.0$ & $0.7 \pm 0.5$ & $0$ & -- & $0$ & $\mathbf{691 \pm 517}$ & $0.1 \pm 0.0$ & $0.7 \pm 0.5$ & $1.7 \pm 1.2$ & $0.0 \pm 0.0$ & $0.7 \pm 0.5$ \\
        21 & $7.0 \pm 1.6$ & $10 \pm 8$ & $1.0 \pm 0.0$ & $\mathbf{440 \pm 16}$ & $2.8 \pm 0.4$ & $1.3 \pm 0.5$ & $9.3 \pm 9.0$ & $3.0 \pm 4.3$ & $1.0 \pm 0.0$ & $296 \pm 11$ & $0.6 \pm 0.3$ & $2.3 \pm 1.9$ & $38 \pm 13$ & $0.0 \pm 0.0$ & $1.0 \pm 0.0$ \\
        \bottomrule
    \end{tabular}
\end{table}

\begin{table}[h]
    \centering
    \caption{Per-Lagrangian benchmark results at budget $B=50$ (mean and error over three repeats), where row numbers refer to Table~\ref{tab:benchmark-classes}. Bold marks the policy with the most viable points (ties bolded jointly).}
    \label{tab:policy-benchmark-per-Lagrangian-b50}
    \vspace{1em}
    \scriptsize
    \setlength{\tabcolsep}{.5pt}
    \begin{tabular}{rccc ccc ccc ccc ccc}
        \toprule
        & \multicolumn{3}{c}{LLaM-small (pretrained)}
        & \multicolumn{3}{c}{LLaM-small (RL)}
        & \multicolumn{3}{c}{LLaM-medium (pretrained)}
        & \multicolumn{3}{c}{LLaM-medium (RL)}
        & \multicolumn{3}{c}{Differential Evolution} \\
        \cmidrule(lr){2-4} \cmidrule(lr){5-7} \cmidrule(lr){8-10} \cmidrule(lr){11-13} \cmidrule(lr){14-16}
        \# & $N_v$ & $R_{100}$ & $N_\sigma$
           & $N_v$ & $R_{100}$ & $N_\sigma$
           & $N_v$ & $R_{100}$ & $N_\sigma$
           & $N_v$ & $R_{100}$ & $N_\sigma$
           & $N_v$ & $R_{100}$ & $N_\sigma$ \\
        \midrule
        1 & $0$ & -- & $0$ & $\mathbf{139 \pm 197}$ & $0.2 \pm 0.0$ & $0.7 \pm 0.9$ & $0.3 \pm 0.5$ & $0.0 \pm 0.0$ & $0.3 \pm 0.5$ & $0$ & -- & $0$ & $0$ & -- & $0$ \\
        2 & $0$ & -- & $0$ & $0$ & -- & $0$ & $0$ & -- & $0$ & $0$ & -- & $0$ & $0$ & -- & $0$ \\
        3 & $0$ & -- & $0$ & $0$ & -- & $0$ & $0$ & -- & $0$ & $0$ & -- & $0$ & $0$ & -- & $0$ \\
        4 & $0$ & -- & $0$ & $0$ & -- & $0$ & $0$ & -- & $0$ & $0$ & -- & $0$ & $0$ & -- & $0$ \\
        5 & $0$ & -- & $0$ & $0$ & -- & $0$ & $0$ & -- & $0$ & $0$ & -- & $0$ & $0$ & -- & $0$ \\
        6 & $1016 \pm 103$ & $0.1 \pm 0.0$ & $4.0 \pm 0.0$ & $2310 \pm 341$ & $0.0 \pm 0.0$ & $4.0 \pm 0.0$ & $607 \pm 181$ & $0.2 \pm 0.0$ & $4.0 \pm 0.0$ & $\mathbf{3893 \pm 646}$ & $0.0 \pm 0.0$ & $3.0 \pm 1.4$ & $2107 \pm 172$ & $0.0 \pm 0.0$ & $4.0 \pm 0.0$ \\
        7 & $0$ & -- & $0$ & $0$ & -- & $0$ & $0$ & -- & $0$ & $\mathbf{562 \pm 795}$ & $0.1 \pm 0.0$ & $1.3 \pm 1.9$ & $0$ & -- & $0$ \\
        8 & $0$ & -- & $0$ & $0$ & -- & $0$ & $0$ & -- & $0$ & $0$ & -- & $0$ & $0$ & -- & $0$ \\
        9 & $0$ & -- & $0$ & $0$ & -- & $0$ & $0$ & -- & $0$ & $0$ & -- & $0$ & $0$ & -- & $0$ \\
        10 & $3.7 \pm 5.2$ & $0.0 \pm 0.0$ & $1.3 \pm 1.9$ & $117 \pm 55$ & $2.1 \pm 2.1$ & $10.3 \pm 3.1$ & $42 \pm 56$ & $0.3 \pm 0.4$ & $1.7 \pm 0.9$ & $\mathbf{277 \pm 202}$ & $0.2 \pm 0.0$ & $2.0 \pm 1.6$ & $42 \pm 25$ & $1.1 \pm 0.8$ & $8.3 \pm 2.5$ \\
        11 & $0$ & -- & $0$ & $\mathbf{382 \pm 540}$ & $0.1 \pm 0.0$ & $1.7 \pm 2.4$ & $1.7 \pm 1.7$ & $0.0 \pm 0.0$ & $0.7 \pm 0.5$ & $0$ & -- & $0$ & $0$ & -- & $0$ \\
        12 & $0$ & -- & $0$ & $417 \pm 589$ & $0.1 \pm 0.0$ & $0.3 \pm 0.5$ & $0$ & -- & $0$ & $\mathbf{1309 \pm 1852}$ & $0.0 \pm 0.0$ & $0.3 \pm 0.5$ & $0$ & -- & $0$ \\
        13 & $280 \pm 92$ & $4.0 \pm 1.0$ & $2.0 \pm 0.8$ & $319 \pm 33$ & $3.8 \pm 1.0$ & $3.0 \pm 0.8$ & $98 \pm 17$ & $3.5 \pm 2.5$ & $2.7 \pm 0.9$ & $\mathbf{429 \pm 36}$ & $3.0 \pm 1.0$ & $1.3 \pm 0.5$ & $17 \pm 2$ & $4.1 \pm 3.0$ & $1.3 \pm 0.5$ \\
        14 & $0.7 \pm 0.9$ & $0.0 \pm 0.0$ & $0.3 \pm 0.5$ & $733 \pm 1034$ & $0.0 \pm 0.0$ & $1.0 \pm 0.0$ & $0$ & -- & $0$ & $\mathbf{1410 \pm 1994}$ & $0.0 \pm 0.0$ & $0.3 \pm 0.5$ & $2.0 \pm 1.4$ & $0.0 \pm 0.0$ & $1.3 \pm 0.5$ \\
        15 & $0$ & -- & $0$ & $\mathbf{475 \pm 17}$ & $0.7 \pm 0.3$ & $3.3 \pm 0.9$ & $16 \pm 16$ & $2.6 \pm 2.6$ & $2.3 \pm 1.7$ & $442 \pm 23$ & $0.3 \pm 0.1$ & $4.0 \pm 0.0$ & $3.7 \pm 2.6$ & $0.0 \pm 0.0$ & $2.0 \pm 1.4$ \\
        16 & $318 \pm 31$ & $3.0 \pm 0.4$ & $2.7 \pm 1.7$ & $452 \pm 53$ & $2.3 \pm 0.6$ & $3.0 \pm 1.4$ & $121 \pm 22$ & $0.7 \pm 0.6$ & $2.0 \pm 0.8$ & $\mathbf{460 \pm 5}$ & $1.7 \pm 0.5$ & $1.0 \pm 0.0$ & $8.0 \pm 7.1$ & $0.0 \pm 0.0$ & $1.0 \pm 0.0$ \\
        17 & $3.7 \pm 5.2$ & $9.1 \pm 0.0$ & $1.3 \pm 1.9$ & $3.7 \pm 2.9$ & $0.0 \pm 0.0$ & $1.0 \pm 0.8$ & $2.7 \pm 2.5$ & $0.0 \pm 0.0$ & $1.7 \pm 1.2$ & $\mathbf{141 \pm 199}$ & $0.2 \pm 0.0$ & $0.3 \pm 0.5$ & $3.3 \pm 2.1$ & $0.0 \pm 0.0$ & $3.0 \pm 1.6$ \\
        18 & $0$ & -- & $0$ & $1.7 \pm 2.4$ & $0.0 \pm 0.0$ & $0.3 \pm 0.5$ & $1.0 \pm 0.8$ & $0.0 \pm 0.0$ & $0.7 \pm 0.5$ & $\mathbf{492 \pm 695}$ & $0.1 \pm 0.0$ & $1.3 \pm 1.9$ & $0$ & -- & $0$ \\
        19 & $0.3 \pm 0.5$ & $0.0 \pm 0.0$ & $0.3 \pm 0.5$ & $0$ & -- & $0$ & $0.3 \pm 0.5$ & $0.0 \pm 0.0$ & $0.3 \pm 0.5$ & $\mathbf{1896 \pm 1880}$ & $0.1 \pm 0.0$ & $2.0 \pm 2.2$ & $4.3 \pm 2.1$ & $0.0 \pm 0.0$ & $1.0 \pm 0.0$ \\
        20 & $1.3 \pm 0.5$ & $0.0 \pm 0.0$ & $1.0 \pm 0.0$ & $0$ & -- & $0$ & $0.3 \pm 0.5$ & $0.0 \pm 0.0$ & $0.3 \pm 0.5$ & $\mathbf{577 \pm 740}$ & $0.5 \pm 0.4$ & $0.7 \pm 0.5$ & $6.0 \pm 4.2$ & $0.0 \pm 0.0$ & $1.0 \pm 0.0$ \\
        21 & $213 \pm 82$ & $1.9 \pm 1.4$ & $1.0 \pm 0.0$ & $397 \pm 7$ & $2.6 \pm 0.5$ & $1.0 \pm 0.0$ & $4.7 \pm 2.6$ & $0.0 \pm 0.0$ & $1.0 \pm 0.0$ & $206 \pm 152$ & $0.9 \pm 0.3$ & $0.7 \pm 0.5$ & $\mathbf{523 \pm 64}$ & $0.0 \pm 0.0$ & $1.0 \pm 0.0$ \\
        \bottomrule
    \end{tabular}
\end{table}
\clearpage
\begin{figure}
    \centering
    \setlength{\tabcolsep}{2pt}
    \begin{tabular}{cc}
      \includegraphics[width=0.48\linewidth]{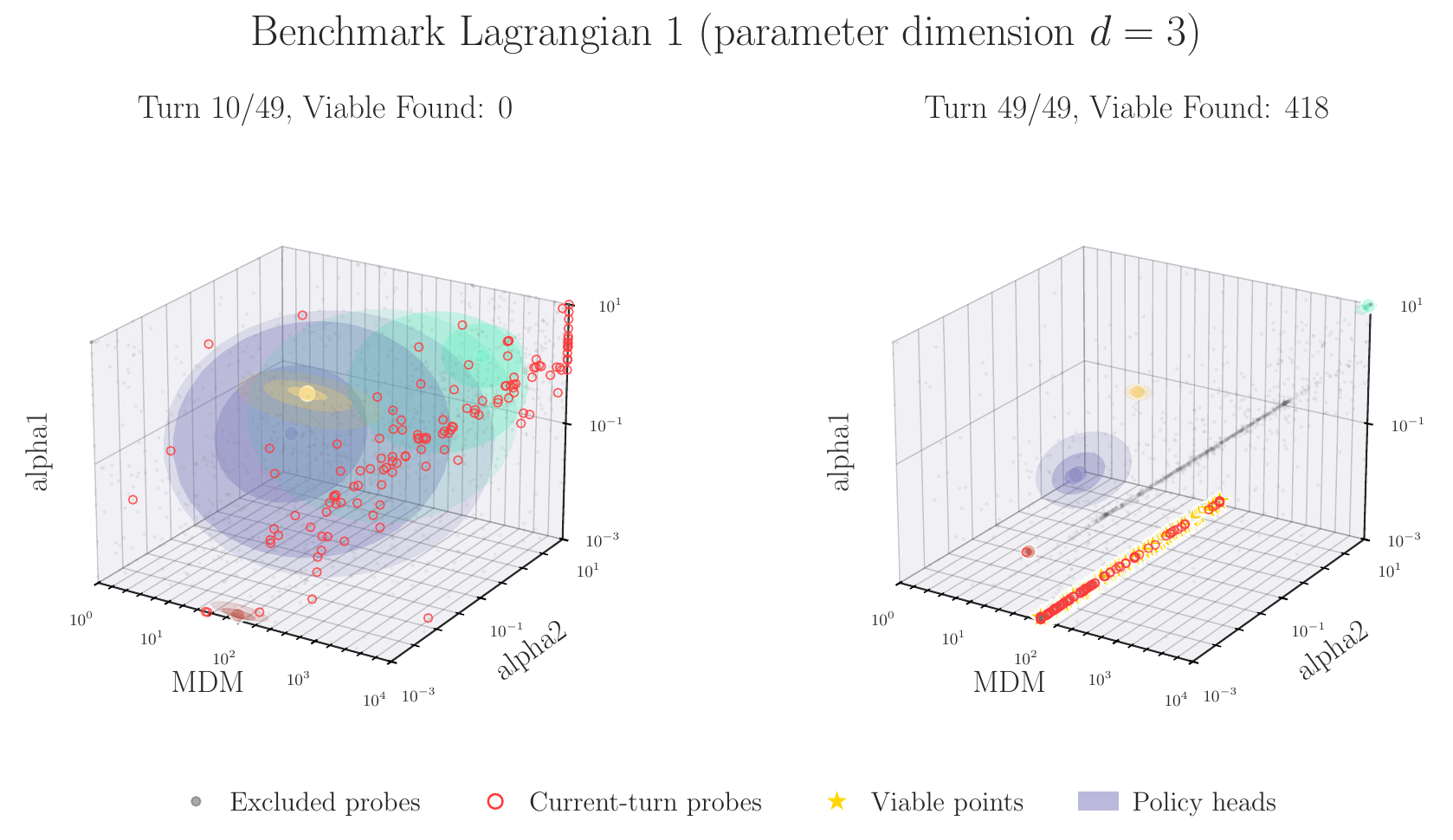} &
      \includegraphics[width=0.48\linewidth]{images/results/benchmark-3d/3d_lg_6.pdf} \\[-0.3em]
      \includegraphics[width=0.48\linewidth]{images/results/benchmark-3d/3d_lg_10.pdf} &
      \includegraphics[width=0.48\linewidth]{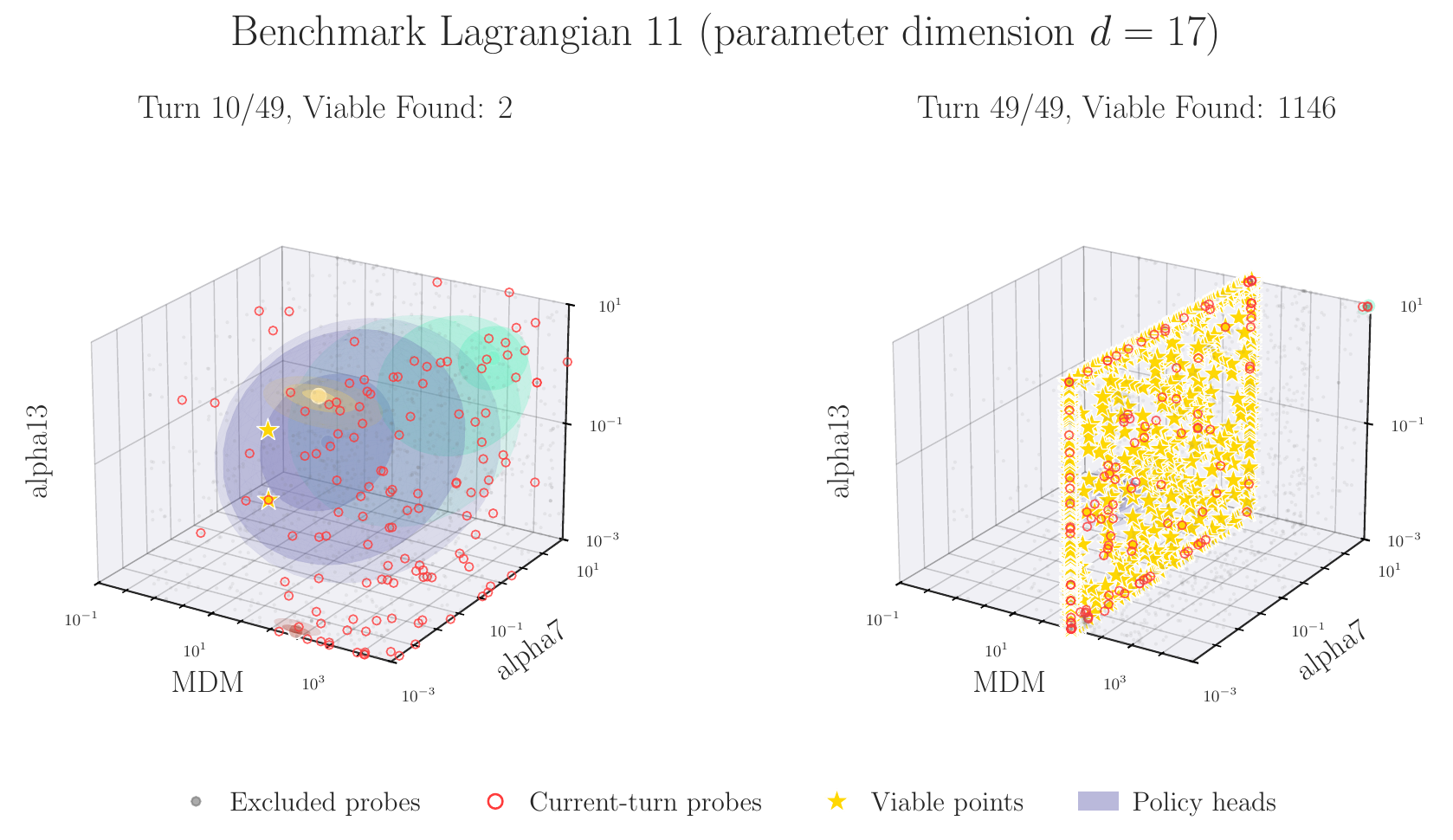} \\[-0.3em]
      \includegraphics[width=0.48\linewidth]{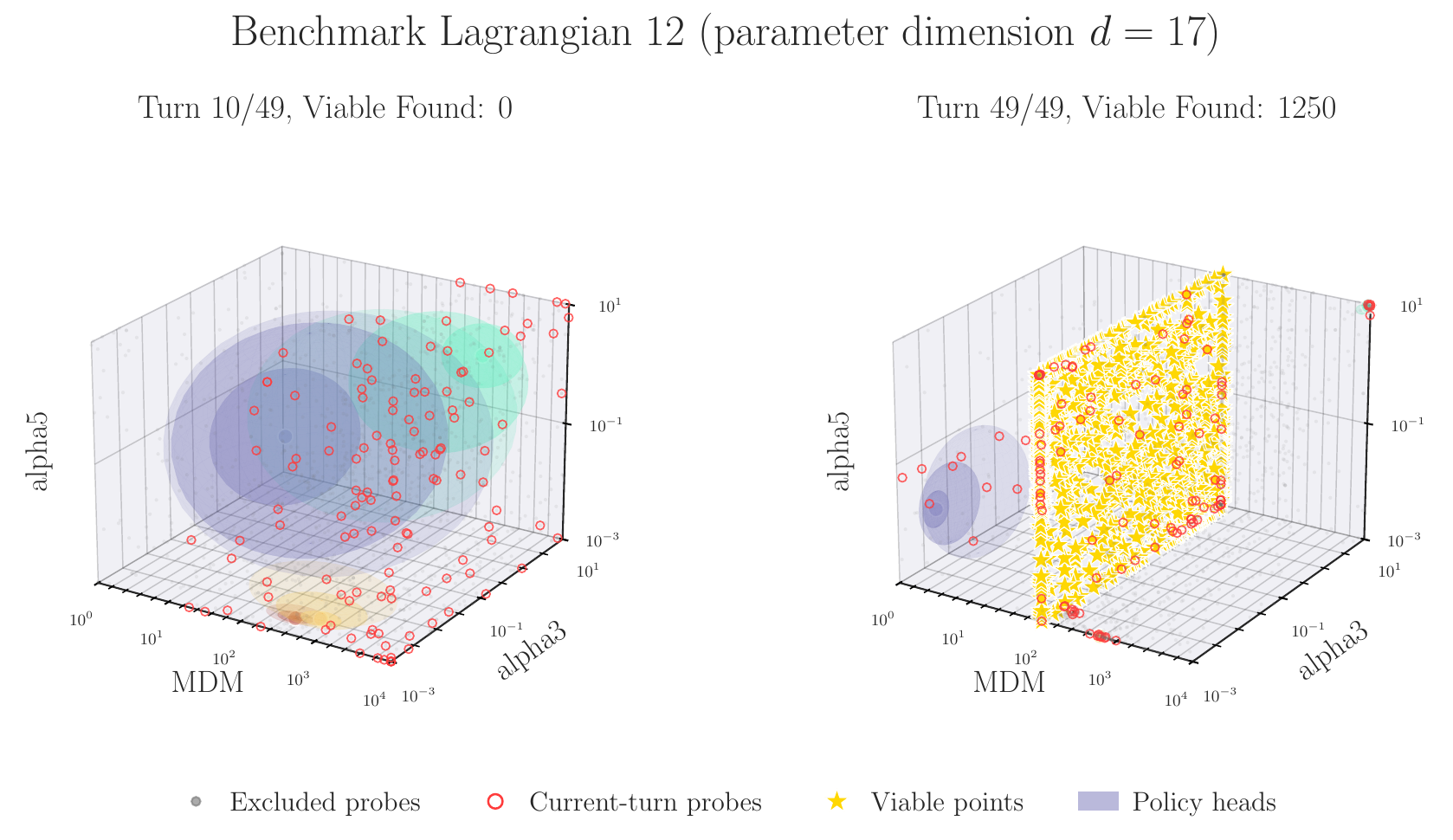} &
      \includegraphics[width=0.48\linewidth]{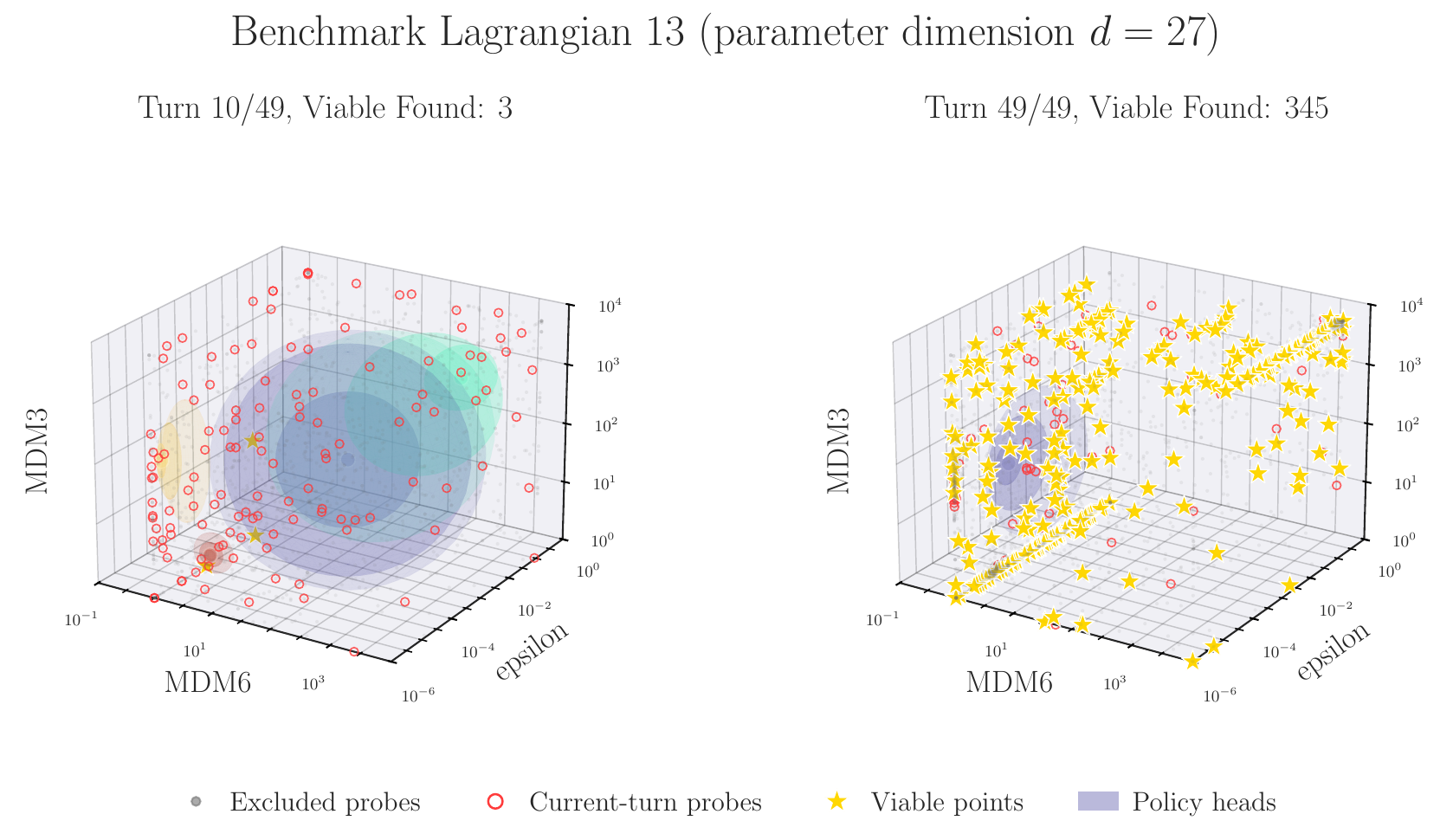}\\[-0.3em]
      \includegraphics[width=0.48\linewidth]{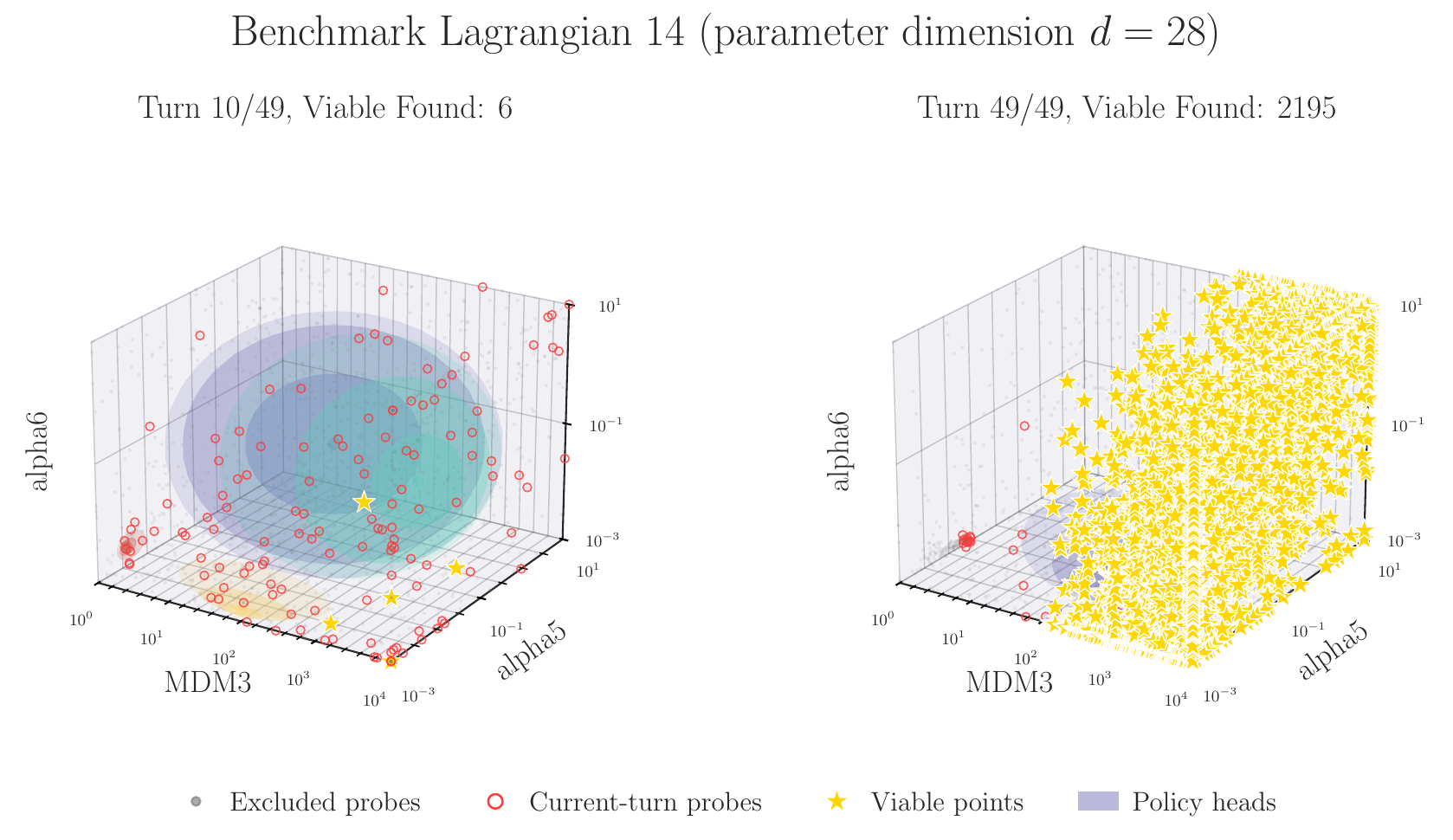} &
      \includegraphics[width=0.48\linewidth]{images/results/benchmark-3d/3d_lg_15.pdf}\\[-0.3em]
    \end{tabular}
    \caption{Example LLaM-small-RL search episodes at budget $B=50$ for
     benchmark Lagrangians (see numbering of Table \ref{tab:benchmark-classes}), shown early in the episode (left) and at the final turn (right). Grey points are excluded probes, red dots are current turn's probes, gold stars are the cumulative viable points found, and the shaded ellipses are the policy's beta distribution heads. The
    search acts in the full $d$-dimensional parameter space. We display the three parameters along which its viable points spread the most. \textbf{Policy heads need not be in the right location when Differential Evolution is chosen by the LLaM.}}
    \label{fig:3d-episodes-1}
\end{figure}
\clearpage
\begin{figure}
    \centering
    \setlength{\tabcolsep}{2pt}
    \begin{tabular}{cc}
      \includegraphics[width=0.48\linewidth]{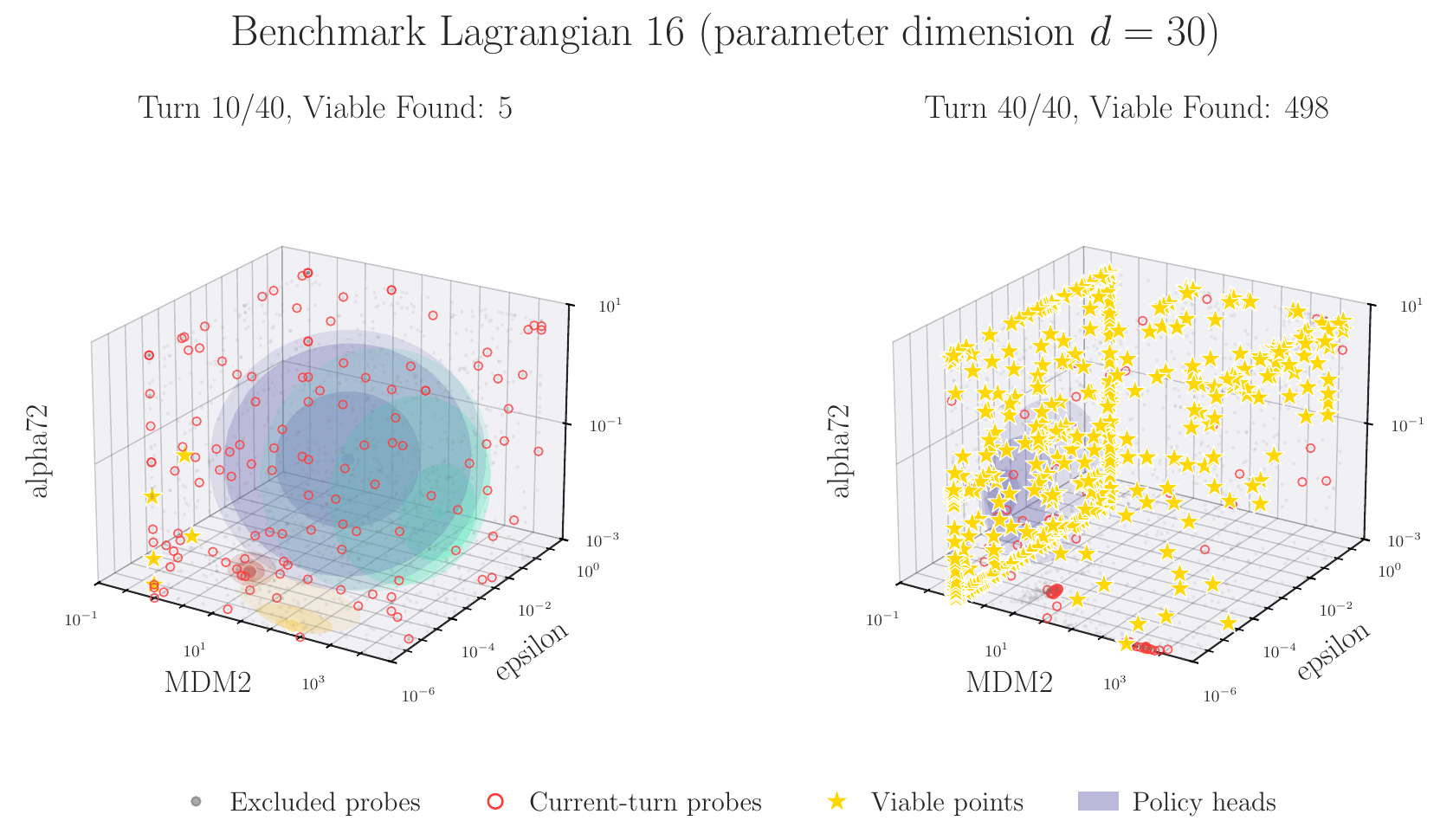} &
      \includegraphics[width=0.48\linewidth]{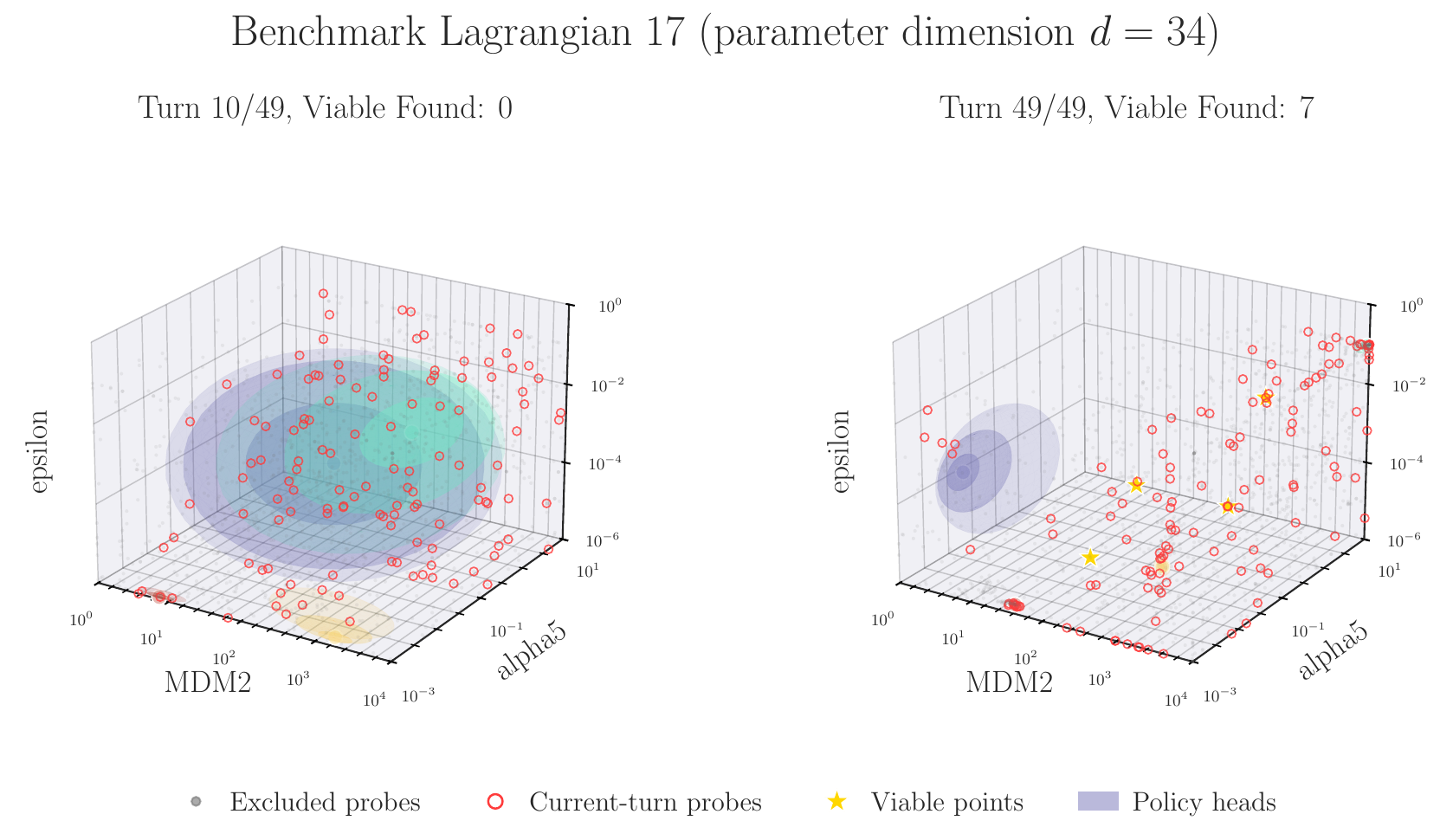}\\[-0.3em]
      \includegraphics[width=0.48\linewidth]{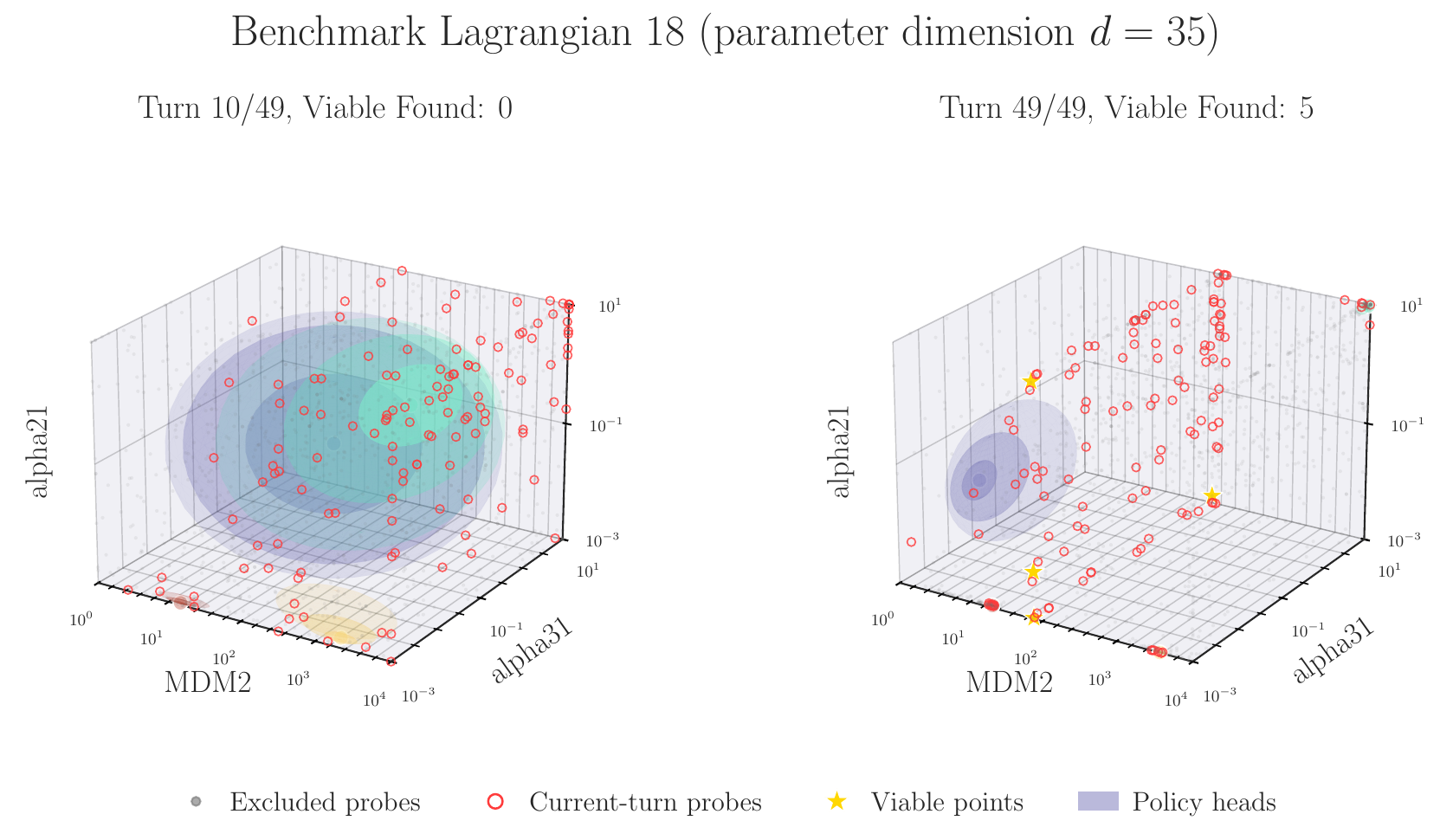} &
      \includegraphics[width=0.48\linewidth]{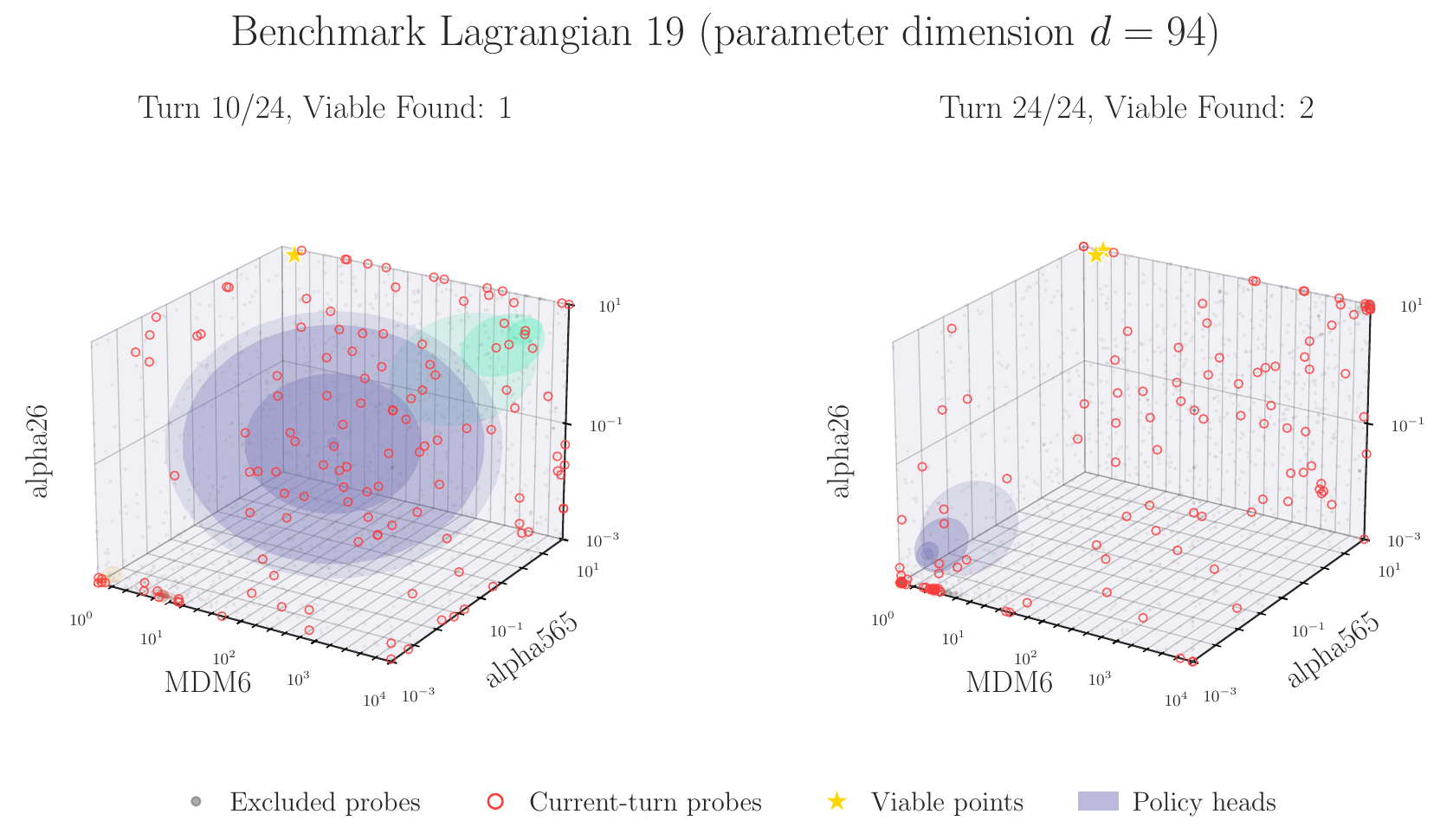}\\[-0.3em]
      \includegraphics[width=0.48\linewidth]{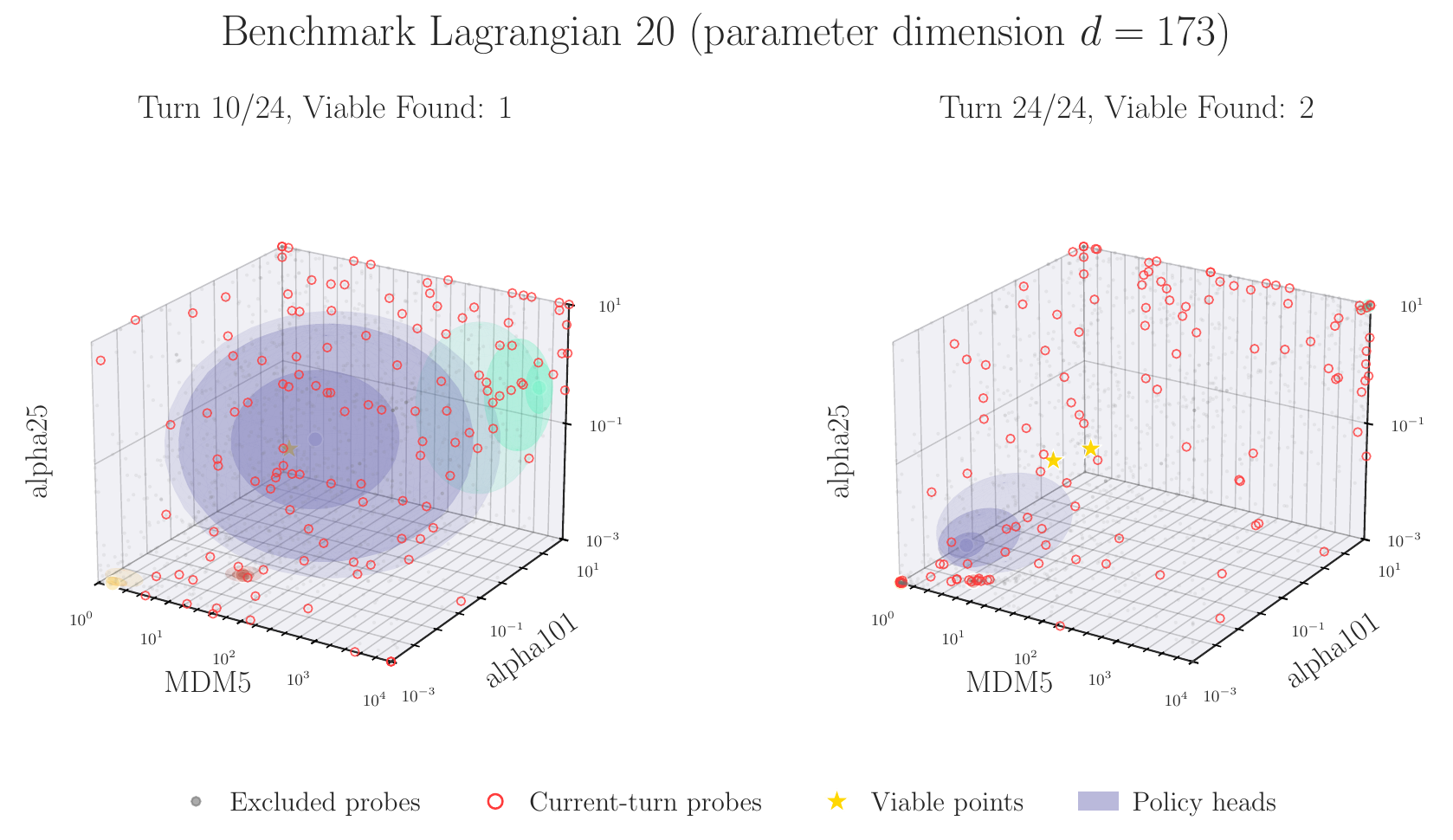} &
      \includegraphics[width=0.48\linewidth]{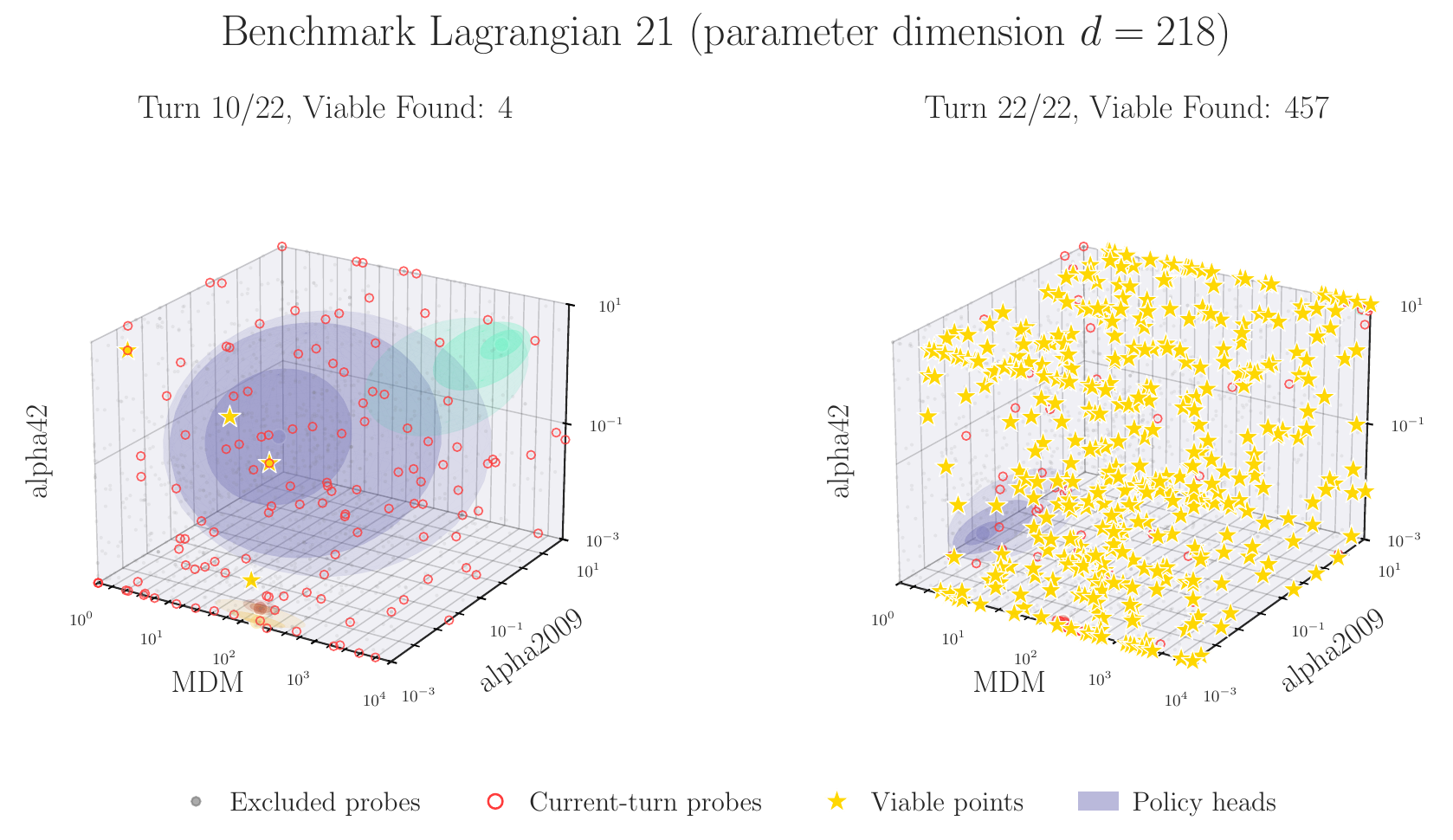}
    \end{tabular}
    \caption{Example LLaM-small-RL search episodes at budget $B=50$ for benchmark Lagrangians (see numbering of Table \ref{tab:benchmark-classes}), shown early in the episode (left) and at the final turn (right). Grey points are excluded probes, red dots are current turn's probes, gold stars are the cumulative viable points found, and the shaded ellipses are the policy's beta distribution heads. The
    search acts in the full $d$-dimensional parameter space. We display the three parameters along which its viable points spread the most. \textbf{Policy heads need not be in the right location when Differential Evolution is chosen by the LLaM.}}
    \label{fig:3d-episodes-2}
\end{figure}
\clearpage
\subsection{Scan Trees}
\label{app:scan-trees}

\begin{figure}[h]
    \centering
    \includegraphics[width=.85\linewidth]{images/results/new-trees/Complex_Scalar_Doublet_CsDh_DM_Z2_3_4_5_tree.pdf}
    \includegraphics[width=.6\linewidth]{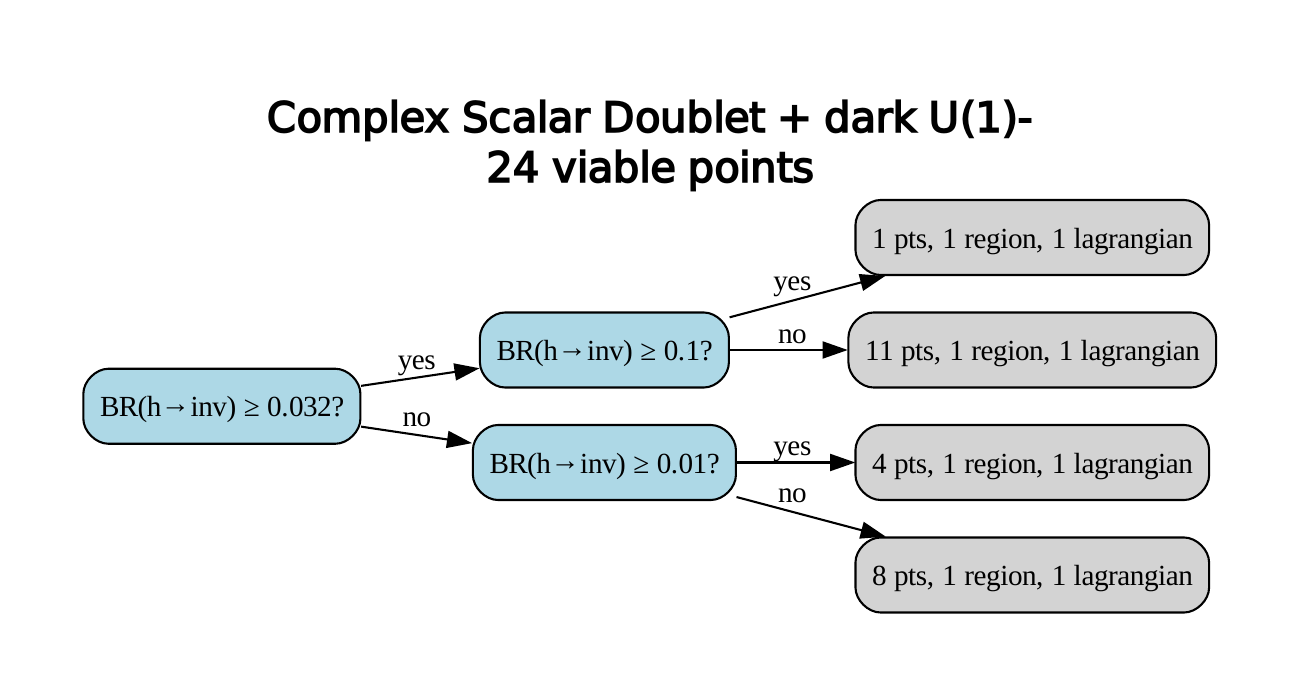}
    \includegraphics[width=.75\linewidth]{images/results/new-trees/Real_Scalar_Singlet_RsSg_DM_Z2_3_4_5_tree.pdf}
    \caption{Three Decision Trees from the One-Multiplet Search Space. None meet the condition for the LLM agent to be applied.}
    \label{fig:tree_threenon-llm}
\end{figure}

\begin{sidewaysfigure}[p]
    \centering
    \includegraphics[width=\linewidth,height=\textwidth,keepaspectratio]{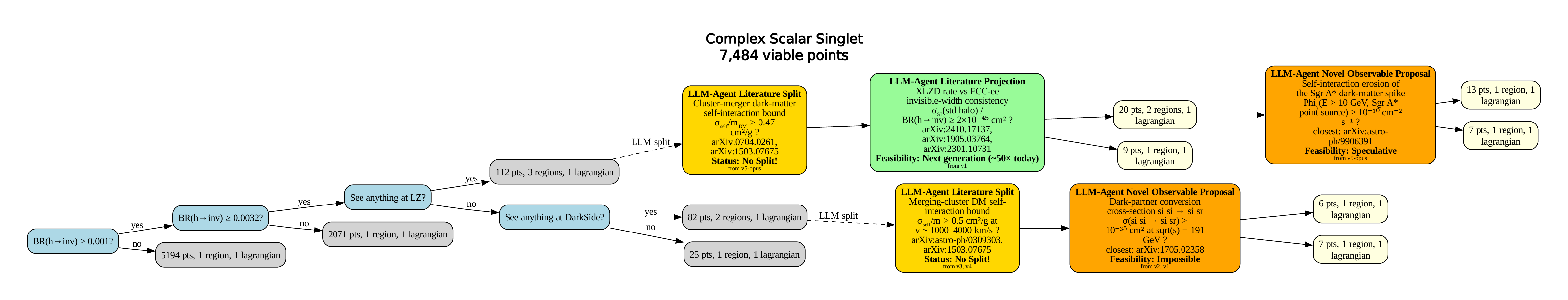}
    \caption{Complex Scalar Singlet Decision Tree.}
    \label{fig:tree_cssg_z2345}
\end{sidewaysfigure}

\begin{sidewaysfigure}[p]
    \centering
    \includegraphics[width=\linewidth,height=\textwidth,keepaspectratio]{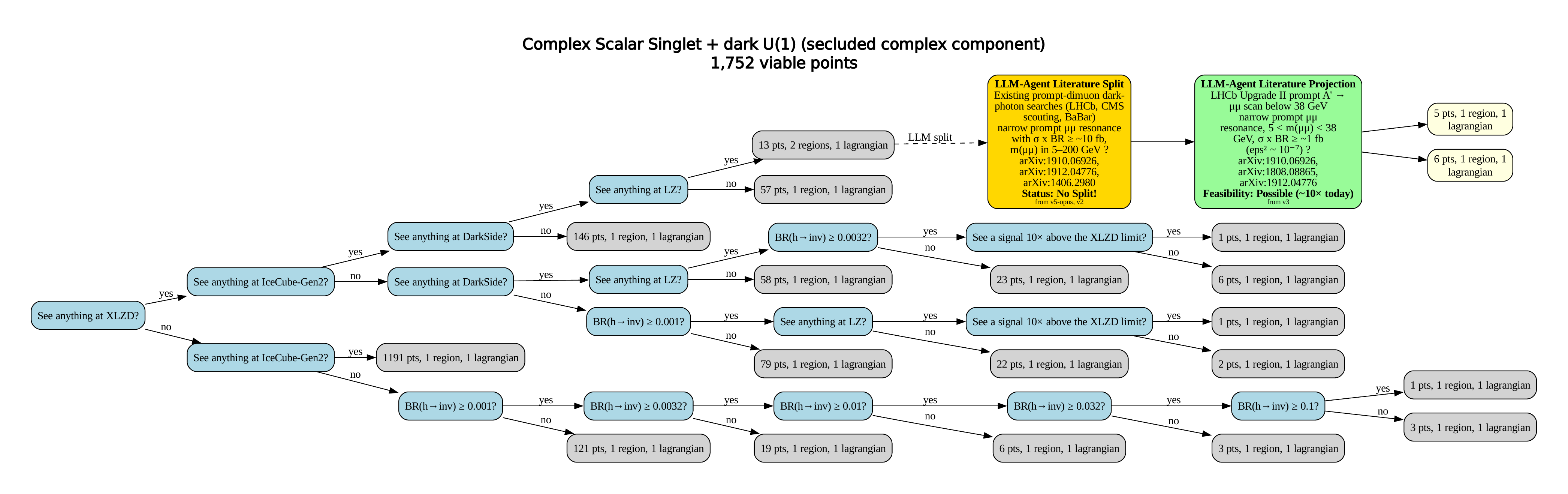}
    \caption{Decision Tree for a Complex Scalar Singlet with Dark $U(1)'$ and a secluded complex component.}
    \label{fig:tree_cssg_u1p_secluded_p_z2}
\end{sidewaysfigure}

\begin{sidewaysfigure}[p]
    \centering
    \includegraphics[width=\linewidth,height=\textwidth,keepaspectratio]{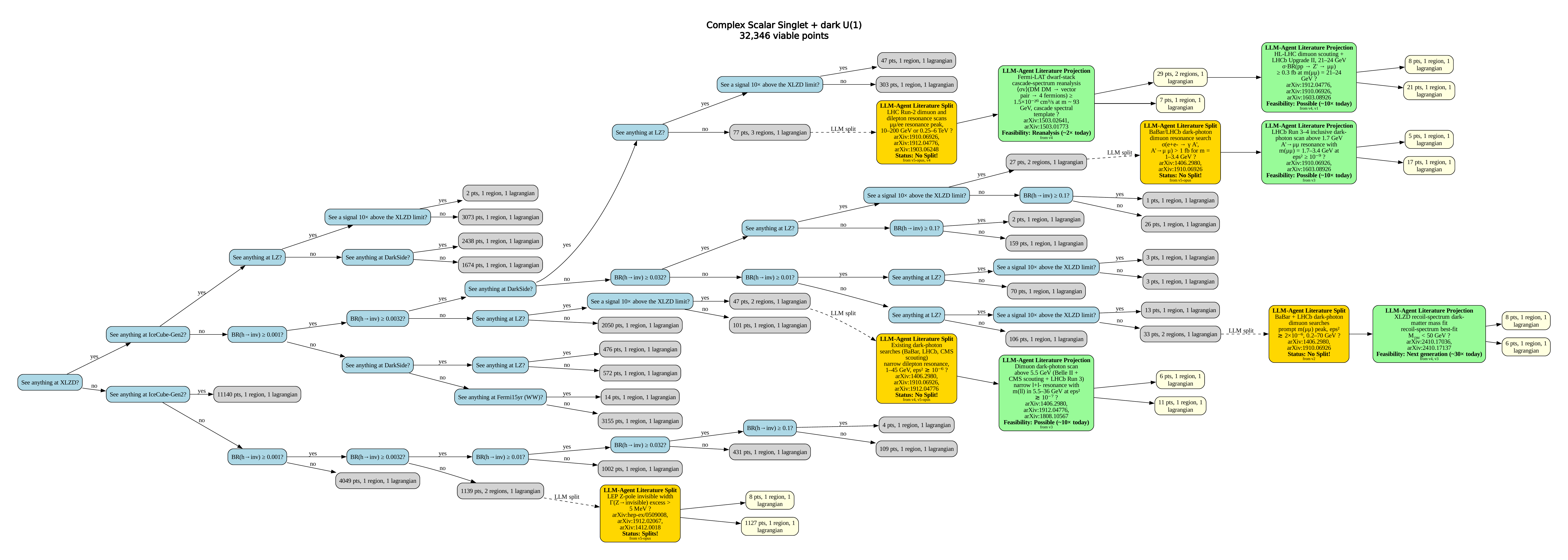}
    \caption{Decision Tree for a Complex Scalar Singlet with Dark $U(1)'$.}
    \label{fig:tree_cssg_u1p_p_z2345}
\end{sidewaysfigure}

Figures \ref{fig:tree_threenon-llm}-\ref{fig:tree_cssg_u1p_p_z2345} are decision trees for all models in our one-scalar multiplet scan for which we find non-zero viable points. The LLM-agent trees are shown when available. The LLM-agent cites \cite{LHCb:2019vmc,CMS:2019buh,BaBar:2014zli,LHCb:2018roe,Ilten:2016tkc,
Belle-II:2018jsg,ALEPH:2005ab,Curtin:2014cca,Fermi-LAT:2015att,Elor:2015tva,
LZ:2024zvo,XLZD:2024nsu,Randall:2008ppe,ATLAS:2023tkt,deBlas:2019rxi,Harvey:2015hha,Markevitch:2003at,Gondolo:1999ef,Tulin:2017ara,ATLAS:2019erb,Janot:2019oyi} throughout the model trees. 

\clearpage

\clearpage

\end{document}